\documentclass[letterpaper,twocolumn,10pt]{article}
\usepackage{usenix-2020-09}

\usepackage[inkscapelatex=false]{svg}

\usepackage{amsmath,amsfonts}
 
\usepackage{amssymb}

\usepackage{pifont} 
\usepackage{enumerate}

\usepackage{tikz}
\usepackage{tkz-euclide}
\tikzset{%
    pics/sema/.style args={#1/#2/#3}{code={%
        \ifstrequal{#2}{0}{%
            \node[circle,minimum width=1mm,draw,fill=#1] {};
        }{%
            \tkzDefPoint(0,0){O}
            \tkzDrawSector[R,fill=#1](O,1mm)(90,90-#2)
            \tkzDrawSector[R,fill=#3](O,1mm)(90-#2,90-360)
    }
    }},
}

\usepackage{xcolor}

\definecolor{armygreen}{rgb}{0.29, 0.33, 0.13}
\definecolor{electricpurple}{rgb}{0.75, 0.0, 1.0}
\definecolor{limegreen}{rgb}{0.75, 1.0, 0.0}
\definecolor{skyblue}{rgb}{0.53, 0.81, 0.98}
\definecolor{goldenrod}{rgb}{0.98, 0.80, 0.20}
\definecolor{babypink}{rgb}{0.96, 0.76, 0.76}
\definecolor{flamingopink}{rgb}{0.99, 0.56, 0.67}
\definecolor{bananamania}{rgb}{0.98, 0.91, 0.71}
\definecolor{cambridgeblue}{rgb}{0.64, 0.76, 0.68}
\definecolor{asparagus}{rgb}{0.53, 0.66, 0.42}
\definecolor{desertsand}{rgb}{0.93, 0.79, 0.69}
\definecolor{tropicalholiday}{HTML}{8ECFC9}
\definecolor{ao(english)}{rgb}{0.0, 0.5, 0.0}
\definecolor{opm1}{HTML}{8152A8}
\definecolor{opm2}{HTML}{C4B8F4}

\usepackage{amssymb}
\usepackage{bbding} 
\newcommand{\cmark}{\textcolor{green!80!black}{\ding{51}}}
\newcommand{\xmark}{\textcolor{red}{\ding{55}}}

\usepackage{makecell}
\usepackage{colortbl} 
\usepackage{adjustbox}

\usepackage{pgf-pie}
\usepackage{graphicx}
\usepackage{subfigure}

\usepackage[switch]{lineno}

\usepackage{xcolor}
\definecolor{babypink}{rgb}{0.96, 0.76, 0.76}
\definecolor{flamingopink}{rgb}{0.99, 0.56, 0.67}
\definecolor{bananamania}{rgb}{0.98, 0.91, 0.71}
\definecolor{cambridgeblue}{rgb}{0.64, 0.76, 0.68}
\definecolor{asparagus}{rgb}{0.53, 0.66, 0.42}
\definecolor{desertsand}{rgb}{0.93, 0.79, 0.69}
\definecolor{tropicalholiday}{HTML}{8ECFC9}
\definecolor{ao(english)}{rgb}{0.0, 0.5, 0.0}

\usepackage{lipsum}
\usepackage{tikz}
\usetikzlibrary{arrows.meta}
\usetikzlibrary{automata,positioning}

\usepackage{threeparttable}
\usepackage{diagbox}

\newenvironment{packeditemize}{
	\begin{list}{$\bullet$}{
			\setlength{\labelwidth}{4pt}
			\setlength{\itemsep}{0pt}
			\setlength{\leftmargin}{\labelwidth}
			\addtolength{\leftmargin}{\labelsep}
			\setlength{\parindent}{0pt}
			\setlength{\listparindent}{\parindent}
			\setlength{\parsep}{0pt}
			\setlength{\topsep}{1pt}}}{\end{list}}

\newenvironment{circitemize}{
	\begin{list}{$\circ$}{
			\setlength{\labelwidth}{4pt}
			\setlength{\itemsep}{0pt}
			\setlength{\leftmargin}{\labelwidth}
			\addtolength{\leftmargin}{\labelsep}
			\setlength{\parindent}{0pt}
			\setlength{\listparindent}{\parindent}
			\setlength{\parsep}{0pt}
			\setlength{\topsep}{1pt}}}{\end{list}}

\renewcommand*{\arraystretch}{1.5}%
\definecolor{tabred}{RGB}{230,36,0}%
\definecolor{tabgreen}{RGB}{0,116,21}%
\definecolor{taborange}{RGB}{250,124,30}%
\definecolor{tabbrown}{RGB}{171,70,0}%
\definecolor{tabyellow}{RGB}{251,253,169}%
\newcommand*{\vcorr}{%
  \vadjust{\vspace{-\dp\csname @arstrutbox\endcsname}}%
  \global\let\vcorr\relax
}%

\usepackage{mathtools}

\usepackage{lscape}
\usepackage[figuresright]{rotating}
\usepackage[graphicx]{realboxes}
\usepackage{adjustbox}
\usepackage{caption}

\usepackage{hyperref}

\usepackage{colortbl} 
\usepackage{xfrac}

\usepackage{appendix}
\usepackage{float}

\usepackage{fbox} 
\usepackage{fancybox}
\usepackage{framed}

\usepackage{multirow}

\def\BibTeX{{\rm B\kern-.05em{\sc i\kern-.025em b}\kern-.08em
    T\kern-.1667em\lower.7ex\hbox{E}\kern-.125emX}}
    
\usepackage{CJKutf8}  
\usepackage{pgfplots}
\usepackage{filecontents}

\usepackage{tabularx,makecell, caption}
\usepackage{enumitem} 

\newcolumntype{L}{>{\arraybackslash}X}

\usepackage{multicol}
\usepackage{blindtext}
\usepackage{etoolbox,xstring,mfirstuc,textcase}

\usepackage{subfiles}
\usepackage[most]{tcolorbox}
\usepackage{arydshln}
\usepackage{tikz}
\usepackage{relsize}

\usepackage{xcolor}
\usepackage{amsthm}
\theoremstyle{plain}       
\newtheorem{thm}{Theorem}

\theoremstyle{definition}

\newtheorem*{prf}{Proof}

\usepackage{siunitx}
\usepackage{comment}
\usepackage{indentfirst} 
\usepackage{framed}

\usepackage[font=small,labelfont=bf,tableposition=top]{caption}
\usepackage{booktabs}
\usepackage{dashbox}

\usepackage{listings}
\usepackage{url}

\usepackage{color}
\usepackage{algorithm,algpseudocode} 

\begin{document}
\title{Split Unlearning\thanks{Accepted by \textcolor{violet}{ACM CCS'25} (\underline{Distinguished Paper Award}).
The peer-reviewed version of this work is published in ACM CCS 2025 and is available at \url{https://dl.acm.org/doi/abs/10.1145/3719027.3744787.}}}

 \author{
 {\rm Guangsheng Yu$^{1,\textcolor{green}{\dag}}$, Yanna Jiang$^{1,}\thanks{Contributed equally to this research.}$\,  Qin Wang$^{1,2}$, Xu Wang$^{1}$, } \\ {\rm Baihe Ma$^{1}$, Caijun Sun$^{3}$, Wei Ni$^1$, Ren Ping Liu$^{1}$} \\
 {\normalsize {$^1$University of Technology Sydney} $|$ $^2$CSIRO Data61 $|$ $^3$Independent} 
 }

\maketitle

\begin{abstract}

We introduce \textit{Split Unlearning}, a novel machine unlearning technology designed for Split Learning (SL), enabling the first-ever implementation of \textbf{S}harded, \textbf{I}solated, \textbf{S}liced, and \textbf{A}ggregated (SISA) unlearning in SL frameworks. 
Particularly, the tight coupling between clients and the server in existing SL frameworks results in frequent bidirectional data flows and iterative training across all clients, violating the ``Isolated'' principle and making them struggle to implement SISA for independent and efficient unlearning.

To address this, we propose \textsc{SplitWiper} with a new \textit{one-way-one-off propagation} scheme, which leverages the inherently ``Sharded'' structure of SL and decouples neural signal propagation between clients and the server, enabling effective SISA unlearning even in scenarios with absent clients.
We further design \textsc{SplitWiper+} to enhance client label privacy, which integrates differential privacy and label expansion strategy to defend the privacy of client labels against the server and other potential adversaries. Experiments across diverse data distributions and tasks demonstrate that \textsc{SplitWiper} achieves \textbf{0\%} accuracy for unlearned labels, and \textbf{8\%} better accuracy for retained labels than non-SISA unlearning in SL. Moreover, the one-way-one-off propagation maintains \textit{constant} overhead, reducing computational and communication costs by \textbf{99\%}. \textsc{SplitWiper+} preserves \textbf{90\%} of label privacy when sharing masked labels with the server.

\end{abstract}


\section{Introduction}
\label{sec:intro}

Split Learning (SL) is one of the latest approaches to distributed learning, suitable for scenarios where clients with limited computational resources, such as smartphones or IoT devices, need to offload part of their training to the servers~\cite{poirot2019split,vepakomma2018split,thapa2022splitfed}. Existing SL frameworks include two configurations: \textit{Vanilla SL} and \textit{U-shaped SL}~\cite{GUPTA20181}. In Vanilla SL, the model is split into two segments: The clients manage the segment receiving the training data and send intermediate values to the server, while the server handles the segment producing the final outputs without accessing raw data. U-shaped SL extends this by eliminating the need for label sharing, protecting both label privacy and data privacy. The model in U-shaped SL is divided into three segments: The clients process the input data and produce output classes, while the server manages the intermediate layers.

\begin{figure}[!t]
    \centering
    \includegraphics[width=1\linewidth]{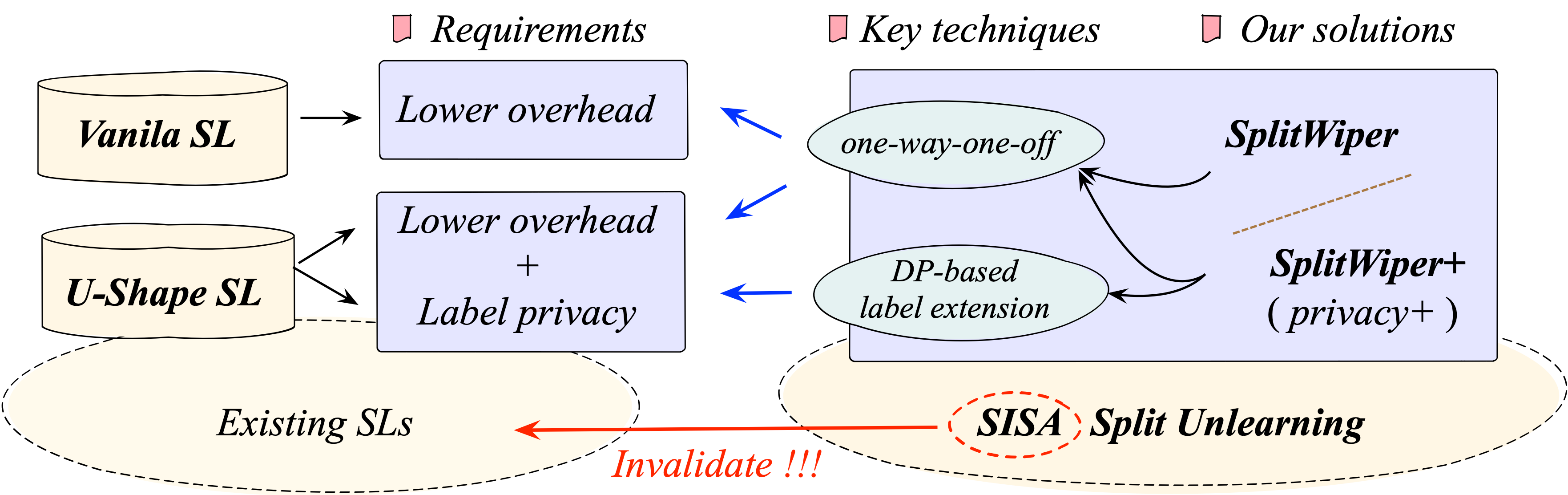}
    \caption{\textbf{Why we need \textsc{SplitWiper}/\textsc{SplitWiper+}?} The ``Isolated'' principle of SISA invalidates existing SLs, hindering efficient and effective split unlearning. Our solution that features a novel one-way-one-off design and DP-based label extension strategy enables SISA unlearning and fulfills the fundamental requirements of SLs.}
\label{fig:motivation}
\vspace{-0.5cm}
\end{figure}

The need for the compliance with regulatory requirements for ``the right to forget'' (e.g., GDPR~\cite{voigt2017eu}), which grants individuals the right to remove their personal data or impact of the data be erased, necessitates the development of effective \textit{machine unlearning}~\cite{sisa} techniques.
However, the requirement becomes challenging within the context of SL, where the model segments across clients and servers need to be jointly updated.
Current methods include retraining the model~\cite{retrain,retrain1} or using alternatives like fine-tuning, gradient ascent, and knowledge distillation~\cite{qu2023learn,nguyen2022survey,distil1Unl}. 
While these alternatives are favored for their higher unlearning \underline{efficiency} (i.e. lower \textit{time consumption}) compared to retraining, which prioritizes \underline{effectiveness} (i.e., \textit{accuracy})~\cite{liu2024fishers}, they all require all data samples during each update iteration. This becomes problematic when some data samples are absent due to contextual constraints. \textbf{S}harded, \textbf{I}solated, \textbf{S}liced, and \textbf{A}ggregated (SISA) methods~\cite{sisa,chen2022graph,chen2022recommendation} address this by segmenting data and training multiple models in parallel, maintaining higher effectiveness and efficiency than non-SISA unlearning methods.

Typical SL solutions~\cite{poirot2019split,vepakomma2018split,thapa2022splitfed,zhang2023privacy} are inadequate for directly applying SISA unlearning in multi-client scenarios, where multiple clients have independent models while sharing a common model on the server. Although clients' training data can be considered ``Sharded'', SL undermines the ``Isolated'' requirement of SISA because: \textit{(1) the server model is tightly coupled with the client models through frequent bidirectional propagation, and (2) the shared server model is iteratively trained across all clients.}
Thus, a new framework is needed to decouple the shared server model from the client models, enabling SISA unlearning in multi-client SL environments.

\smallskip
\noindent\textbf{Contributions.}
In this paper, we address these gaps in enabling SISA unlearning in multi-client SL environments where unlearning can thus be conducted smoothly even with absent clients (cf. Fig.~\ref{fig:motivation}). 

\smallskip
\noindent$\triangleright \quad$
\underline {We introduce \textsc{SplitWiper}} (Sec.\ref{sec:sisa_design}), a new SL framework that implements the SISA guideline, achieving SISA unlearning capabilities: (i) \textit{selective involvement of only clients who request unlearning} and (ii) \textit{effective unlearning while maintaining comparable utility for retained data}.
\textsc{SplitWiper} features an efficient and streamlined interaction between clients and the server by incorporating a new low-effort pre-training phase on clients, after which the clients' weights are frozen (one-off). 
It allows iterative updates to focus only on the server's weights without requiring feedback loops or updates to be returned to clients (one-way).
This novel \textit{one-way-one-off} protocol architecture for propagating neural signals from clients to the server forms the foundation for enabling SISA unlearning in \textsc{SplitWiper}.
Our solution removes the requirement for all clients to engage in unlearning processes by \textbf{I}solating training on each client, allowing for the formation of distinct \textbf{S}hards or \textbf{S}lices of data that are \textbf{A}ggregated exclusively on the server.

\noindent$\triangleright \quad$
\underline {We further introduce \textsc{SplitWiper+}} (Sec.\ref{sec:label-protection}), a privacy-enhancing variation of \textsc{SplitWiper}, to address the privacy concerns associated with label sharing during the learning and unlearning tasks. 
The ``Isolated'' principle of SISA invalidates traditional privacy-preserving methods, such as non-label-sharing U-shaped SL~\cite{vepakomma2018split}.
\textsc{SplitWiper+} incorporates a novel \textit{label expansion} strategy, marking the first approach to mask the quantity of training labels and their semantics. 
By expanding, shuffling, and anonymizing real labels, while protecting the outputs of clients' models with differential privacy (DP), this strategy prevents adversaries from exploiting the exact number of labels and launching effective attacks with supervised learning techniques.
This innovative design effectively mitigates privacy leakage when clients' labels are exposed to the server during propagation.

\noindent$\triangleright \quad$
\underline {We conduct extensive experiments} (Sec.\ref{sec:evaluation}) in various settings where the clients either share the same training task with various data distributions or engage in different tasks, exploring the framework's adaptability and performance across a range of scenarios.
The results show that \textsc{SplitWiper} achieves complete unlearning accuracy (\textbf{0\%}) while improving retained accuracy by \textbf{8\%}, engaging only the clients who request unlearning to minimize involvement and reduce overhead. \textsc{SplitWiper} maintains \textit{constant} overhead, cutting off computational and communication costs by over \textbf{99\%} compared to existing SL frameworks. Additionally, \textsc{SplitWiper+} preserves over \textbf{90\%} of client label privacy from the server.


\section{Related Work}
\label{sec:related_work}
 
\noindent\textbf{Unlearning in distributed environments.} Machine unlearning has not been fully explored within current distributed learning frameworks, including SL. While efficient unlearning strategies have gained attention in Federated Learning (FL), which also involves collaboration between clients and a central server.
Some researchers have worked on accelerating retraining across all clients, the most robust unlearning method in FL, using optimized algorithms like distributed Newton-type updates~\cite{9796721}, SGA-EWC~\cite{9964015}, and SFU~\cite{li2023subspace}. Solutions like FedEraser~\cite{liu2021federaser} reduce the impact on non-unlearning-involved clients through calibration but still require cooperation from some clients. However, in SL, where client resources are often limited, non-requesting clients may lack the incentive or ability to contribute, especially if they are uninterested in or unable to participate in the unlearning process due to network issues, such as low bandwidth, high latency, or intermittent connectivity. This presents a significant challenge in SL compared to other distributed environments like FL, where developing unlearning methods that accommodate these constraints remains an unresolved issue.

 \smallskip
\noindent\textbf{SISA guideline for SL unlearning.} SISA~\cite{sisa} is a state-of-the-art guideline for robust and efficient machine unlearning, enabling rapid adaptation to data removal requests without requiring full model retraining. By strategically partitioning data into distinct ``shards'' linked to different model versions, the SISA guideline allows for the targeted unlearning of specific data shards without requiring the participation of all shards.
This method has been successfully applied in various domains, including unlearning in graph neural networks~\cite{chen2022graph} and large language models~\cite{chen2023unlearn}.
However, SISA unlearning faces significant challenges in multi-client SL scenarios due to the interdependence between client model segments and the shared server model segment, violating the Isolated requirement of SISA.
Consequently, the direct application of conventional SISA is impractical for existing SL scenarios.

\smallskip
\noindent\textbf{Label privacy under SL unlearning.}
Beyond unlearning, privacy concerns regarding intermediate outputs and labels in SL have been widely studied. U-shaped SL~\cite{GUPTA20181} avoids label propagation to the server by placing the output layer on the client side but violates SISA's Isolated requirement by necessitating the return of intermediate values across all clients, disabling the independent unlearning and adding overhead to resource-limited clients.
If labels are shared but privacy is required, existing research primarily addresses privacy risks by directly adding noise and obfuscating labels~\cite{ghazi2021deep, gao2024label} or further protecting intermediate outputs transmitted to the server.
Wu et al.~\cite{wu2023federated} introduced a DP mechanism to perturb these transmitted outputs, while Li et al.~\cite{li2021label} developed the Marvell method to quantify privacy leakage and optimize noise structures. However, these methods overlook the exposure of sensitive information through label quantity and semantics. Xiao et al.~\cite{xiao2021mixing} obfuscated both labels and outputs via multiple linear combinations to reduce association with original data but still failed to conceal label quantity, potentially exposing sensitive information. 
Protecting the quantity of labels is critical in label-sharing SL, as revealing the exact number of labels allows adversaries to exploit supervised learning techniques that achieve higher accuracy by relying on this information, significantly increasing the risk of inference attacks and unauthorized model reconstruction.

\smallskip
\noindent\textbf{Ours.} Our design overcomes those challenges (Table~\ref{tab_label_privacy}). \textsc{SplitWiper} adapts SISA for multi-client SL, starting with low-effort pre-training on the clients followed by freezing their weights. This allows for one-way-one-off propagation to the server, enabling efficient unlearning without involving non-unlearning-participating clients. Additionally, \textsc{SplitWiper+} enhances label privacy by expanding, shuffling, and anonymizing real labels, with intermediate outputs expanded and protected using a DP-based mechanism.

\begin{table}[t]
\centering
\caption{
\textbf{Comparison of existing label privacy methods.}
Our work provides a practical privacy-preserving solution beyond existing methods by protecting label semantics, label quantity, and intermediate values while enabling effective SISA split unlearning.}
\label{tab_label_privacy}
\renewcommand{\arraystretch}{1.3} 
\resizebox{\linewidth}{!}{
\begin{tabular}{|c|cccc|}

 \diagbox{\textbf{Method}}{\textbf{Protection}}  & \textbf{\makecell{Label \\ Semantics}} & \textbf{\makecell{Label \\ Quantity}} & \textbf{\makecell{Intermediate \\ Values}} & \textbf{\makecell{Fit for SISA \\ Split Unlearning}}  \\
\midrule

\cellcolor{blue!10}  RRWithPrior~\cite{ghazi2021deep}   &   \cmark &  \xmark & \xmark  & \xmark \\

\cellcolor{blue!10}  LPSC~\cite{gao2024label}   &   \cmark &  \xmark & \xmark  & \xmark \\

\cellcolor{blue!10}  DP-based Defense~\cite{wu2023federated}   &   \xmark &  \xmark & \cmark  & \xmark \\

\cellcolor{blue!10}  Marvell~\cite{li2021label}   &  \xmark &  \xmark & \cmark  & \xmark \\

\cellcolor{blue!10}  MALM~\cite{xiao2021mixing}   &   \cmark & \xmark & \cmark  & \xmark \\

\cellcolor{blue!10}  U-shaped SL~\cite{GUPTA20181}   &   \cmark &  \cmark & \xmark  & \xmark \\

\midrule

\textbf{Ours}   &   \cmark &  \cmark &  \cmark  & \cmark \\


\end{tabular}
}
\vspace{-0.1in}
\end{table}


\section{Preliminaries}
\label{sec:preliminary}

We outline the key background knowledge relevant to SL and machine unlearning.
By reviewing existing techniques and identifying their limitations, we set the stage for the challenges addressed in our proposed frameworks.

\subsection{Split Learning}

SL is a distributed ML technique designed to train neural networks without sharing raw data between $K$ clients, each owning datasets {$D_{o}^{1}, D_{o}^{2}, \ldots, D_{o}^{K}$}, and the server that holds no data~\cite{GUPTA20181,singh2019detailed,wu2023split}. A typical SL solution involves two key factors:

\begin{packeditemize}

\item\textbf{Topology.}
A network is divided into two parts, referred to as \textit{client}-side and \textit{server}-side. Consider the network is a sequence of layers $\{L_1,\ldots, L_n\}$. The topology step involves splitting this sequence into two subsets: 
The client-side layers for $K$ clients are denoted as $\{\{L_1^1, \ldots, L_m^1\}$,  $\ldots$, $\{L_1^K, \ldots, L_m^K\}\}$
and the server-side layers $\{L_{m+1}, \ldots, L_n\}$. The clients and server initialize their respective portions of the network randomly.

\item\textbf{Training.} 
Each client $k$ processes its local data through the initial layers of the model, producing an intermediate output $H_k^{(t)} = \mathcal{F}_o^k(X_k; W_{k}^{(t)})$ through its layers $\{L_1^k, \ldots, L_m^k\}$, where $W_{k}^{(t)}$ represents the weights of client $k$ at iteration $t$. This output is sent to the server, which continues the forward propagation through its layers $\{L_{m+1}, \ldots, L_n\}$ and updates its own weights $W_s^{(t)}$ based on the gradients: $W_s^{(t+1)} = W_s^{(t)} - \eta \Phi\left(\frac{\partial \mathcal{L}_1}{\partial W_s^{(t)}}, \frac{\partial \mathcal{L}_2}{\partial W_s^{(t)}}, \dots, \frac{\partial \mathcal{L}_K}{\partial W_s^{(t)}}\right)
$, where $\frac{\partial \mathcal{L}_k}{\partial W_s^{(t)}}$ is the gradient contribution of the loss values from client $k$ and $\Phi(\cdot)$ is a customized aggregation function. The server then sends the gradients to the clients, allowing them to update their weights: $W_{k}^{(t+1)} = W_{k}^{(t)} - \eta \frac{\partial \mathcal{L}}{\partial W_{k}^{(t)}}$.

\end{packeditemize}

The above processes enable training without sharing raw data or model details between clients and the server. In multi-client, single-server scenarios, two primary training methods are used: synchronizing model weights between clients after each epoch or alternating training epochs between clients and the server without synchronization~\cite{GUPTA20181}.

\smallskip
\noindent\textbf{Incremental training process.}
This process has been utilized by many learning frameworks, including FL~\cite{tan2022federated, nguyen2021federated} and SL, where a model is continuously updated by gradually incorporating new data over time. Each iteration of back-propagation during model training needs to be sent back to every client involved in the training process and builds incrementally on the previous ones, embodying the incremental nature of SL~\cite{9562559}. This process enables the model to dynamically learn and adapt as new data becomes available.

The challenges of applying SISA in multi-client SL scenarios stem from the incremental training process inherent in these setups.
This iterative process undermines the isolated requirement of SISA due to the tight coupling between the server and client model segments through regular bidirectional propagation and the shared server model being incrementally trained across all clients.
The knowledge to be unlearned, reflected in $W_{k}^{(t)}$, would have also influenced $W_{k'}^{(t')}$ for all $k' \neq k$ and $t' \geq t$, necessitating the recalculation of corresponding changes in subsequent updates, thereby making it impossible to apply SISA unlearning~\cite{liu2024decentralized, liu2024fishers}.

\subsection{Machine Unlearning}\label{subsec: machine_unlearning}

Machine unlearning~\cite{shaik2024exploring,liu2024survey,nguyen2022survey,qu2023learn} involves removing the influence of specific data subsets from a trained model while preserving its performance on the remaining data. The goal is to eliminate the impact of certain labels on the model's predictions without compromising accuracy on the retained data.
Let $D = \{ \left( X_{i}, Y_{q} \right) \}_{i=1,q=1}$ represent the dataset, where $X_{i}$ is the $i$-th training sample and $Y_{q} \in \{ 1, \cdots, Q \}$ is its label. Unlearning removes the influence of subset $D_u \subset D$ with labels $Y_u$ while preserving performance on the remaining data $D_r = D \backslash D_u$ with labels $Y_o$, ensuring $Y_u$ is forgotten and predictions $Y_o$ remain accurate.

\smallskip
\noindent\textbf{Retraining.} The most direct and effective, though not efficient, method of implementing unlearning is retraining~\cite{retrain,retrain1}. In this method, the revoked sample is deleted, and the model is retrained on the original dataset minus the deleted sample, denoted as $D_r = D \setminus {((X_i, Y_q)} \subset D_u)$. While this method is effective and straightforward, the computational overhead becomes prohibitive for complex models and large datasets.

\smallskip
\noindent\textbf{Alternatives.}
Referring to fine-tuning, gradient ascent, and knowledge distillation, 
These unlearning methods are proposed to improve unlearning efficiency compared to retraining, but they still consider retraining as the baseline for unlearning effectiveness~\cite{qu2023learn,nguyen2022survey,distil1Unl}, including:
\begin{packeditemize}
    \item \textit{Fine-tuning.} Adjusting the model parameters by further training on $D_r$ to deliberately induce catastrophic forgetting of $D_u$, typically using \( W^{(t+1)} = W^{(t)} - \eta \nabla \mathcal{L}_{r}(W^{(t)}) \), where \( \mathcal{L}_{r} \) is the loss function computed on the retained data.

    \item \textit{Gradient ascent.} Updating the weights in the direction of the gradient of a loss function to maximize the objective: $W^{(t+1)} = W^{(t)} + \eta \|\nabla \mathcal{L}(W^{(t)}) - \nabla \mathcal{L}_r(W^{(t)})\|$.

    \item \textit{Knowledge distillation.} Trading-off between a preserver that aids to retain knowledge and a destroyer that aids in forgetting knowledge: $\mathcal{L}=\alpha \mathcal{L}_u + (1-\alpha) \mathcal{L}_r$.
\end{packeditemize}

\smallskip
\noindent\textbf{SISA.}
SISA~\cite{sisa} is a guideline designed to enable efficient and independent unlearning for those who request it, without affecting other participants.
In SISA, the original training set $D_{o}$ is divided into $K$ disjoint shards \{$D_{o}^{1}, D_{o}^{2}, \ldots, D_{o}^{K}$\}, each used to train a separate model \{$\mathcal{F}_{o}^{1}, \mathcal{F}_{o}^{2}, \ldots, \mathcal{F}_{o}^{K}$\}. These models generate individual predictions, which are then aggregated to produce a global prediction. When an unlearning request is made, only the shard containing the specific data sample needs to be retrained, allowing the unlearning process to be conducted independently for the requesting party, without impacting other clients or the overall system.

\smallskip
\noindent\textbf{Why SISA split unlearning.}
In multi-client SL, the ability to unlearn data despite client absence due to network issues is crucial. Existing frameworks rely on non-SISA unlearning methods, which require retraining or similar processes from the unlearning request point. While these methods prioritize either effectiveness or efficiency, SISA enables independent unlearning for requesting clients without burdening resource-limited devices in the SL network.
SISA is seamlessly compatible, allowing prevalent unlearning methods to be applied individually on each client\footnote{For simplicity, we assmue retraining being used on each client/shard.}, making it the most practical solution to address limitations in non-SISA methods.


\section{SISA Split Unlearning}
\label{sec:sisa_design}

We define the problem and establish clear objectives for split unlearning.
We present \textsc{SplitWiper}, a SISA-based framework designed to address existing limitations, offering a feasible and scalable solution for split unlearning.

\subsection{Problem Definition}

In the context of SL, the training dataset $D_o$ is distributed among $K$ clients and a server, functioning as a training service provider. Throughout this paper, we assume a horizontally non-IID data distribution, where each client possesses a subset of $D_o$ characterized by distinct feature distributions and imbalanced label distributions.

\smallskip
\noindent\textbf{Multi-client settings.}
We focus on multi-client scenarios.
For a trained model $\mathcal{F}_o$, we can divide it into a client-side part $\mathcal{F}_o^k$ for client $k$, and a server-side part $\mathcal{F}_o^s$. Each client $k$ owns its $\mathcal{F}_o^k$ in parallel and shares a single $\mathcal{F}_o^s$. We define the evaluation of model performance from the perspective of client $k$ as $\mathcal{T}(Concat(\mathcal{F}_o^k, \mathcal{F}_o^s), D_o^k)$, where $\mathcal{T}$ represents the evaluation function. Each client functions as an individual user managing its local data, aiming to enhance the prediction accuracy of its pre-trained local model by leveraging the server’s computational capabilities with the shared server-side model segment. Consequently, the performance assessment focuses on the server's perspective, which aggregates the independent evaluations, $\mathcal{T}(Concat(\mathcal{F}_o^k, \mathcal{F}_o^s), D_o^k),\forall k$.

 \smallskip
\noindent\textbf{Independent unlearning.} 
Accordingly, any client $k$ that raises an unlearning request should be conducted only between this particular client and the server, with no needs of other clients $k'$ $(k' \neq k)$ being involved in the unlearning process. This means the server should unlearn the unlearned features of client $k$ from the server-side part $\mathcal{F}_o^s$ upon the intermediate outputs of the updated client-side part $\mathcal{F}_o^k$. Formally, to respond to the unlearning request for a series of data samples $D_u^k$ raised by client $k$, the entire SL network needs to satisfy an unlearned model $\mathcal{F}_u$ trained on $D_u=D_o\setminus\lbrace D_u^k \rbrace$.

 \smallskip
\noindent\textbf{Challenges of SISA split unlearning.}
The split unlearning process faces a conflict between achieving independent unlearning for each client and maintaining overall effectiveness, which we term the \textit{Dilemma of Independence and Effectiveness}, particularly when clients operate independently with limited training resources.
Non-SISA methods, depending on full client participation to update shared models, are vulnerable to client absence during unlearning and impose significant overhead on both the initiating client $k$ and other clients $k'$ ($k' \neq k$). 
This method forces disinterested clients $k'$ to expend resources solely to meet the needs of client $k$.

\begin{figure*}[!htbp]
\setlength\belowcaptionskip{-0.2cm}
        \centering
        \includegraphics[width=1\linewidth]{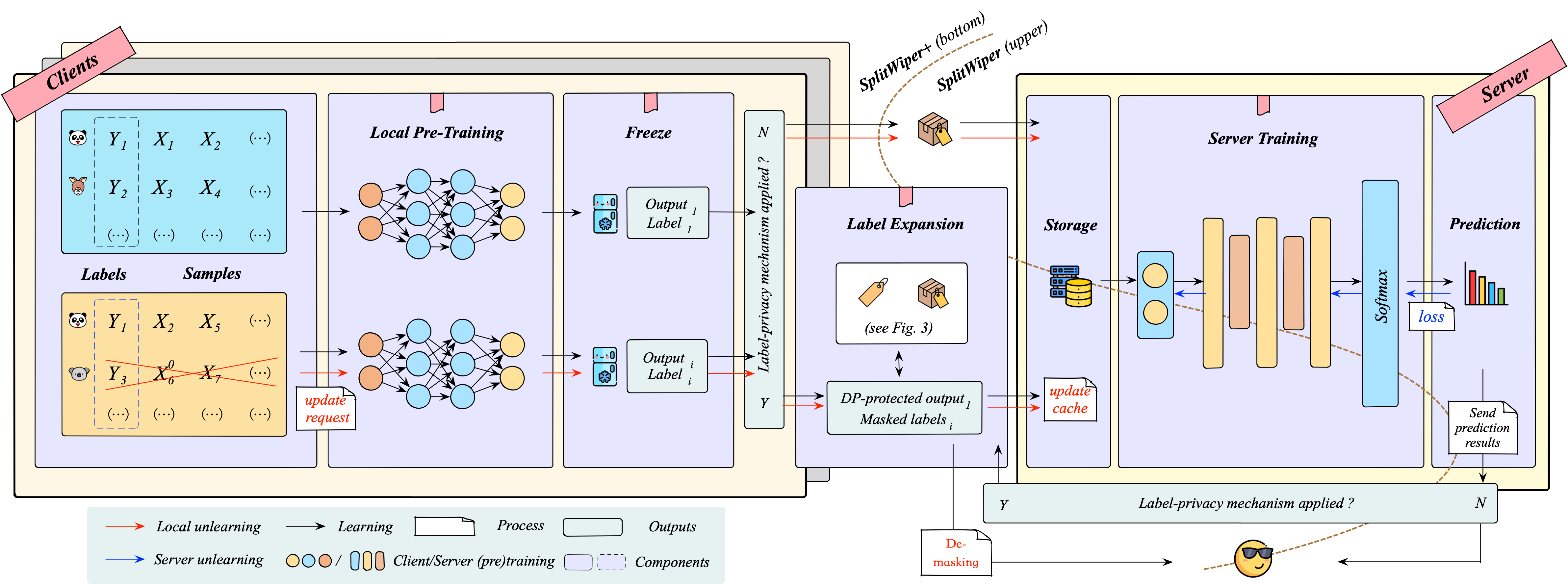}
        \caption{\textbf{Architectural Design:}
        In \textsc{SplitWiper} (Sec.\ref{sec:sisa_design}), each client $k$ trains its local model $\mathcal{F}_o^k$ on its dataset $D_o^k$, treating them as SISA shards without additional partitioning. After training, clients freeze the weights and send outputs with labels to the server for caching (one-off), while the server continues training $\mathcal{F}_o^s$ using stored values without returning to the clients (one-way).
        \textsc{SplitWiper+} (Sec.\ref{sec:label-protection}) enhances privacy by using a label expansion strategy to convert real labels into masked ones, which clients then share with the server for label-protected training.}
        \label{fig_framework}
\end{figure*}

The pursuit of SISA split unlearning is to devise an unlearning algorithm fulfilling the following goals:

\begin{packeditemize}

    \item \textbf{\textcolor{violet}{G1}: Independent unlearning.} 
    This goal ensures that unlearning affects only the requesting client and server, leaving other clients ($k’ \neq k$) and their resources unaffected.
    
\smallskip
    \item \textbf{\textcolor{violet}{G2}: Effective unlearning with retained utility.}
    This is to ensure that an unlearning process thoroughly removes the influence of the unlearned data from the model, while also maintaining the predictive accuracy for the retained data. 

\end{packeditemize}



\subsection{\textsc{SplitWiper}}\label{subsec:framework}

We propose \textsc{SplitWiper} with a novel one-way-one-off propagation scheme, enabling SISA split unlearning to achieve \textcolor{violet}{\textbf{G1}} and \textcolor{violet}{\textbf{G2}}. 
The \textbf{S}harded training of  \textsc{SplitWiper} involves multiple clients, each acting as a distinct shard with different data samples. 
Our one-way-one-off propagation ensures \textbf{I}solated training by eliminating backward gradient propagation (one-way) and sending each client’s intermediate output to the server only once (one-off). 
This confines the impact of each shard to its corresponding client, with no information shared between different clients' models. Within each client, data can be further divided into \textbf{S}lices by embedding SISA. On the server side, \textbf{A}ggregation combines stored intermediate outputs from clients, refining the server’s model and enabling efficient unlearning across multi-client SL.

\subsubsection{Learning in \textsc{SplitWiper}}
The \textsc{SplitWiper} framework (cf. Fig.\ref{fig_framework}) consists of the following three learning phases (cf. Algorithm~\ref{algo:sisa-sl} in \textbf{Appendix}~\ref{appendix_algo}): client model \textit{pre-training}, client model \textit{freezing and caching}, and server \textit{model training}. 


\smallskip
\noindent\textbf{Client model pre-training} (Line~\ref{alg_1:pre-training_start} - Line~\ref{alg_1:pre-training_end}).
The concept of SISA is applied in this phase in a sense that the non-IID distributed datasets \{$D_o^{1}, D_o^{2}, \ldots, D_o^{k}$\} owned by any client $k$ are inherently treated as the shards of SISA. 
By default, \textsc{SplitWiper} does not involve shard partitioning, as the server lacks access to any client's dataset and only provides training services for outsourcing tasks. Consequently, each client $k$ pre-trains its local model $\mathcal{F}_o^k$ on its own dataset $D_o^k$ independently and in parallel. This setup allows clients to use entirely different datasets and training tasks, as long as the dimensions of their outputs match the server's input requirements to ensure smooth integration and communication.


\noindent\textbf{Client model freezing and caching} (Line~\ref{alg_1:freezing}).
The new one-way-one-off propagation is crucial by freezing clients' weights and caching them on the server.
After client model training, the outputs of the last layer (the cut or intermediate layer) and corresponding labels are shared with the server for caching, while weights are frozen to prevent further computation and communication during server training. 
In \textsc{SplitWiper}, mechanisms such as dropout layers or batch normalization are managed solely by the server, ensuring client model determinism. 
This allows for consistent sharing of intermediate outputs, which the server can store without concerns about client model variability.

\noindent\textbf{Server model training}  (Line~\ref{alg_1:training_start} - Line~\ref{alg_1:training_end}).
As the central server that hosts the final output layer of the entire network, the server’s training process for its model, $\mathcal{F}_o^s$, functions similarly to the aggregation function $\Phi(\cdot)$ in SISA. By receiving and storing intermediate outputs from each client, the server accelerates the training of its model, which is typically much larger than the client models. With the client-side weights frozen, gradients are no longer back-propagated to the clients, restricting the incremental training process to the server. 
This ensures that subsequent updates do not propagate to all clients, implementing the one-way-one-off propagation scheme. 
This method lays a foundation for enabling unlearning with absent clients while significantly reducing overhead.



\subsubsection{Unlearning in \textsc{SplitWiper}}


Our \textbf{\textcolor{violet}{G1}} ensures that only the clients requesting unlearning need to participate, while others remain uninvolved.
We give below the workflow of SISA split unlearning in \textsc{SplitWiper} (cf. Algorithm~\ref{algo:sul-labels} in \textbf{Appendix}~\ref{appendix_algo})
The algorithm takes as input the number of epochs for client model updating $ N $, the number of epochs for server model updating $ M $, and the initiating client $k$ who proposes the requirement for unlearning, and outputs the updated predicted label on the $Concat(\mathcal{F}_u^k, \mathcal{F}_u^s)$. 
It works in three steps as follows:

\begin{packeditemize}

\item \textbf{Step 1}: \textbf{Client weight unfreezing and dataset modification} (Line \ref{alg_2:unfreezing_start} - Line \ref{alg_2:unfreezing_end}).
In this step, client $k$ that initiates the retraining request unfreezes its weights and removes the intended unlearned samples from its dataset, $D_r^k=D_o^k\setminus\lbrace D_u^k \rbrace$. 
This ensures the local model is prepared for unlearning by excluding any influence from the unlearned samples.

\item \textbf{Step 2}: \textbf{Client model unlearning and freezing} (Line \ref{alg_2:unlearning_start} - Line \ref{alg_2:unlearning_end}).
The initiating client $k$ retrains its model for $ N $ epochs and then freezes the updated weights. 
Subsequently, it sends intermediate outputs and corresponding labels to the server for further training on the server.
This step removes residual traces of the unlearned samples from the local model of client $k$ and ensures that the outputs provided to the server reflect the updated model state.

\item \textbf{Step 3}: \textbf{Server model updating} (Line \ref{alg_2:updating_start} - Line \ref{alg_2:updating_end}).
The server updates its storage with the intermediate outputs of client $k$, trains the server model for $M$ epochs using the stored values, and then obtains the updated output distribution.
This step integrates the updates of client $k$ into the server model, ensuring consistent unlearning while maintaining utility for unaffected data.

\end{packeditemize}

Given that \textbf{\textcolor{violet}{G2}} (effective unlearning with retained utility) is a fundamental requirement for any viable unlearning method, this approach also satisfies \textbf{\textcolor{violet}{G1}} (independent unlearning). Clients freeze their weights and transfer intermediate values to the server, where they are securely stored. These values are selectively unfrozen only when necessary to perform the unlearning process, ensuring independent executions without involving or disrupting other clients.



\section{\textsc{SplitWiper+}: Privacy-Preserving Variation via Label Expansion}
\label{sec:label-protection}



The one-way-one-off propagation mechanism halts learning propagation at the server side and freezes the clients' parameters after their outputs are stored for the first time. 
While this design ensures efficiency and privacy for the shared data, further advancements in label privacy can be explored, particularly in scenarios where labels cannot be directly shared with the server, as seen in U-shaped SL, which is incompatible with the one-way-one-off propagation mechanism and therefore unable to support SISA split unlearning.
So, this further introduces a key challenge if preserving labels private is a strict requirement:

\textit{How to keep labels private while still enabling their sharing with the server for the unlearning process, without compromising the integrity of \textbf{\textcolor{violet}{G1}} and \textbf{\textcolor{violet}{G2}}?}

\smallskip
\noindent\textbf{Player model.}
Two types of players are considered in this label-privacy-preserving scenario.


\begin{packeditemize}
    \item \textbf{Internal users.} This includes clients ($\mathbb{C}_{inter}$) and servers ($\mathbb{S}_{inter}$) involved in training, modeled as \textit{honest-but-curious}~\cite{li2021label}. They adhere to the protocol without tampering, such as sending incorrect features or causing failures. Each client ${C}_{inter} \in \mathbb{C}_{inter}$ knows the shared labels among participants but keeps their complete label set $\mathbb{Y}_{inter}$ private. Clients also aim to hide the quantity and semantics of their labels from the server $\mathbb{S}_{inter}$. This model emphasizes privacy preservation in internal label sharing.

    \item \textbf{External attackers.} An external attacker $\mathbb{A}_{ext}$ is considered to know the distribution of the original dataset fed into the input layer at the targeted client $\mathbb{C}_{target}$, and the posterior distribution of the model at the server side. The attacker $\mathbb{A}_{ext}$ aims to use a shadow dataset that mimics these distributions to construct an attack model for conducting inference attacks~\cite{10.1145/3460120.3484756}. This player model focuses on \textit{the level of privacy preservation in the unlearning capability} of the framework.
\end{packeditemize}

Considering the player model and keeping labels private while achieving efficiency and effectiveness, we introduce an additional objective beyond \textbf{\textcolor{violet}{G1}} and \textbf{\textcolor{violet}{G2}} as follows.
\begin{packeditemize}
\item \textbf{\textcolor{violet}{G3}: Privacy-preserving label sharing.} 
This goal emphasizes that the learning and unlearning processes should not disclose any label information from the client side to the server, including the quantity and semantics of the labels. 
\end{packeditemize}

\smallskip
\noindent\textbf{Protecting label quantity matters.}
In \textbf{\textcolor{violet}{G3}}, protecting the label privacy goes beyond protecting their semantics to include the quantity of labels, which is crucial for data privacy. 
By concealing the exact number of labels, adversaries are prevented from applying supervised learning methods that rely on knowing the number of classes for high accuracy. Instead, they are forced to use unsupervised learning techniques, lacking clear objectives and explicit class labels and generally resulting in lower accuracy~\cite{osisanwo2017supervised, love2002comparing}.
Protecting label quantity thus reduces the risk of unauthorized model reconstruction and inference attacks, strengthening the overall privacy and security of the learning process.

\smallskip
We introduce \textsc{SplitWiper+} featuring a label expansion strategy combined with a DP-based masking scheme that protects the semantic information and quantity of labels shared with the server to achieve \textbf{\textcolor{violet}{G3}}. This method is crucial when the server's trustworthiness is uncertain, as it conceals real labels, protecting client privacy and data security. By effectively masking both the quantity and specifics of the labels, clients can securely transmit masked labels to the server, mitigating privacy and security concerns.

\begin{figure}[t]
\setlength\belowcaptionskip{-0.4cm}
    \centering
    \includegraphics[width=1\linewidth]{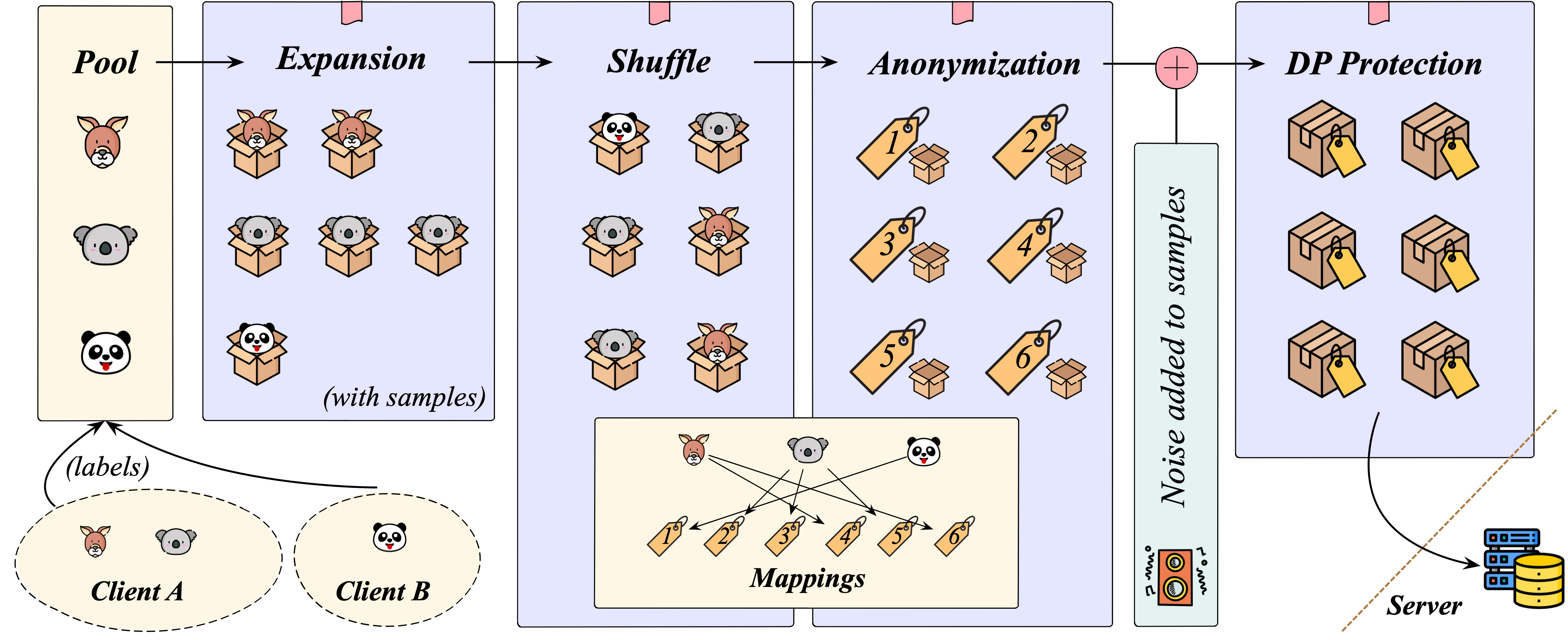}
    \caption{\textbf{Label Expansion:} Our proposed strategy preserves label quantity and semantics by expanding, shuffling, and anonymizing them, with intermediate values expanded via a DP mechanism.}
    \label{fig:anonymized_label}
\end{figure}

Compared to conventional U-shaped SL tailored for label privacy, our new DP-based method in \textsc{SplitWiper+} requires only one-way communication between clients and the server. 
This significantly reduces the communication frequency between clients and the server, enhancing efficiency and lowering operational costs.
We show our label expansion-and-masking scheme in Fig.\ref{fig:anonymized_label}.

\subsection{Label Expansion in \textsc{SplitWiper+}}
\label{sec:label_expansion}

In \textsc{SplitWiper+}, the label expansion strategy plays a pivotal role in masking both the semantics and quantity of labels, ensuring compliance with \textbf{\textcolor{violet}{G3}}.
By incorporating secure label consensus building among all clients and implementing label expansion and anonymization, this strategy effectively obscures the true identity and count of shared labels.
It provides the foundation for secure and private label sharing between clients and the server, addressing privacy concerns about label information leakage in \textsc{SplitWiper}.

We consider a scenario with $Q$ real labels, represented as $\mathbb{Y} = {Y_1, \cdots, Y_Q}$, where each client $k$ holds a subset $\mathbb{Y}_k = {Y_1, \cdots, Y_{Q_k}}$. These subsets collectively form the entire label set, such that 
\begin{equation}
\setlength\abovedisplayskip{3pt}   \setlength\belowdisplayskip{3pt}
    \bigcup_{k=1}^{K}\mathbb{Y}_k = \mathbb{Y}. 
\end{equation}

The corresponding intermediate values for label $Y_q$ on client $k$ are denoted as $V_q^k$.
In Vanilla SL, where label privacy is not a concern, the data exchanged between client $k$ and the server consists of pairs of labels and their intermediate values $\{Y_q, V_q^k\}$. To address the risks of label leakage, \textsc{SplitWiper+} modifies this method by sharing masked labels and their DP-protected intermediate values $\{Y_q^*, {V_q^*}^k\}$ instead.

To address the potential risk of sensitive label information being leaked during label sharing, we introduce a method that allows for label sharing without revealing the full set of labels $\mathbb{Y}_k$ held by client $k$ (cf. Algorithm~\ref{algo:label-sharing} in \textbf{Appendix}~\ref{appendix_algo}). Given the non-IID data distribution across clients, implementing a unified label expansion strategy in \textsc{SplitWiper+} requires additional communication among clients. This strategy only requires clients to agree on all labels used in the SL task, without the need to exchange the actual data samples. Since internal users are assumed to be honest but curious, direct label sharing among clients could still expose sensitive information, making our method crucial for preserving privacy.

\smallskip
\noindent\textbf{Secure label consensus building} (Line \ref{alg_3:label_sharing_start}-\ref{alg_3:label_sharing_end}). 
The communication among clients facilitates consensus among all clients on the complete set of real labels $\mathbb{Y}$ by allowing them to sequentially contribute their labels to a communal pool. 
The process is governed by two probabilistic thresholds:
\begin{packeditemize}
    \item $\lambda_1$: The probability that client $k$ will select and add labels from its set $\mathbb{Y}_k$ that are absent in the communal pool $\mathbb{Y}$.
    \item $\lambda_2$: The probability that client $k$ will continue to participate honestly in the sharing process after assessing whether $\mathbb{Y}_k$ is already included in $\mathbb{Y}$.
\end{packeditemize}

These probabilistic controls over label sharing and continued participation ensure that adversaries cannot infer the complete label set $\mathbb{Y}_k$ of any individual client from publicly shared information during the process. 
Even in scenarios where an untrustworthy client leaks information about the process, the adversary—potentially a curious server—remains unable to determine the specific labels associated with each client.

\smallskip
\noindent\textbf{Label expansion and anonymization} (Line \ref{alg_3:label_expansion_start}-\ref{alg_3:label_expansion_end}).  
Once the complete set $\mathbb{Y}$ is aggregated, clients collaborate to assign an expansion factor $\gamma_q$ to each label $Y_q$. This factor, $\gamma_q$, can be a randomly selected integer within a predefined range $\Gamma_p$ for each $Y_q$, or be a constant integer $\gamma > 1$ and uniformly applied across all labels, i.e., $\gamma_q = \gamma,  \forall q $.
Each label, $Y_q$, is then expanded to $\gamma_q$ instances, forming the initial expanded label set $\mathbb{Y}_{exp} = \{Y_1^*, \cdots,Y_e^*,\cdots Y_E^*\} $, where $E = \sum_{q=1}^{Q}{\gamma_q}$. 
This expansion process masks the total count $Q$ of real labels, preventing the server from inferring specific client tasks or data types based on label quantity.
The elements within $\mathbb{Y}_{exp}$ are shuffled and re-indexed to anonymize label semantics. While we use a basic pseudonymization technique, our framework can also accommodate methods like hashing~\cite{ribeiro2019privacy} or encryption~\cite{zhang2020batchcrypt, deshpande2021sypse} based on specific needs. The mapping between $\mathbb{Y}$ and $\mathbb{Y}_{exp}$ is preserved in $\mathcal{G}_Y$, ensuring that labels are expanded and anonymized without losing semantic integrity.

\smallskip
Since the information shared between client $k$ and the server consists of pairs of labels and their corresponding intermediate values $\{Y_q, V_q^k \}$, when $Y_q$ is expanded into $\gamma_q$ masked labels, the corresponding $V_q^k$ also need to be expanded to match the masked labels. During this expansion, we introduce a DP mechanism to obfuscate $V_q^k$ for further privacy protection (cf. Algorithm~\ref{algo:label-expansion}).

\smallskip
\noindent\textbf{DP-based expansion-and-masking scheme} (Line \ref{alg_4:expansion_start}-\ref{alg_4:expansion_end}).  
Following the pre-determined label expansion map $\mathcal{G}_Y$, Client $k$ takes the pairs $\{ Y_q, V_q^k \}$ obtained from its local model $\mathcal{F}_o^k$ as input. 
Each $Y_q$ is then expanded into $\gamma_q$ masked labels, and $V_q^k$ is correspondingly expanded into $\gamma_q$ DP-protected intermediate values, resulting in the set $\mathbb{U}_{exp} = \bigcup \{Y_e^*, {V_e^*}^k \}$. 
Let the total collection of DP-protected intermediate values across all clients as $\mathbb{V}_{exp} = \{{V_1^*}, \cdots, {V_{e}^*},\cdots, {V_E^*}\} $, keeping consistency with $\mathbb{Y}_{exp}$. 
The expansion of the intermediate values is carefully documented in $\mathcal{G}_V$, establishing a mapping that pairs expanded labels with their corresponding intermediate values, so we have $(\mathcal{G}_Y, \mathcal{G}_V)$ satisfying:
\begin{equation}
\setlength\abovedisplayskip{3pt}   \setlength\belowdisplayskip{3pt}
    (\mathcal{G}_Y, \mathcal{G}_V): \{ Y_q,V_q\} \rightarrow \bigcup  \{ Y_e^*,{V_e^*}\}.
    \label{equation_4_2_1_1}
\end{equation}
The mapping $(\mathcal{G}_Y, \mathcal{G}_V)$ represents a one-to-many expansion. Without collusion between clients and the server, $(\mathcal{G}_Y, \mathcal{G}_V)$ remains confidential, preventing the server from knowing the actual quantity or specific content of labels and intermediate values used in the training task.

\subsection{Masked Label Sharing in \textsc{SplitWiper+}}

Building on the proposed label expansion strategy, we examine how \textsc{SplitWiper+} implements masked label sharing to enhance privacy, distinguishing it from \textsc{SplitWiper}.

\smallskip
\noindent\textbf{Masked label sharing.}
In \textsc{SplitWiper+}, after freezing the pre-trained model, clients employ an expansion-and-masking scheme before sharing their data with the server. In contrast, \textsc{SplitWiper} involves clients directly transmitting the obtained data $\{Y_q, V_q^k\}$ to the server after freezing the pre-trained model, without performing any additional masking step. 
The transformation of $\{Y_q, V_q^k\}$ into $\{Y_e^*, {V_e^*}^k\}$ enables clients to transmit masked labels and their corresponding DP-protected intermediate values to the server, ensuring that the original information remains concealed.

Theorem~\ref{theorem_DP-based label anonymization} (cf. \textbf{Appendix}~\ref{appendix_proof} for proof) establishes that with the DP protection, the server cannot discern whether any two received labels are expanded from the same real label based on the corresponding intermediate values.
In other words, the server cannot infer the quantity and content of real labels based on the received information, ensuring the privacy protection of the genuine labels used by the client.

\begin{thm}
\label{theorem_DP-based label anonymization} 

Consider that $\mathbb{V}_{exp} = \{{V_{1}^*}, \cdots, {V_{E}^*}\}$ is expanded by $\mathbb{V} =  \{{V}_{1}, \cdots, {V}_{Q}\}$ using a given method $\mathcal{G}_V$, where $\forall {V}_{q} \in \mathbb{V}$, $ \gamma_q $ copies with $\varepsilon$-DP-based noise are generated as $ \{ {V_{a_1}^*} , \cdots, {V_{a_{\gamma_s}}^*} \} \in \mathbb{V}_{exp}$.
The probability that any two elements in $\mathbb{V}_{exp}$ correspond to the same element in $\mathbb{V}$ satisfies $2\varepsilon$-DP.

\end{thm}

\smallskip
\noindent\textbf{Label demasking.}
The server in \textsc{SplitWiper+} proceeds with the next stage of training based on the received masked labels and intermediate values.
The final training output on the server should yield masked predicted labels, which does not hinder the server-side training since it utilizes the same masked labels for learning. 

We introduce the label demasking on the client side when a client needs to utilize the complete model, i.e., $Concat(\mathcal{F}_o^k, \mathcal{F}_o^s)$. This step converts the predicted masked labels received from the server back into their original.

Given that $\mathcal{G}_Y$ is one-to-many label expansion, its inverse, $\mathcal{G}_Y^{-1}$, operates as a many-to-one function, defined as:
\begin{equation}
\setlength\abovedisplayskip{3pt}   \setlength\belowdisplayskip{3pt}
    Y_q = \mathcal{G}_Y^{-1} ( Y_e^*).
    \label{equation_4_2_1_2}
\end{equation}
Each masked label corresponds to a real label, and since the mapping $\mathcal{G}_Y$ is shared among all clients, they can easily demask the server’s predicted results using the inverse mapping $\mathcal{G}_Y^{-1}$. This demasking process ensures the integrity and applicability of the model’s predictions. The demasked labels are then used to compute the performance $\mathcal{T}(Concat(\mathcal{F}_o^k, \mathcal{F}_o^s), D_o^k)$, ensuring evaluation is based on accurate and relevant data, reflective of true model performance in real-world settings.

\smallskip
\noindent\textbf{Label protection in unlearning process.}
When the initiating client $k$ requests unlearning, only client $k$ participates in the subsequent update process. 
Other clients do not receive any additional updated information. 
The server, upon receiving the updated masked labels and corresponding DP-protected intermediate values shared by client $k$, gains insight into the aspects of the unlearning process. 
However, since the labels are masked, the server cannot accurately infer the quantity or category of the real labels involved in the unlearning from the changes in the number of masked labels. 
Thus, we deem that the privacy associated with the unlearning labels remains protected, effectively safeguarding sensitive information from inadvertent disclosure.

\smallskip
\noindent\textbf{Natural barrier against shadow inference.}
Shadow inference attacks involve an adversary using a shadow dataset to mimic the training and unlearning processes of the target model, ultimately allowing them to infer which data points have been unlearned, thereby compromising the privacy of the dataset~\cite{10.1145/3460120.3484756}. The proposed label expansion strategy addresses this vulnerability and strengthens client privacy by implementing two key methods:

\begin{packeditemize} 
    \item \textbf{Training on private datasets.} When training on private datasets, an external attacker with partial access to a victim’s data cannot construct a complete shadow dataset. While~\cite{10.1145/3460120.3484756} assumes the attacker knows all labels in the victim’s dataset, this is unrealistic unless using a well-known public dataset. Balancing positive and negative cases and diversifying shadow models is crucial for generalization in the attack model. Without full label knowledge, inferring the complete input-output mapping is challenging, restricting the attacker’s ability to replicate the training environment and conduct effective inference.
    
    \item \textbf{Training on various tasks.} When clients are engaged in different tasks, such as CIFAR-10/MNIST, the attacker's challenge increases, as they cannot easily reconstruct a shadow dataset that accurately reflects the diverse tasks of the clients. This further diminishes the attacker's ability to predict the model's behavior and the impact of unlearning.

\end{packeditemize}
For simplicity in our experiments (cf. Sec.\ref{subsec: exp_setting}), we use public datasets to mimic this process and intentionally ignore the potential risk of inference posed by using public datasets.

\smallskip
\noindent\textbf{\textsc{SplitWiper} as a specific case of \textsc{SplitWiper+}.}
The label expansion strategy provides inactive privacy protection when expansion factor $\gamma_q = \gamma = 1,\forall{q}$, in which case \textsc{SplitWiper+} becomes \textsc{SplitWiper} with no DP applied. For ease of comparison, throughout this paper, we will refer to \textsc{SplitWiper} as the case where \(\gamma = 1\) and \textsc{SplitWiper+} as the case where \(\gamma > 1\) is applied. 
These two will be differentiated whenever label privacy is relevant in the experiments; otherwise, \textsc{SplitWiper} will be used by default.


Label expansion in \textsc{SplitWiper+} introduces additional computational overhead compared to \textsc{SplitWiper} and increases the size of the data transmitted from clients to the server, thereby increasing communication costs. 
Thanks to one-way-one-off propagation, these additional overheads are incurred only once, preserving the significant advantages over existing SL methods, particularly in client-side efficiency. 
A detailed analysis of client-side and server-side overhead complexities are shown in \textbf{Appendix}~\ref{appendix_complexity}, where Table~\ref{tab_overhead} and Table~\ref{tab_overhead_server} respectively present a comparison of the client-side and server-side overhead among \textsc{SplitWiper}, \textsc{SplitWiper+}, and existing Vanilla and U-shaped SL frameworks.

\section{Evaluation}
\label{sec:evaluation}

In this section, we empirically evaluate \textsc{SplitWiper} of \textsc{SplitWiper+}, examining its compliance with key objectives: \textbf{\textcolor{violet}{G1}} (independent unlearning), \textbf{\textcolor{violet}{G2}} (effective unlearning with retained utility), and \textbf{\textcolor{violet}{G3}} (privacy-preserving label sharing).

\subsection{Experimental Settings}\label{subsec: exp_setting} 

Our experiments run on NVIDIA PCIe A100, 2 $\times$ 40GB. We use Python 3.7.16 and PyTorch 2.3.1 to build and train our framework.
We evaluate the effectiveness, efficiency, and privacy level of \textsc{SplitWiper} and \textsc{SplitWiper+} with $K=5$ clients and a single server, where client data is distributed in a non-IID manner. 

After learning tasks, a subset of clients, denoted as $\mathbb{C}_{u}$, initiates unlearning requests. In this evaluation, we set $\lvert \mathbb{C}_{u} \rvert = 1$, where $\lvert \cdot \rvert$ represents the $\ell_1$ norm, indicating that only one client requests unlearning. The remaining clients, who do not initiate unlearning requests, are denoted as $\mathbb{C}_{o}$.

\textsc{SplitWiper} and \textsc{SplitWiper+} are versatile, accommodating a variety of unlearning scenarios including class-, client-, and sample-level tasks. 
The experiments specifically focus on class-level unlearning tasks for clarity of presentation~\cite{chundawat2023can}. Further experiments on sample-level unlearning tasks are presented in \textbf{Appendix}~\ref{appendix_exp_sample unlearning}. In our settings, $\mathbb{C}_{u}$ selects a subset of label, denoted as $\mathbb{Y}_{u}$ for unlearning, aiming to effectively erase all related information without affecting other clients $\mathbb{C}_{o}$, or other labels $\mathbb{Y}_{o}$. 
We set $\lvert \mathbb{Y}_{u} \rvert = 1$, i.e., only a specific label is required for unlearning.
The unlearned label set, $\mathbb{Y}_{u}$, is exclusively possessed by $\mathbb{C}_{u}$, aligned with the non-IID data distribution setting and enabling a clearer demonstration of unlearning effectiveness.

\smallskip
\noindent\textbf{Dataset and data distribution.}
The experiments are conducted under various settings based on the dataset used and the degree of non-IID distribution among the clients:
\begin{packeditemize}
    \item \textbf{Clients with same task.} Clients operate on the same dataset but under different conditions:
        \begin{circitemize}
            \item \textit{Homogeneous labels but heterogeneous volumes.} All clients possess every label except $Y_{u}$, but the volume of labels varies among clients.
            \item \textit{Partially overlapping labels.} Clients have different types and quantities of labels with some overlap, i.e., all labels except $Y_{u}$ are randomly distributed among 1-3 clients.
            \item \textit{Heterogeneous labels.} Each client has a unique set of labels, with no overlap. 
        \end{circitemize}
    \item \textbf{Clients with different tasks.} Clients utilize different datasets for separate tasks, with distinct labels among the clients.
\end{packeditemize}
In each setting, the training and testing datasets are distributed among the clients using the same method to ensure consistency between the clients' training and testing data.

For experiments where clients perform the same task, we use CIFAR-10~\cite{nguyen2021dataset, jin2023poster} and CIFAR-100~\cite{song2019privacy, gowal2021improving} (widely utilized in computer vision and neural networks). CIFAR-10 consists of $60,000$ $32\times32$ color images distributed across $10$ classes, while CIFAR-100 includes $100$ classes with $600$ images per class. In the CIFAR-10 experiments, the designated label $Y_{u}$ is set to ``bird,'' and in CIFAR-100, it is set to ``aquarium-fish.''
For experiments where clients perform different tasks, $3$ clients utilize the CIFAR-10 dataset, while $2$ clients use the MNIST dataset~\cite{cohen2017emnist}, which comprises $60,000$ $28\times28$ grayscale images of handwritten digits. In this setup, the label ``bird'' from CIFAR-10 remains designated as $Y_{u}$.

\begin{table*}[t]
\setlength\abovecaptionskip{3pt} 
\centering
\setlength\tabcolsep{2pt}
\renewcommand\arraystretch{1.15}
\caption{Evaluation on \textbf{Efficiency} and \textbf{Effectiveness} over \textcolor{violet}{VGG} across Different SL Frameworks}
\label{tab_time}

\resizebox{\linewidth}{!}{
\begin{tabular}{|cc|cc|  ccccccc|  cccccccccc|}
\toprule

\multicolumn{4}{c}{} &   \multicolumn{7}{c}{\textbf{(Client-side) Training / \textcolor{violet}{Transmission} Time (s)}} &  \multicolumn{10}{c}{\textbf{Accuracy (\%)}} \\

\cmidrule{6-11}\cmidrule{13-21}

&&&   &   & \multicolumn{2}{c}{\textbf{Existing SL}} & \textbf{\textsc{SplitWiper}} & \multicolumn{3}{c}{\textbf{\textsc{SplitWiper+}}} & &   \multicolumn{2}{c}{\textbf{Existing SL}} &  \textbf{\textsc{SplitWiper}} & \multicolumn{6}{c|}{\textbf{\textsc{SplitWiper+}}} \\

\cmidrule{6-7} \cmidrule{9-11} \cmidrule{13-14}  \cmidrule{16-21} 

&&&  &  &  U-shaped & Vanilla & $\gamma=1$ & \textbf{$\gamma=2$} & \textbf{$\gamma=3$} & \textbf{$\gamma=\Gamma_p$} & &  U-shaped & Vanilla   & $\gamma=1$ & \multicolumn{2}{c}{\textbf{$\gamma=2$}} & \multicolumn{2}{c}{\textbf{$\gamma=3$}}  & \multicolumn{2}{c|}{\textbf{$\gamma=\Gamma_p$}} \\

\midrule

 \multirow{24}*{\rotatebox{90}{\textbf{\makecell{Clients with same task}}}} &  \multirow{8}*{\rotatebox{90}{\textbf{\makecell{Homo labels\\Hetero volumes}}}} & \multirow{4}*{\rotatebox{90}{\makecell{CIFAR-10}}} & \multirow{2}*{L.} & \cellcolor{teal!9}$\mathbb{C}_{u}$  & 74.36/\textcolor{violet}{216.40} & 66.03/\textcolor{violet}{261.63} & \cellcolor{yellow!15}2.23/\textcolor{violet}{9.04} &\cellcolor{yellow!15} 9.73/\textcolor{violet}{18.08} & \cellcolor{yellow!15}15.85/\textcolor{violet}{27.13}& \cellcolor{yellow!15}9.08/\textcolor{violet}{18.08} & 
 \cellcolor{teal!9}$\mathbb{Y}_{u}$ & 95.40 & 92.70 & \cellcolor{blue!8} 99.80 & \cellcolor{blue!8} 98.10 & \cellcolor{blue!8} 98.70 & \cellcolor{blue!8} 97.40 & \cellcolor{blue!8} 97.90 & \cellcolor{blue!8} 99.40 & \cellcolor{blue!8} 99.70 \\
\end{tabular}
}

\begin{tablenotes}
      \scriptsize
      \item[] \textit{Notation:}  \ding{172} \textbf{Data distribution}; \ding{173} \textbf{Dataset}; \ding{174} \textbf{Scenario} (L. for learning, U. for unlearning); \ding{175} \textbf{Client type} ($\mathbb{C}_{u}$ for clients involved in unlearning, $\mathbb{C}_{o}$ for clients excluded from unlearning); \ding{176} \textbf{Label type} ($\mathbb{Y}_{u}$ for to-be-unlearned labels, $\mathbb{Y}_{o}$ for retained labels).
     \end{tablenotes}

\vspace{-0.3cm}
\end{table*}

\begin{table*}[t]
\setlength\abovecaptionskip{3pt} 
\centering
\setlength\tabcolsep{2pt}
\renewcommand\arraystretch{1.15}
\caption{Evaluation on \textbf{Efficiency} and \textbf{Effectiveness} over \textcolor{violet}{ResNet18} across Different SL Frameworks}
\label{tab_time_resnet}

\resizebox{\linewidth}{!}{
\begin{tabular}{|cc|cc|  ccccccc|  cccccccccc|}
\toprule

\multicolumn{4}{c}{} &   \multicolumn{7}{c}{\textbf{(Client-side) Training / \textcolor{violet}{Transmission} Time (s)}} &  \multicolumn{10}{c}{\textbf{Accuracy (\%)}} \\

\cmidrule{6-11}\cmidrule{13-21}

&&&   &   & \multicolumn{2}{c}{\textbf{Existing SL}} & \textbf{\textsc{SplitWiper}} & \multicolumn{3}{c}{\textbf{\textsc{SplitWiper+}}} & &   \multicolumn{2}{c}{\textbf{Existing SL}} &  \textbf{\textsc{SplitWiper}} & \multicolumn{6}{c|}{\textbf{\textsc{SplitWiper+}}} \\

\cmidrule{6-7} \cmidrule{9-11} \cmidrule{13-14}  \cmidrule{16-21} 

&&&  &  &  U-shaped & Vanilla & $\gamma=1$ & \textbf{$\gamma=2$} & \textbf{$\gamma=3$} & \textbf{$\gamma=\Gamma_p$} & &  U-shaped & Vanilla   & $\gamma=1$ & \multicolumn{2}{c}{\textbf{$\gamma=2$}} & \multicolumn{2}{c}{\textbf{$\gamma=3$}}  & \multicolumn{2}{c|}{\textbf{$\gamma=\Gamma_p$}} \\

\midrule

 \multirow{24}*{\rotatebox{90}{\textbf{\makecell{Clients with same task}}}} &  \multirow{8}*{\rotatebox{90}{\textbf{\makecell{Homo labels\\Hetero volumes}}}} & \multirow{4}*{\rotatebox{90}{\makecell{CIFAR-10}}} & \multirow{2}*{L.} & \cellcolor{blue!9}$\mathbb{C}_{u}$  & 67.12/\textcolor{violet}{241.35} & 63.09/\textcolor{violet}{245.70} & \cellcolor{teal!11}2.07/\textcolor{violet}{8.55} &\cellcolor{teal!11} 9.53/\textcolor{violet}{16.80} & \cellcolor{teal!11}12.77/\textcolor{violet}{25.50}& \cellcolor{teal!11}8.83/\textcolor{violet}{24.75} & 
 \cellcolor{blue!9}$\mathbb{Y}_{u}$ & 90.90 & 88.60 & \cellcolor{yellow!14} 89.70 & \cellcolor{yellow!14} 89.20 & \cellcolor{yellow!14} 90.30 & \cellcolor{yellow!14} 88.30 & \cellcolor{yellow!14} 90.40 & \cellcolor{yellow!14} 89.90 & \cellcolor{yellow!14} 91.20 \\
\end{tabular}
}

\begin{tablenotes}
      \scriptsize
      \item[] \textit{Notation:}  \ding{172} \textbf{Data distribution}; \ding{173} \textbf{Dataset}; \ding{174} \textbf{Scenario} (L. for learning, U. for unlearning); \ding{175} \textbf{Client type} ($\mathbb{C}_{u}$ for clients involved in unlearning, $\mathbb{C}_{o}$ for clients excluded from unlearning); \ding{176} \textbf{Label type} ($\mathbb{Y}_{u}$ for to-be-unlearned labels, $\mathbb{Y}_{o}$ for retained labels).
     \end{tablenotes}

\vspace{-0.2cm}
\end{table*}

\smallskip
\noindent\textbf{Model architecture.}
We utilize the architecture based on a simplified VGG~\cite{bao2024efficient} and a ResNet18~\cite{he2016deep} to construct the client-side and server-side models for our split (un)learning experiments. 
In this configuration, the first convolutional layer of VGG or ResNet18 serves as the client model $\mathcal{F}_o^k$, while the remaining other layers comprise the server model $\mathcal{F}_o^s$. 
Given that VGG and ResNet have negligible differences in their first convolutional layer, we let the number of parameters in the client model $\mathcal{F}_o^k$ be consistent across both architectures.
The number of epochs for client-side pre-training is set to $N = 10$. The number of server-side model training and retraining epochs is set to $M = 30$ for experiments involving CIFAR-10 and $M = 50$ for experiments utilizing CIFAR-100.

In scenarios where clients train on the same task, each client's model, $\mathcal{F}_o^k, \forall k$, adheres to a uniform structure with the same input and output sizes across all clients. 
Conversely, in scenarios with clients engaged in different tasks, the input size of each $\mathcal{F}_o^k$ is tailored based on the specific training dataset $D_o^k$ used locally. 
However, the output size remains consistent to meet the input requirements of the server model $\mathcal{F}_o^s$. 
This design ensures that despite the diverse data types and tasks, all client models can interface correctly with the centralized server model within SL.

\smallskip
\noindent\textbf{Benchmark and experimental groups.}
We use non-SISA unlearning on Vanilla and U-shaped SL~\cite{gao2023pcat} as benchmarks to evaluate \textsc{SplitWiper}. To ensure fairness, both benchmarks are initialized with pre-trained results from \textsc{SplitWiper}. The client and server components are trained for the same $M$ epochs as \textsc{SplitWiper} during learning and retraining for unlearning requests. By standardizing initial conditions and training durations, we systematically compare the efficiency and effectiveness of \textsc{SplitWiper} against the benchmarks.

To assess the impact of various label expansion strategies in \textsc{SplitWiper} and \textsc{SplitWiper+}, we categorize experiments based on different expansion factors. Specifically, we define four experimental groups: three groups with uniform label expansions of $\gamma = 1$, $2$, or $3$, and one group with a variable expansion factor, $\gamma = \Gamma_p$. For $\gamma = 1$ (\textsc{SplitWiper}), labels are shuffled and re-indexed without altering their actual count, similar to the Vanilla SL approach where real labels are shared. For $\gamma = 2$ or $\gamma = 3$ (\textsc{SplitWiper+}), clients expand labels to $\gamma$ instances and apply differential privacy ($\varepsilon$-DP) with $\varepsilon = 0.1$ or $0.2$ to protect intermediate values.

In the variable scenario ($\gamma = \Gamma_p$), the expansion factor $\gamma_q$ for each label $Y_q$ is randomly selected from a predefined range, $\Gamma_p = [1, 3]$. This configuration enables us to examine the effects of variability in the expansion factor on both label privacy and system performance.

\smallskip
\noindent\textbf{Performance metrics.}
\textbf{\textcolor{violet}{G1}} represents a significant advancement by enabling absent clients and involving only those requesting unlearning, laying the groundwork for SISA in SL. While not easily assessed quantitatively, \textbf{\textcolor{violet}{G1}} serves as the foundational experiment setting and premise for achieving \textbf{\textcolor{violet}{G2}} and \textbf{\textcolor{violet}{G3}}.
We analyze the unlearning effectiveness and efficiency of existing SL frameworks, \textsc{SplitWiper}, and \textsc{SplitWiper+}. We measure label accuracy during learning and unlearning phases for effectiveness, and track training and transmission times for efficiency. We focus on client-side overhead and use a 100 Mbps bandwidth setting to simulate realistic conditions. 

To further evaluate \textbf{\textcolor{violet}{G3}}, we assess the robustness of our privacy-preserving method. We examine the server's ability to infer real labels from masked labels and intermediate outputs. In \textsc{SplitWiper+}, the server uses clustering algorithms on received data pairs to group similar labels. The metric is the probability of correctly matching masked labels to real ones.

\subsection{Unlearning Efficiency and Overhead}

\begin{figure*}[t]
\setlength\abovecaptionskip{3pt} 
    \centering
    \subfigure[Training time over VGG\label{fig_training_time_epoch}]{
    \includegraphics[width=0.23\linewidth]{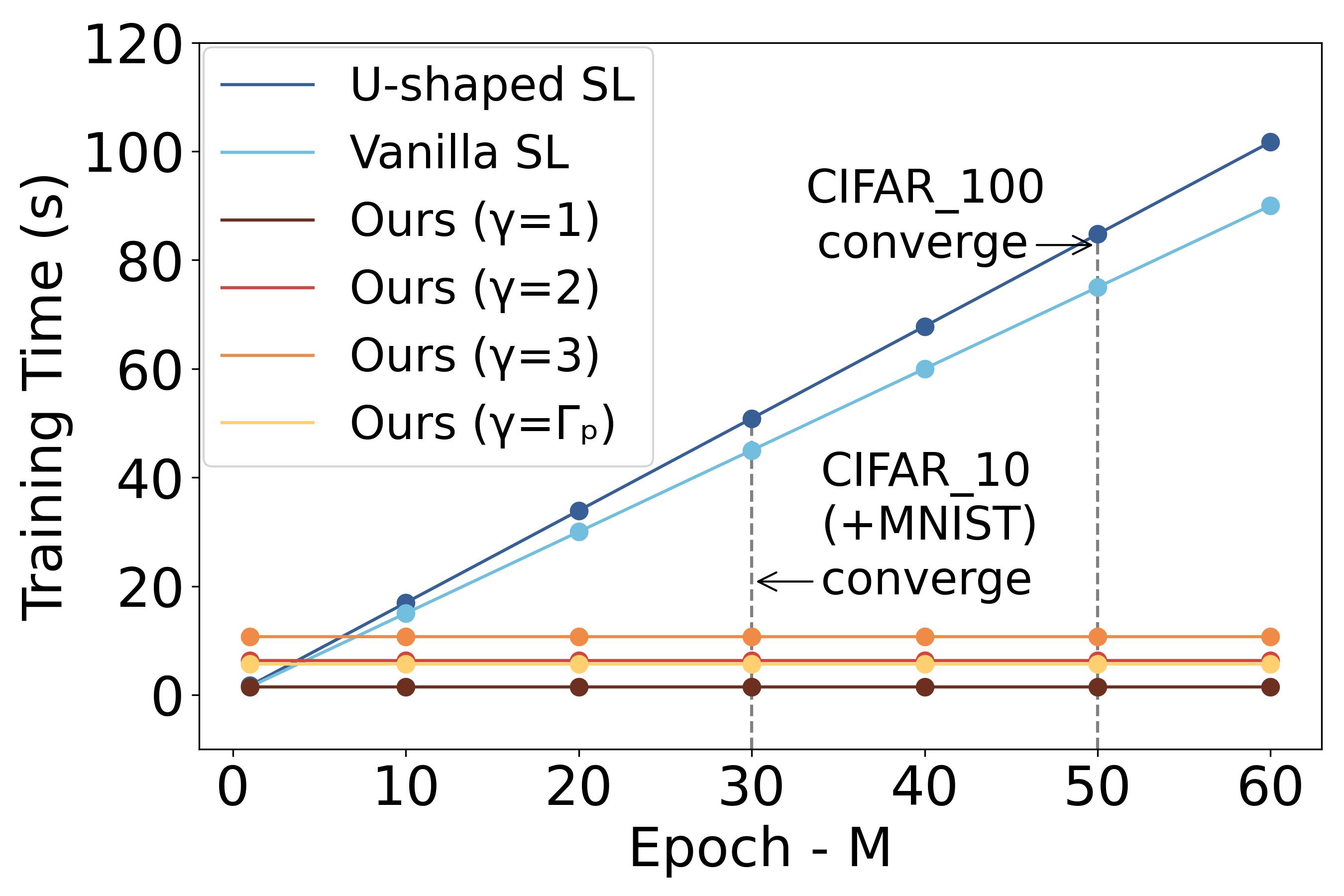}
    }
    \subfigure[Transmission time over VGG\label{fig_transmission_time_epoch}]{
    \includegraphics[width=0.23\linewidth]{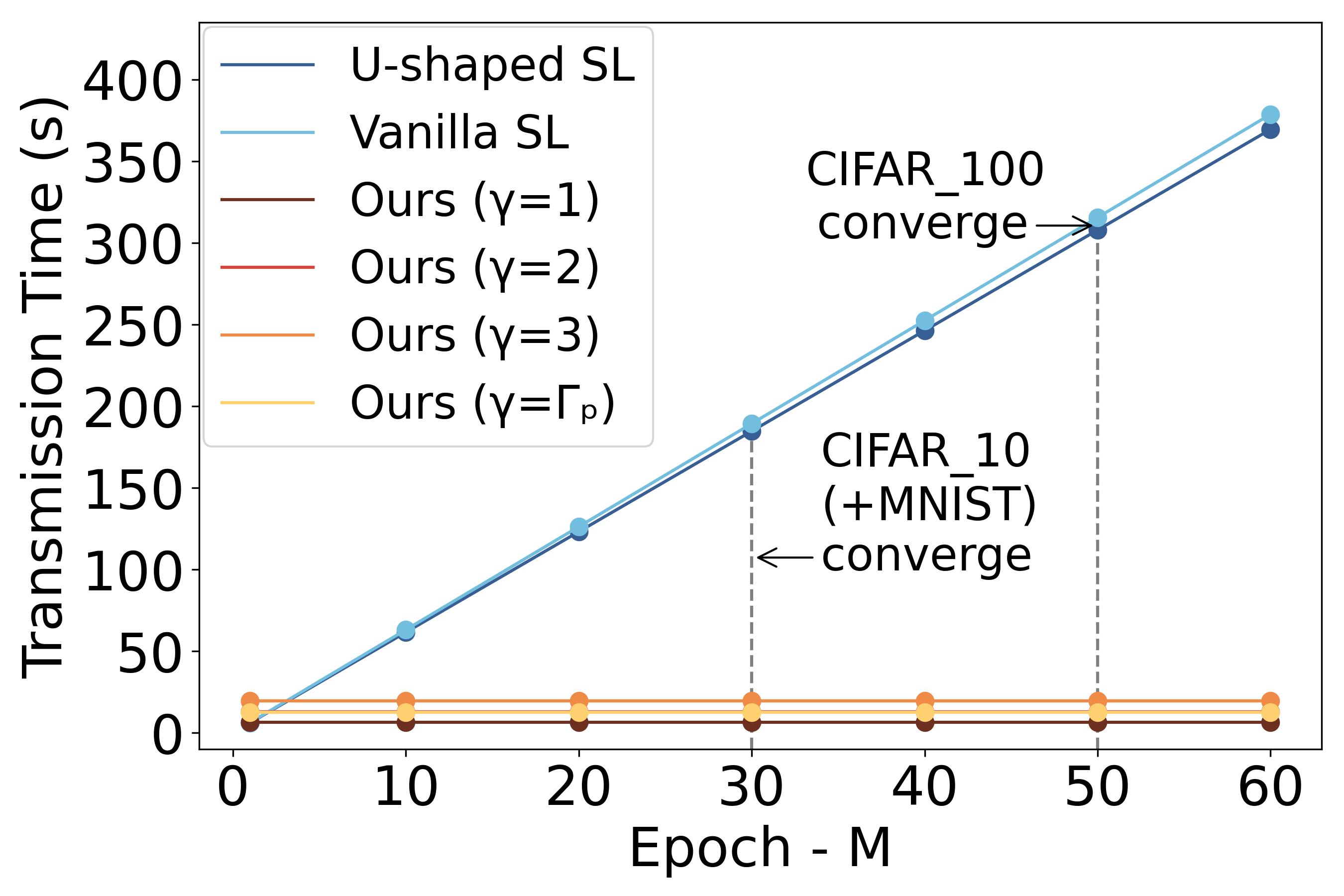}
    }
    \subfigure[Training time over ResNet18\label{fig_training_time_epoch_ResNet18}]{
    \includegraphics[width=0.23\linewidth]{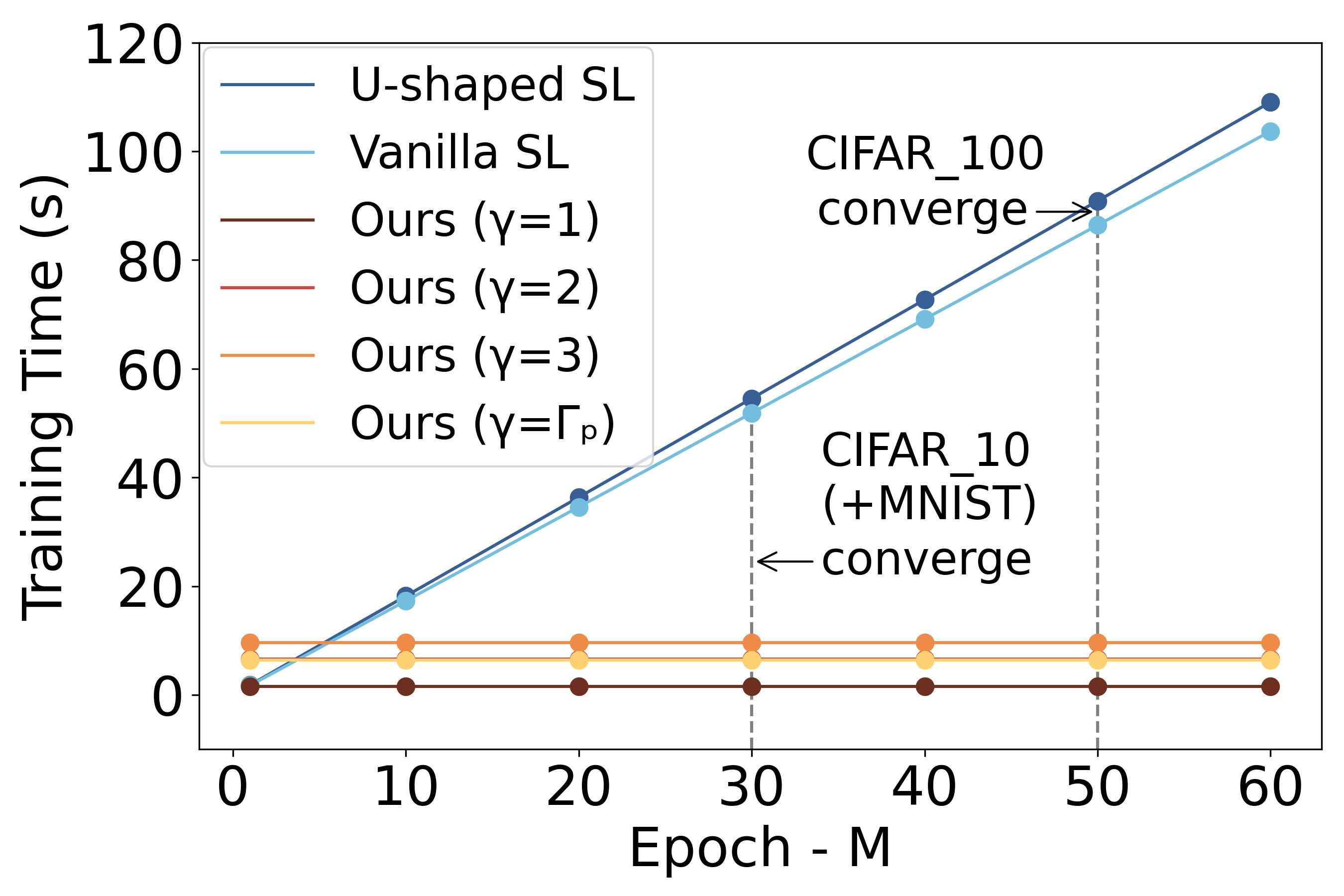}
    }
    \subfigure[Transmission time over ResNet18\label{fig_transmission_time_epoch_ResNet18}]{
    \includegraphics[width=0.23\linewidth]{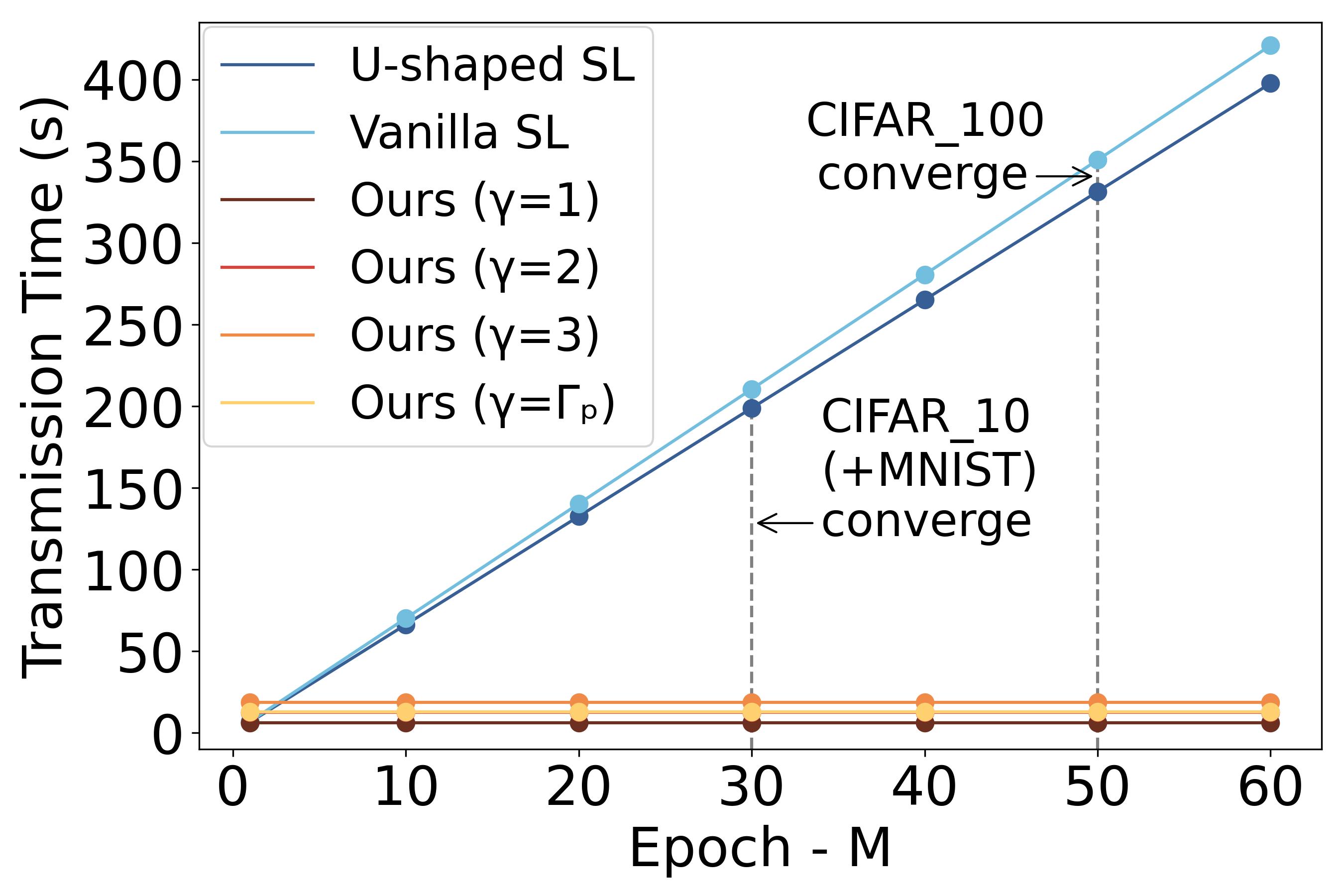}
    }
    \DeclareGraphicsExtensions.
    \caption{Client-side time consumption across different SL frameworks.}
    \label{fig_Time_with_epoch}
\end{figure*}

\begin{figure*}[t]
\setlength\abovecaptionskip{3pt} 
    \centering
    \subfigure[Training and transmission times across varies $\gamma$ over VGG \label{fig_Time_gamma}]{
    \includegraphics[width=0.23\linewidth]{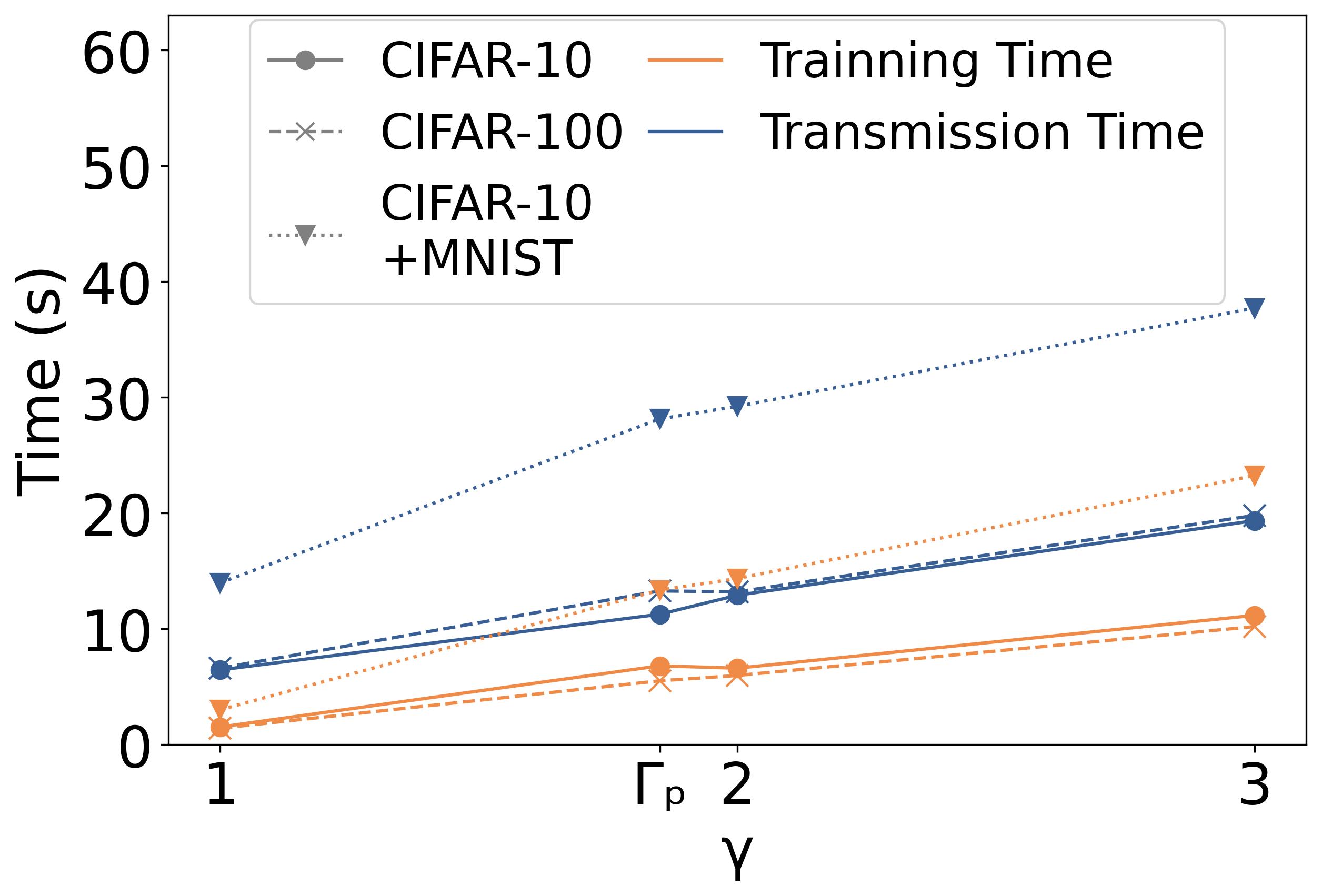}
    }
    \subfigure[Training and transmission times across varies $\gamma$ over ResNet18  \label{fig_Time_gamma_ResNet18}]{
    \includegraphics[width=0.23\linewidth]{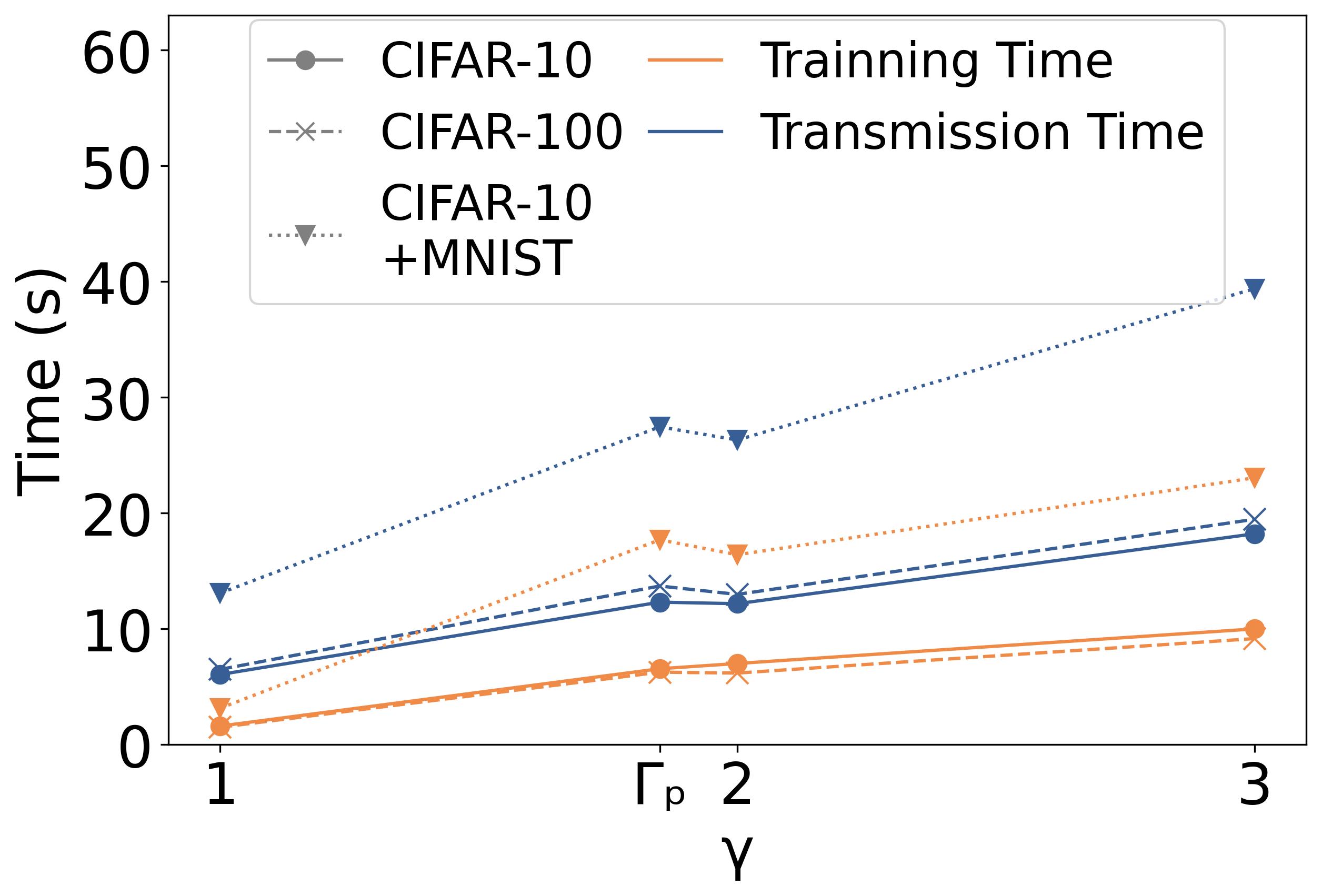}
    }
    \subfigure[Test accuracy across varies $\varepsilon$ under different $\gamma$ settings over VGG\label{fig_Accuracy_epsilon}]{
    \includegraphics[width=0.23\linewidth]{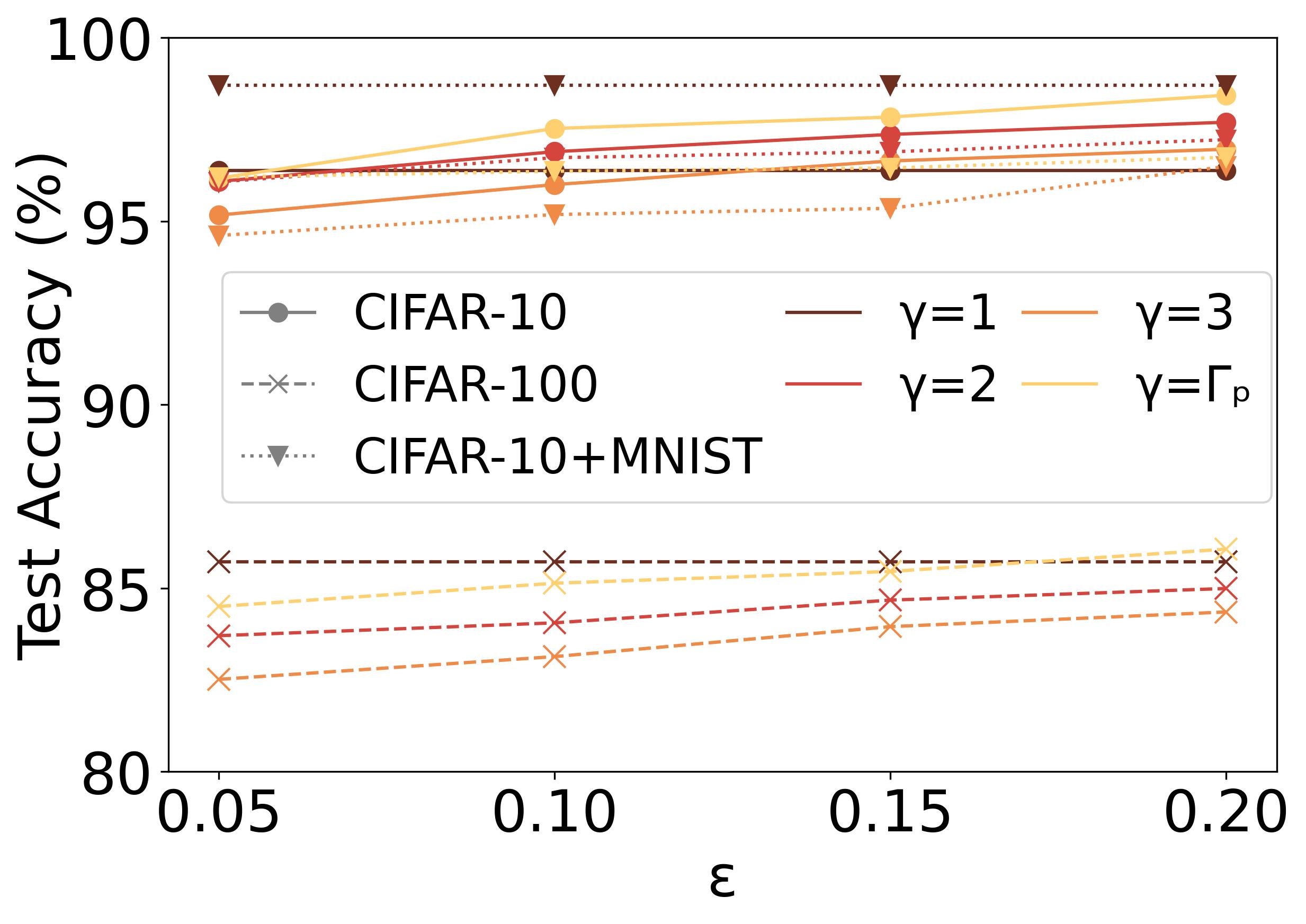}
    }
    \subfigure[Test accuracy across varies $\varepsilon$ under different $\gamma$ settings over ResNet18 \label{fig_Accuracy_epsilon_ResNet18}]{
    \includegraphics[width=0.23\linewidth]{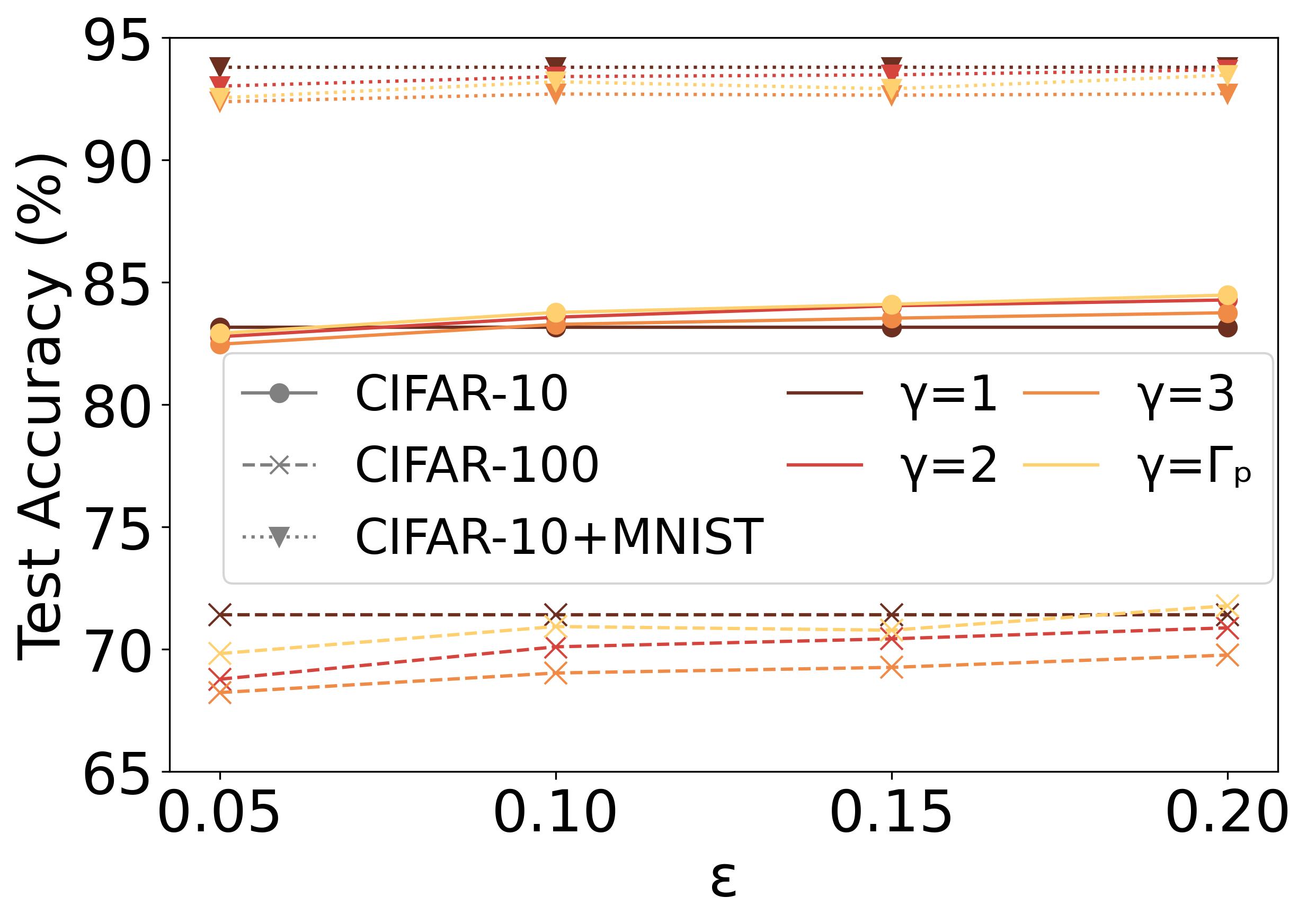}
    }
    \DeclareGraphicsExtensions.
    \caption{Impacts of the expansion factor $\gamma$ and privacy budget $\varepsilon$ on \textsc{SplitWiper} and \textsc{SplitWiper+} in terms of client-side training and transmission time consumption and test accuracy.}
    \label{fig_protection}
\vspace{-0.2in}

\end{figure*}

The left sides of Table~\ref{tab_time} and Table~\ref{tab_time_resnet} show the training and transmission times for clients in various SL frameworks over VGG and ResNet18, focusing on clients with unlearning requests ($\mathbb{C}_{u}$) and other clients ($\mathbb{C}_{o}$) across both learning and unlearning scenarios. Since the degree of DP-based noise addition does not affect time costs, we exclude the influence of $\varepsilon$ from our analysis, focusing solely on the effects of different label expansion factors $\gamma$.

The results highlight the efficacy of \textsc{SplitWiper} in achieving \textbf{\textcolor{violet}{G1}} (independent unlearning). Training and transmission times in \textsc{SplitWiper} are significantly lower than those in Vanilla and U-shaped SL across all data distributions and scenarios. Crucially, during unlearning, only the requesting clients, $\mathbb{C}_{u}$, incur additional costs, while other clients, $\mathbb{C}_{o}$, remain unaffected, as the server uses stored data from the learning phase to handle unlearning, minimizing overall client-side expenses and disruptions. The time overhead for the label expansion strategy discussion between clients (cf. Algorithm~\ref{algo:label-sharing}) is negligible, measured in milliseconds, and does not significantly impact our analysis of training and transmission times.

In \textsc{SplitWiper} and \textsc{SplitWiper+}, clients only need to train the pre-trained model over local data once. They then mask labels according to the label expansion strategy and apply DP protection to the corresponding intermediate values before transmitting them to the server in a one-way propagation manner. 
Subsequent training phases focus exclusively on the server, disconnecting from client-side computations. 
Therefore, the training and transmission times for clients are fixed values dependent on $\gamma$ and remain unaffected by the number of training epochs $M$. 
In contrast, in Vanilla and U-shaped SL, where each epoch involves client participation, the computational and transmission times increase linearly with $M$, as listed in Table~\ref{tab_overhead} (cf. \textbf{Appendix}~\ref{appendix_complexity}).

Fig.~\ref{fig_Time_with_epoch} shows the average training and transmission times for all clients across different SL frameworks as training epochs $M$ increase. At $M=1$, the times for Vanilla and U-shaped SL are nearly identical to those of \textsc{SplitWiper} with $\gamma=1$. As $\gamma$ increases (\textsc{SplitWiper+}), computational and communication costs rise, but they typically stay below $5$ due to client-specified balancing. Even in practical settings with $M=30$ for CIFAR-10 and CIFAR-10/MNIST, and $M=50$ for CIFAR-100, \textsc{SplitWiper} retains a clear advantage, especially during unlearning, where only the requesting clients ($\mathbb{C}_{u}$) incur costs, while others ($\mathbb{C}_{o}$) do not.

To illustrate the impact of the label expansion factor $\gamma$ on client-side overhead in \textsc{SplitWiper+}, we plot the average training and transmission times for all clients across various $\gamma$ values in Fig.\ref{fig_Time_gamma} and Fig.\ref{fig_Time_gamma_ResNet18}. Both figures, for VGG and ResNet18, show a roughly linear increase in overhead as $\gamma$ grows.
The case $\gamma = \Gamma_p$ represents scenarios where $\gamma_q$ is randomly selected as an integer from $[1, 3]$, with an average $\gamma_q$ of $1.9$. This introduces slightly higher overhead due to variability in $\gamma_q$ compared to a uniform expansion factor.
Since CIFAR-100 and CIFAR-10 are similar in size ($\sim 177$ MB), their computational and transmission overheads are nearly identical. However, using CIFAR-10 and MNIST in scenarios with differing client tasks significantly increases training and transmission times.

\vspace{-0.1em}
\begin{center}
\fbox{%
    \begin{minipage}{0.95\linewidth}

    \textbf{Takeaway (faster).} \textsc{SplitWiper}/\textsc{SplitWiper+} mitigate overhead growth with one-way-one-off propagation, ensuring unrelated clients remain unaffected during unlearning. While \textsc{SplitWiper+} introduces label expansion overhead, it matches \textsc{SplitWiper} in reducing computational and communication costs, achieving a 97-99\% reduction in unlearning overhead compared to existing SLs.
    \end{minipage}
}
\end{center}

\subsection{Unlearning Effectiveness and Accuracy}
\label{subsection:exp_acc}

The right sides of Tables~\ref{tab_time}–\ref{tab_time_resnet} show the accuracy of the labels to be unlearned ($Y_{u}$) and other labels ($Y_{o}$) under both learning and unlearning scenarios across the test set. All frameworks, including existing ones and ours with varying $\gamma$ and $\varepsilon$, effectively satisfy \textbf{\textcolor{violet}{G2}} (effective unlearning with retained utility). They achieve 
\textbf{zero} accuracy for $Y_{u}$ during unlearning (consistent with retraining results~\cite{chundawat2023can,chundawat2023zero,tarun2023fast}) while maintaining the accuracy of $Y_{o}$.
CIFAR-100, with more labels and fewer training examples per label, presents a greater learning challenge than CIFAR-10, resulting in lower accuracy across all frameworks. Additionally, ResNet18 performs worse than VGG on CIFAR-100, as VGG proves better suited for this task in our experiments.

\begin{figure*}[t]
\setlength\abovecaptionskip{3pt} \setlength\belowcaptionskip{-0.3cm}
    \centering
    \subfigure[Perfect clustering accuracy with elbow method.\label{fig_kmeans_elbow_perfect}]{
    \includegraphics[width=0.23\linewidth]{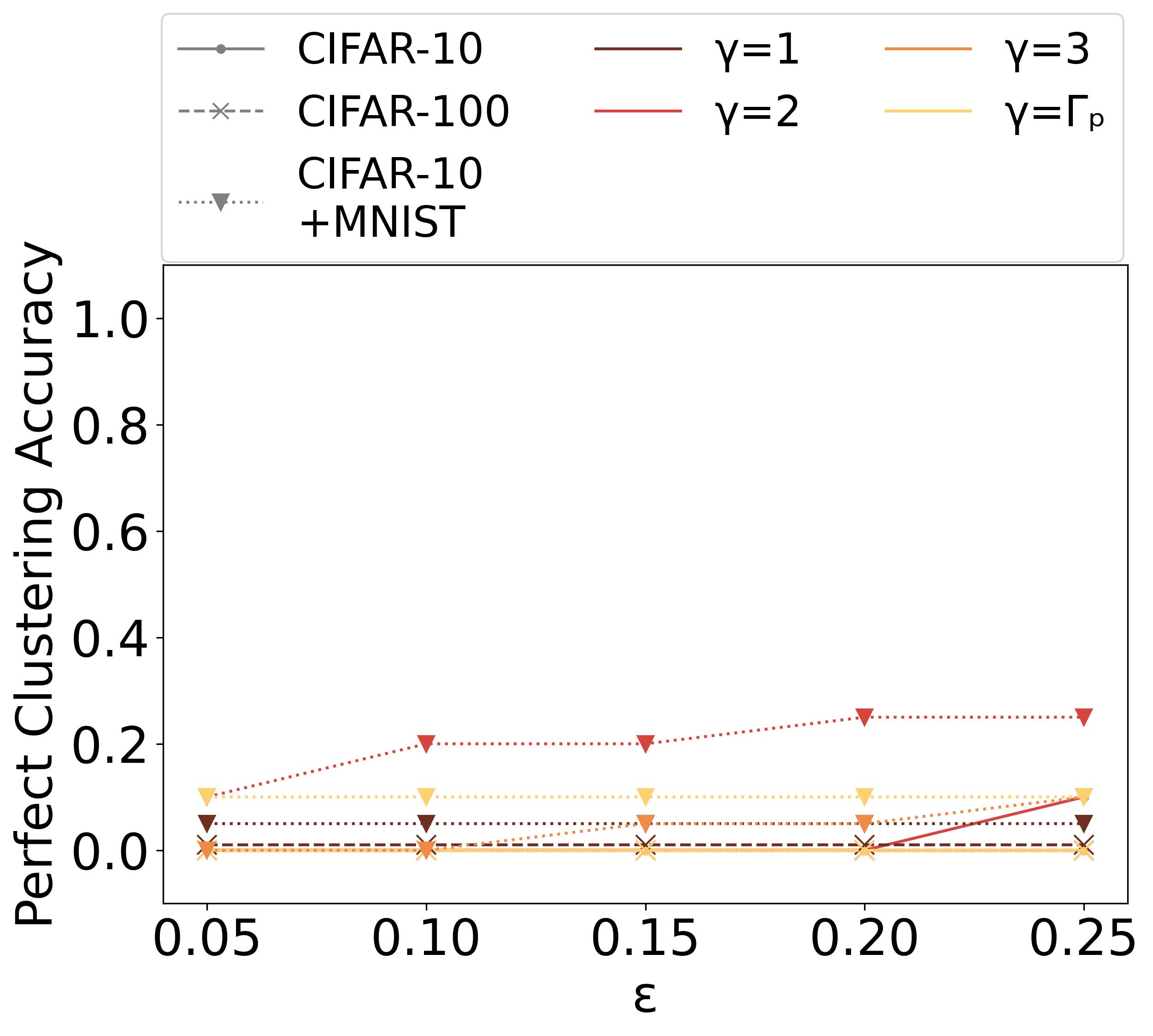}
    }
    \subfigure[Inclusive clustering accuracy with elbow method.\label{fig_kmeans_elbow_inclusive}]{
    \includegraphics[width=0.23\linewidth]{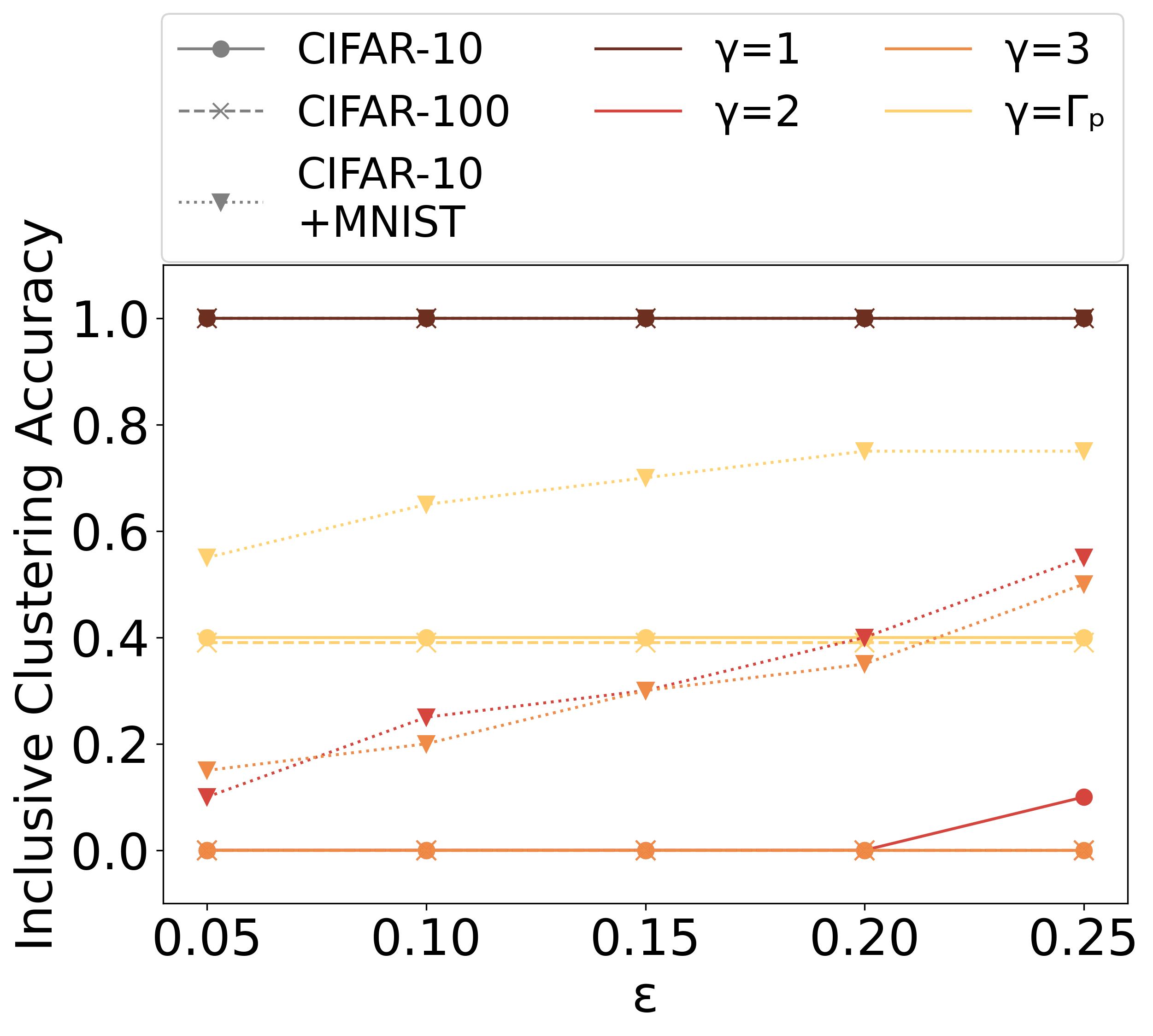}
    }
    \subfigure[Perfect clustering accuracy with real labels number $Q$. \label{fig_kmeans_real_perfect}]{
    \includegraphics[width=0.23\linewidth]{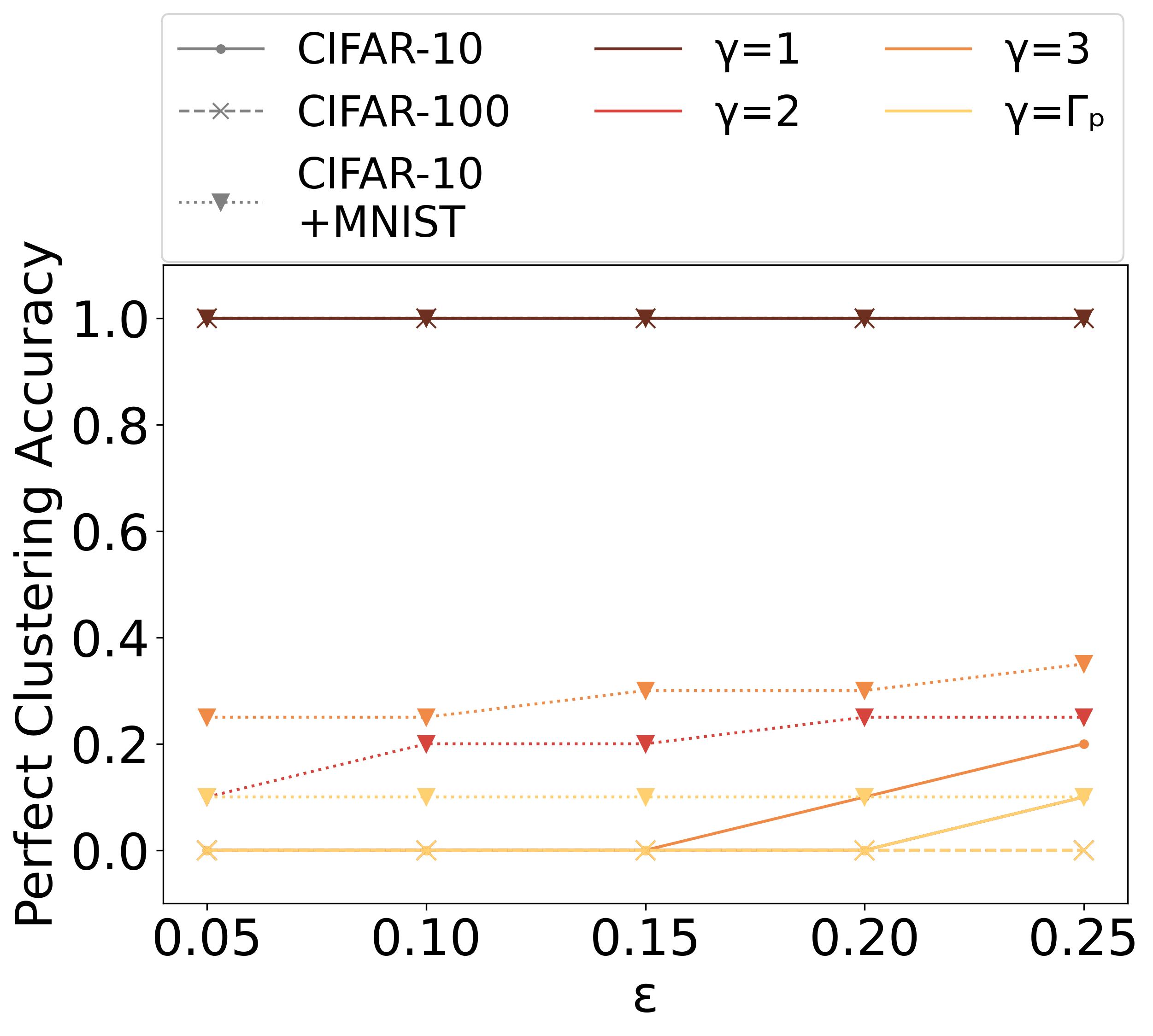}
    }
    \subfigure[Inclusive clustering accuracy with with real labels number $Q$. \label{fig_kmeans_real_inclusive}]{
    \includegraphics[width=0.23\linewidth]{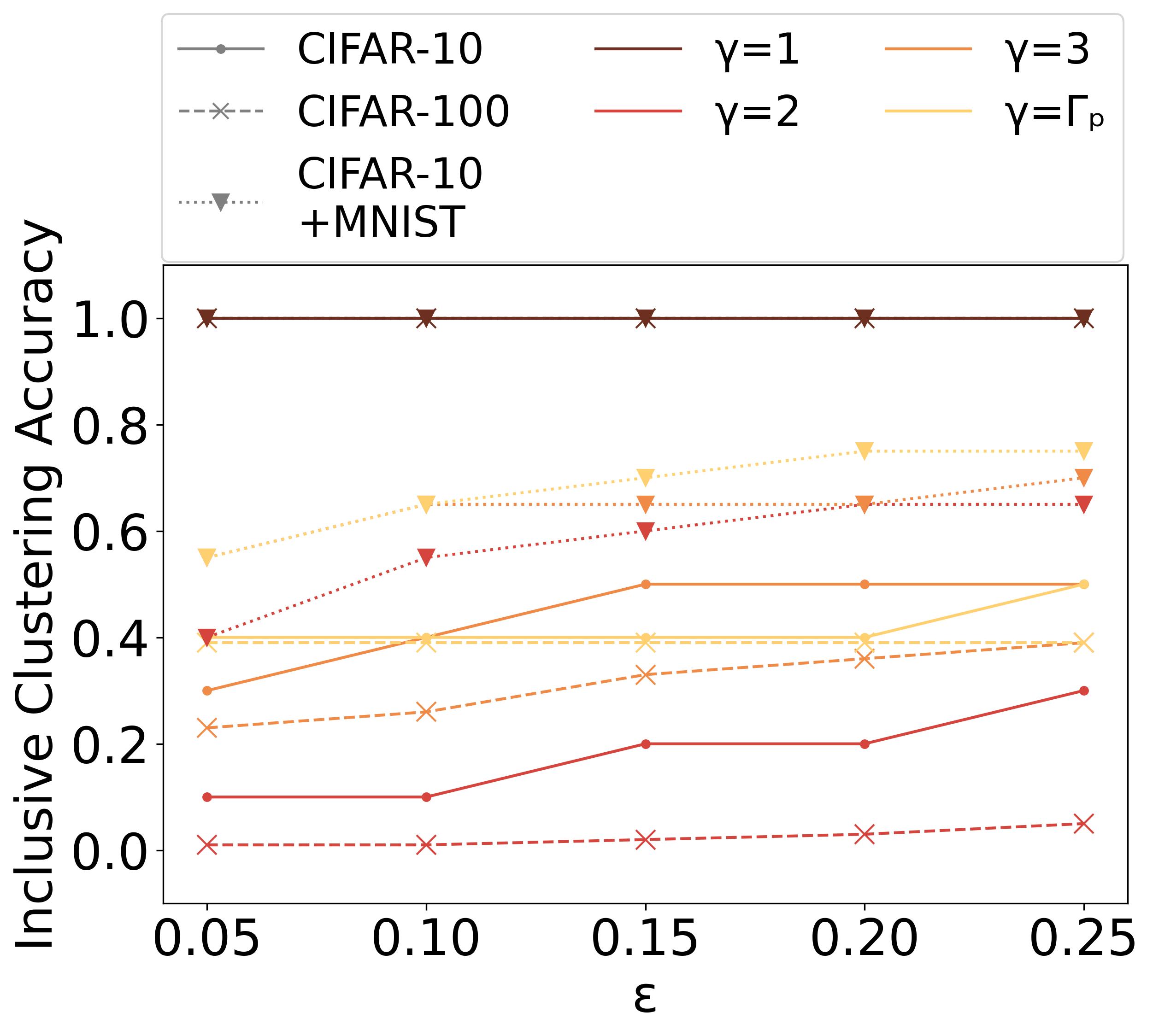}
    }
    \DeclareGraphicsExtensions.
    \caption{KMeans inclusive clustering accuracy and perfect clustering accuracy with Variable $\gamma$ and $\varepsilon$, using the elbow method or the real label number $Q$ as the cluster parameter. The low perfect clustering accuracy with both the elbow method and using $Q$ as the parameter indicates the effective preservation of label-related privacy in \textsc{SplitWiper+}. }
    \label{fig_kmeans}
\end{figure*}

Compared to Vanilla and U-shaped SL, \textsc{SplitWiper} achieves notable accuracy improvements for both $Y_{u}$ and $Y_{o}$ in the learning scenario. It surpasses Vanilla SL by 4.77\% on CIFAR-10, 4.37\% on CIFAR-100, and 2.98\% in CIFAR-10 and MNIST scenarios. Against U-shaped SL, it shows gains of 4.18\%, 3.82\%, and 3.54\% in the same respective scenarios.
\textsc{SplitWiper}’s key innovation is freezing client parameters after pre-training, preventing them from being influenced during server-side training. Shared server model back-propagation ($\mathcal{F}_o^s$) can cause overfitting or hinder learning in a client’s model ($\mathcal{F}_o^k$). Freezing client parameters enhances efficiency, reduces overhead, and minimizes cross-client interference, improving the accuracy of $Concat(\mathcal{F}_o^k, \mathcal{F}_o^s)$.

Fig.\ref{fig_Accuracy_epsilon} and Fig.\ref{fig_Accuracy_epsilon_ResNet18} compare \textsc{SplitWiper} and \textsc{SplitWiper+}, showing the effects of $\varepsilon$ and $\gamma$ on accuracy. While increasing $\gamma$ slightly reduces accuracy, it remains acceptable. In \textsc{SplitWiper+} with $\gamma > 1$, higher $\varepsilon$ reduces noise and improves test accuracy. With $\varepsilon=0.2$, \textsc{SplitWiper+} can even outperform \textsc{SplitWiper} ($\gamma=1$) due to minimal noise and label expansion functioning as data augmentation\cite{takahashi2019data, shorten2019survey}, enhancing accuracy and generalization.
Specifically, \textsc{SplitWiper+} with $\varepsilon=0.2$ and $\gamma=\Gamma_p$ achieves an average accuracy gain of 0.37\% on CIFAR-10 and 0.12\% on CIFAR-100, particularly improving $\mathbb{Y}_o$ while maintaining comparable performance on $\mathbb{Y}_u$. This highlights \textsc{SplitWiper+}’s ability to strengthen privacy protection without compromising utility.

\vspace{-0.5em}
\begin{center}
\fbox{%
\begin{minipage}{0.95\linewidth}
\textbf{Takeaway (more effective).}
\textsc{SplitWiper}/\textsc{SplitWiper+} effectively unlearn upon request while preserving the accuracy of retained data. The unlearning process in \textsc{SplitWiper} is over 8\% more effective than existing SL, with \textsc{SplitWiper+} further enhancing accuracy by an additional 0.37\%.

\end{minipage}
}
\end{center}

\subsection{Label-related Privacy Analysis}
\label{subsection:exp_privacy}

Considering \textbf{\textcolor{violet}{G3}} (privacy-preserving label sharing), we evaluate the server's ability to infer real label characteristics, such as semantic information and label quantity, from the received masked labels and corresponding intermediate outputs.
Given that the server lacks access to a local dataset and the shared labels are anonymized, it faces significant constraints in conducting semantic inference attacks on labels.

We focus our analysis on the server's capability to deduce the actual number of labels used by clients, as the protection of label quantity is a pivotal advancement in our approach, effectively preventing attackers from leveraging the exact number of labels to perform high-accuracy supervised learning attacks. 
The inference of label quantity requires unraveling the critical mapping mechanism $(\mathcal{G}_Y, \mathcal{G}_V): \{ Y_q,V_q^k\} \rightarrow \{ Y_x^*,{V_x^*}^k\}$, the key difference between \textsc{SplitWiper+} and \textsc{SplitWiper}. It highlights the enhanced privacy of the former in protecting label information against server-based inference attacks.

In \textsc{SplitWiper+}, the server, with attempt to infer the actual label quantity, is limited to employing clustering algorithms on the received data pairs $\{ Y_x^*,{V_x^*}^k\}$. These algorithms group labels with closer intermediate outputs into the same cluster, hypothesizing that these labels are expanded from the same original label.
We assess the effectiveness of this clustering-based label matching attack using two metrics:
\begin{packeditemize}
    \item \textbf{Inclusive clustering accuracy.} This accuracy indicates the probability that labels derived from the same real label are inclusively grouped within the same cluster formed by the clustering algorithm. It underscores the containment relationship where a single cluster may encompass an entire real label group, but not necessarily align exactly with it.
    \item \textbf{Perfect clustering accuracy.} This accuracy indicates the proportion of instances where the clustered groups formed by the clustering algorithm match exactly with the true label expansion groups, reflecting a flawless alignment.
\end{packeditemize}
Inclusive clustering accuracy evaluates whether masked labels originating from the same label are grouped correctly based on their proximity in feature space. In contrast, perfect clustering accuracy assesses both the separation between groups and the algorithm’s ability to distinguish them. These metrics provide insights into the effectiveness of \textsc{SplitWiper+} in preserving label-related privacy while maintaining the structure of masked labels during clustering.

We employ the KMeans clustering algorithm~\cite{bradley1998refining, syakur2018integration}, a standard method renowned for its robustness and widespread use in detecting patterns in ungrouped data.
Given that the real number of labels, $Q$, is confidential to the server, we utilize the elbow method~\cite{yuan2019research} to determine an appropriate number of clusters $K$. 
Additionally, to illustrate the potential privacy breach in \textsc{SplitWiper+}, we also conduct experiments where $Q$ is used directly as the parameter $K$ for KMeans. 
This scenario simulates the impact of a hypothetical leakage of $Q$, revealing how such exposure could compromise the privacy measures implemented within \textsc{SplitWiper+} and further emphasizing the critical importance and value of concealing the explicit label quantity.

Given that the pre-trained models on each client are consistent across both VGG and ResNet18 architectures—owing to the limited number of layers on resource-constrained clients and the application of training techniques such as dropout and batch normalization on the server side—the intermediate values exhibit negligible variation. Consequently, we do not differentiate between VGG and ResNet18, focusing instead on the impact of intermediate values on label privacy. 

As illustrated in Fig.\ref{fig_kmeans_elbow_perfect}, when $Q$ (the true number of clusters) is unknown and the server estimates the optimal number of clusters using the elbow method, perfect clustering accuracy remains low, not exceeding 10\% for $\gamma = 3$. This indicates that \textsc{SplitWiper+}, employing a label expansion strategy with $\gamma = 3$, preserves over 90\% of label-related privacy, effectively protecting sensitive information against inference attacks. Furthermore, as $\gamma$ increases or with non-uniform expansion factors ($\gamma = \Gamma_p$), perfect clustering accuracy decreases, providing enhanced privacy protection.

In contrast, the inclusive clustering accuracy (Fig.\ref{fig_kmeans_elbow_inclusive}), is notably higher than the perfect clustering accuracy, due to its less stringent criteria. 
As more masked labels are clustered together by KMeans, the likelihood of satisfying the inclusive clustering accuracy increases. 
Under conditions of higher $\gamma$ or with $\gamma=1$, when more masked labels are grouped into the same cluster, the inclusive clustering accuracy rises. 
However, this scenario primarily indicates that masked labels derived from the same original label are grouped together, without providing the means to differentiate between distinct groups of labels. 
While this may reveal minimal related information, it is insufficient for practical use.

When the number of real labels $Q$ is disclosed to the server (Fig.\ref{fig_kmeans_real_perfect}), clustering-based label matching attacks achieve higher perfect clustering accuracy, increasing by an average of 16\% compared to when $Q$ remains confidential. 
The increase highlights the additional privacy risks posed by the exposure of label counts, underscoring the necessity of implementing privacy-preserving measures in \textsc{SplitWiper+}. 
Furthermore, a larger $\gamma$ amplifies the risks associated with the leakage of $Q$. Higher expansion factors result in an increased number of masked labels derived from each original label, thereby providing additional information that facilitates clustering attacks when $Q$ is known.

\begin{center}
\fbox{%
\begin{minipage}{0.95\linewidth}
\textbf{Takeaway (label privacy retained with one-way-one-off).}
\textsc{SplitWiper+} protects label privacy and intermediate values through expansion and DP-based obfuscation, preventing server inference attacks via clustering techniques.  It achieves over 90\% privacy preservation when the true label count is concealed and approximately 70\% even when the label count is exposed.
\end{minipage}
}
\end{center}



\section{Conclusion}
\label{sec:conclusion}
We achieved the first practical \textit{Split Unlearning} framework with \textsc{SplitWiper} and \textsc{SplitWiper+}, which designed a one-way-one-off propagation to decouple propagation between clients and the server, enabling SISA unlearning with absent clients. 
\textsc{SplitWiper+} further enhanced client label privacy against the server by employing differential privacy. 

Experiments across diverse data distributions and tasks demonstrated that \textsc{SplitWiper} effectively achieved unlearning by involving only the requesting clients, resulting in \textbf{0\%} unlearning accuracy, and \textbf{8\%} better accuracy on retained data than non-SISA unlearning on existing SL frameworks. \textsc{SplitWiper} maintained \textit{constant} overhead and reduced computational and communication costs by over \textbf{99\%}. \textsc{SplitWiper+} preserved over \textbf{90\%} of label privacy, effectively protecting sensitive information from the server.

\begin{table}[t]
\centering
\footnotesize
\caption{Notation}\label{tal: notation}
\label{tab_notation}
\vspace{-0.1in}
\renewcommand{\arraystretch}{1.1} 
\resizebox{\linewidth}{!}{
\begin{tabular}{c|c|p{7.2cm}}
\toprule

\multirow{16}{*}{\rotatebox{90}{\textbf{(Un)Learning}}} 
& \cellcolor{blue!10} client-$k$  & The $k$-th client \\
& \cellcolor{blue!10} $D_o^k$  & Original dataset owned by client-$k$ \\
& \cellcolor{blue!10} $D_u^k$    & Updated dataset (client-$k$) for unlearning \\
& \cellcolor{blue!10} $(X_i, Y_i)$    & The $i$-th data sample and its corresponding label \\
& \cellcolor{blue!10} $\mathcal{F}_o^k / \mathcal{F}_o^s$    & Original model maintained by client-$k$ / the server \\
& \cellcolor{blue!10} $\mathcal{F}_u^k / \mathcal{F}_u^s$    & Updated model (client-$k$ / server) for unlearning \\
& \cellcolor{blue!10} $W_k^{(t)}$    & Model weights of client-$k$ at iteration $t$ \\
& \cellcolor{blue!10} $W_s^{(t)}$    & Model weights of the server at iteration $t$ \\
& \cellcolor{blue!10} $\mathcal{L}_k$  & Loss function of client-$k$ \\
& \cellcolor{blue!10} $\mathcal{T}$  & Evaluation function that assesses models \\
& \cellcolor{blue!10} $N$     & Average epochs of pre-training for clients in training/retraining \\
& \cellcolor{blue!10} $M$     & Average epochs of training for server in training/retraining \\
& \cellcolor{blue!10} $\mathbb{C}_u$  & Set of clients involved during unlearning \\
& \cellcolor{blue!10} $\mathbb{C}_o$    & Set of clients not involved during unlearning \\

\midrule

\multirow{8}{*}{\rotatebox{90}{\textbf{Label Expansion}}} 
& \cellcolor{yellow!15} $Q$ & Number of real labels \\
& \cellcolor{yellow!15} $\mathbb{Y}$    & Complete set of real labels  \\
& \cellcolor{yellow!15} $\mathbb{Y}_k$    & Set of labels owned by client-$k$  \\
& \cellcolor{yellow!15} $\mathbb{Y}_{exp}$ & Set of expanded labels \\
& \cellcolor{yellow!15} $\mathcal{G}_Y$& Corresponding map between $\mathbb{Y}$  and $\mathbb{Y}_{exp}$\\
& \cellcolor{yellow!15} $\gamma_q$ & Label expansion factor for the $q$-th label, $Y_q$\\
& \cellcolor{yellow!15}$V_q^k$ & Intermediate values of $Y_q$ on client-$k$\\ 
& \cellcolor{yellow!15}  $\varepsilon$ & Privacy budget of the DP mechanism \\
\bottomrule
\end{tabular}
}

\end{table}

\bibliographystyle{unsrt}
\bibliography{bib}

@article{voigt2017eu,
  title={The {EU} general data protection regulation ({GDPR})},
  author={Voigt, Paul and Von dem Bussche, Axel},
  journal={A Practical Guide, 1st Ed., Cham: Springer International Publishing},
  year={2017},
}

@inproceedings{retrain,
  title={Efficient two-stage model retraining for machine unlearning},
  author={Kim, Junyaup and Woo, Simon S},
  booktitle={IEEE/CVF Conference on Computer Vision and Pattern Recognition (CVPR)},
  year={2022}
}

@article{retrain1,
  title={Model sparsity can simplify machine unlearning},
  author={Liu, Jiancheng and Ram, Parikshit and Yao, Yuguang and Liu, Gaowen and Liu, Yang and SHARMA, PRANAY and Liu, Sijia and others},
  journal={Advances in Neural Information Processing Systems (NIPS)},
  year={2024}
}

@INPROCEEDINGS{9796721,
  author={Liu, Yi and Xu, Lei and Yuan, Xingliang and Wang, Cong and Li, Bo},
  booktitle={IEEE Conference on Computer Communications (INFOCOM)}, 
  title={The Right to be Forgotten in Federated Learning: An Efficient Realization with Rapid Retraining}, 
  year={2022},
}

@article{shaik2024exploring,
  title={Exploring the landscape of machine unlearning: A comprehensive survey and taxonomy},
  author={Shaik, Thanveer and Tao, Xiaohui and Xie, Haoran and Li, Lin and Zhu, Xiaofeng and Li, Qing},
  journal={IEEE Transactions on Neural Networks and Learning Systems (TNNLS)},
  year={2024},
}

@article{liu2024survey,
  title={A survey on federated unlearning: Challenges, methods, and future directions},
  author={Liu, Ziyao and Jiang, Yu and Shen, Jiyuan and Peng, Minyi and Lam, Kwok-Yan and Yuan, Xingliang and Liu, Xiaoning},
  journal={ACM Computing Surveys (CUSR)},
  year={2024},
}

@article{nguyen2022survey,
  title={A survey of machine unlearning},
  author={Nguyen, Thanh Tam and Huynh, Thanh Trung and Nguyen, Phi Le and Liew, Alan Wee-Chung and Yin, Hongzhi and Nguyen, Quoc Viet Hung},
  journal={arXiv preprint arXiv:2209.02299},
  year={2022}
}

@article{qu2023learn,
  title={Learn to unlearn: A survey on machine unlearning},
  author={Qu, Youyang and Yuan, Xin and Ding, Ming and Ni, Wei and Rakotoarivelo, Thierry and Smith, David},
  journal={arXiv preprint arXiv:2305.07512},
  year={2023}
}

@INPROCEEDINGS{sisa,
  author={Bourtoule, Lucas and Chandrasekaran, Varun and Choquette-Choo, Christopher A. and Jia, Hengrui and Travers, Adelin and Zhang, Baiwu and Lie, David and Papernot, Nicolas},
  booktitle={IEEE Symposium on Security and Privacy (SP)}, 
  title={Machine Unlearning}, 
  year={2021},
}

@inproceedings{thapa2022splitfed,
  title={Splitfed: When federated learning meets split learning},
  author={Thapa, Chandra and Arachchige, Pathum Chamikara Mahawaga and Camtepe, Seyit and Sun, Lichao},
  booktitle={AAAI Conference on Artificial Intelligence (AAAI)},
  year={2022}
}

@ARTICLE{9562559,
  author={Chen, Mingzhe and Gündüz, Deniz and Huang, Kaibin and Saad, Walid and Bennis, Mehdi and Feljan, Aneta Vulgarakis and Poor, H. Vincent},
  journal={IEEE Journal on Selected Areas in Communications (JSAC)}, 
  title={Distributed Learning in Wireless Networks: Recent Progress and Future Challenges}, 
  year={2021},
}

@inproceedings{chundawat2023can,
  title={Can bad teaching induce forgetting? unlearning in deep networks using an incompetent teacher},
  author={Chundawat, Vikram S and Tarun, Ayush K and Mandal, Murari and Kankanhalli, Mohan},
  booktitle={AAAI Conference on Artificial Intelligence (AAAI)},
  year={2023}
}

@inproceedings{cohen2017emnist,
  title={EMNIST: Extending MNIST to handwritten letters},
  author={Cohen, Gregory and Afshar, Saeed and Tapson, Jonathan and Van Schaik, Andre},
  booktitle={International Joint Conference on Neural Networks (IJCNN)},
  year={2017},
}

@article{liu2024decentralized,
  title={Decentralized Federated Unlearning on Blockchain},
  author={Liu, Xiao and Li, Mingyuan and Wang, Xu and Yu, Guangsheng and Ni, Wei and Li, Lixiang and Peng, Haipeng and Liu, Renping},
  journal={arXiv preprint arXiv:2402.16294},
  year={2024}
}

@article{liu2024fishers,
  title={Fishers Harvest Parallel Unlearning in Inherited Model Networks},
  author={Liu, Xiao and Li, Mingyuan and Wang, Xu and Yu, Guangsheng and Ni, Wei and Li, Lixiang and Peng, Haipeng and Liu, Renping},
  journal={arXiv preprint arXiv:2408.08493},
  year={2024}
}

@inproceedings{10.1145/3460120.3484756,
author = {Chen, Min and Zhang, Zhikun and Wang, Tianhao and Backes, Michael and Humbert, Mathias and Zhang, Yang},
title = {When Machine Unlearning Jeopardizes Privacy},
year = {2021},
booktitle = {ACM SIGSAC Conference on Computer and Communications Security (CCS)},
}

@article{li2021label,
  title={Label leakage and protection in two-party split learning},
  author={Li, Oscar and Sun, Jiankai and Yang, Xin and Gao, Weihao and Zhang, Hongyi and Xie, Junyuan and Smith, Virginia and Wang, Chong},
  journal={International Conference on Learning Representations (ICLR)},
  year={2022}
}

@article{xiao2021mixing,
  title={Mixing activations and labels in distributed training for split learning},
  author={Xiao, Danyang and Yang, Chengang and Wu, Weigang},
  journal={IEEE Transactions on Parallel and Distributed Systems (TPDS)},
  year={2021},
}

@article{wu2023federated,
  title={Federated split learning with data and label privacy preservation in vehicular networks},
  author={Wu, Maoqiang and Cheng, Guoliang and Ye, Dongdong and Kang, Jiawen and Yu, Rong and Wu, Yuan and Pan, Miao},
  journal={IEEE Transactions on Vehicular Technology (TVT)},
  year={2023},
}

@inproceedings{chen2022graph,
  title={Graph unlearning},
  author={Chen, Min and Zhang, Zhikun and Wang, Tianhao and Backes, Michael and Humbert, Mathias and Zhang, Yang},
  booktitle={ACM SIGSAC Conference on Computer and Communications Security (CCS)},
  year={2022}
}

@inproceedings{chen2022recommendation,
  title={Recommendation unlearning},
  author={Chen, Chong and Sun, Fei and Zhang, Min and Ding, Bolin},
  booktitle={Proceedings of the ACM Web Conference (WWW)},
  year={2022}
}

@article{chen2023unlearn,
  title={Unlearn what you want to forget: Efficient unlearning for {LLM}s},
  author={Chen, Jiaao and Yang, Diyi},
  journal={The Conference on Empirical Methods in Natural Language Processing (EMNLP)},
  year={2024}
}

@inproceedings{he2016deep,
  title={Deep residual learning for image recognition},
  author={He, Kaiming and Zhang, Xiangyu and Ren, Shaoqing and Sun, Jian},
  booktitle={IEEE Conference on Computer Vision and Pattern Recognition (CVPR)},
  year={2016}
}

@article{nguyen2021federated,
  title={Federated learning for internet of things: A comprehensive survey},
  author={Nguyen, Dinh C and Ding, Ming and Pathirana, Pubudu N and Seneviratne, Aruna and Li, Jun and Poor, H Vincent},
  journal={IEEE Communications Surveys \& Tutorials},
  year={2021},
}

@article{chundawat2023zero,
  title={Zero-shot machine unlearning},
  author={Chundawat, Vikram S and Tarun, Ayush K and Mandal, Murari and Kankanhalli, Mohan},
  journal={IEEE Transactions on Information Forensics and Security (TIFS)},
  year={2023},
}

@article{tarun2023fast,
  title={Fast yet effective machine unlearning},
  author={Tarun, Ayush K and Chundawat, Vikram S and Mandal, Murari and Kankanhalli, Mohan},
  journal={IEEE Transactions on Neural Networks and Learning Systems (TNNLS)},
  year={2023},
}

@article{ghazi2021deep,
  title={Deep learning with label differential privacy},
  author={Ghazi, Badih and Golowich, Noah and Kumar, Ravi and Manurangsi, Pasin and Zhang, Chiyuan},
  journal={Advances in Neural Information Processing Systems (NeurIPS)},
  year={2021}
}

@inproceedings{gao2024label,
  title={Label Privacy Source Coding in Vertical Federated Learning},
  author={Gao, Dashan and Wan, Sheng and Gu, Hanlin and Fan, Lixin and Yao, Xin and Yang, Qiang},
  booktitle={Joint European Conference on Machine Learning and Knowledge Discovery in Databases (ECML PKDD)},
  year={2024},
}

@inproceedings{ribeiro2019privacy,
  title={Privacy protection with pseudonymization and anonymization in a health IoT system: results from ocariot},
  author={Ribeiro, S{\'e}rgio Lu{\'\i}s and Nakamura, Emilio Tissato},
  booktitle={IEEE International Conference on Bioinformatics and Bioengineering (BIBE)},
  pages={904--908},
  year={2019},
  organization={IEEE}
}

@article{lu2022label,
  title={Label-only membership inference attacks on machine unlearning without dependence of posteriors},
  author={Lu, Zhaobo and Liang, Hai and Zhao, Minghao and Lv, Qingzhe and Liang, Tiancai and Wang, Yilei},
  journal={International Journal of Intelligent Systems},
  volume={37},
  number={11},
  pages={9424--9441},
  year={2022},
  publisher={Wiley Online Library}
}

@article{love2002comparing,
  title={Comparing supervised and unsupervised category learning},
  author={Love, Bradley C},
  journal={Psychonomic bulletin \& review},
  volume={9},
  number={4},
  pages={829--835},
  year={2002},
  publisher={Springer}
}

@article{osisanwo2017supervised,
  title={Supervised machine learning algorithms: classification and comparison},
  author={Osisanwo, FY and Akinsola, JET and Awodele, O and Hinmikaiye, JO and Olakanmi, O and Akinjobi, J and others},
  journal={International Journal of Computer Trends and Technology (IJCTT)},
  volume={48},
  number={3},
  pages={128--138},
  year={2017}
}

@article{tan2022federated,
  title={Federated learning from pre-trained models: A contrastive learning approach},
  author={Tan, Yue and Long, Guodong and Ma, Jie and Liu, Lu and Zhou, Tianyi and Jiang, Jing},
  journal={Advances in Neural Information Processing Systems (NeuIPS)},
  year={2022}
}

@article{li2023subspace,
  title={Subspace based federated unlearning},
  author={Li, Guanghao and Shen, Li and Sun, Yan and Hu, Yue and Hu, Han and Tao, Dacheng},
  journal={arXiv preprint arXiv:2302.12448},
  year={2023}
}

@inproceedings{deshpande2021sypse,
  title={Sypse: Privacy-first Data Management through Pseudonymization and Partitioning.},
  author={Deshpande, Amol},
  booktitle={CIDR},
  year={2021}
}

@inproceedings{zhang2020batchcrypt,
  title={$\{$BatchCrypt$\}$: Efficient homomorphic encryption for $\{$Cross-Silo$\}$ federated learning},
  author={Zhang, Chengliang and Li, Suyi and Xia, Junzhe and Wang, Wei and Yan, Feng and Liu, Yang},
  booktitle={USENIX annual technical conference (ATC)},
  pages={493--506},
  year={2020}
}

@inproceedings{liu2021federaser,
  title={Federaser: Enabling efficient client-level data removal from federated learning models},
  author={Liu, Gaoyang and Ma, Xiaoqiang and Yang, Yang and Wang, Chen and Liu, Jiangchuan},
  booktitle={IEEE/ACM International Symposium on Quality of Service (IWQOS)},
  pages={1--10},
  year={2021},
  organization={IEEE}
}

@article{takahashi2019data,
  title={Data augmentation using random image cropping and patching for deep CNNs},
  author={Takahashi, Ryo and Matsubara, Takashi and Uehara, Kuniaki},
  journal={IEEE Transactions on Circuits and Systems for Video Technology},
  year={2019},
}

@article{shorten2019survey,
  title={A survey on image data augmentation for deep learning},
  author={Shorten, Connor and Khoshgoftaar, Taghi M},
  journal={Journal of big data},
  volume={6},
  number={1},
  pages={1--48},
  year={2019},
  publisher={Springer}
}

@inproceedings{syakur2018integration,
  title={Integration k-means clustering method and elbow method for identification of the best customer profile cluster},
  author={Syakur, Muhammad Ali and Khotimah, B Khusnul and Rochman, EMS and Satoto, Budi Dwi},
  booktitle={IOP Conference Series: Materials Science and Engineering},
  volume={336},
  pages={012017},
  year={2018},
  organization={IOP Publishing}
}

@inproceedings{bradley1998refining,
  title={Refining initial points for k-means clustering.},
  author={Bradley, Paul S and Fayyad, Usama M},
  booktitle={ICML},
  volume={98},
  pages={91--99},
  year={1998},
  organization={Citeseer}
}

@article{yuan2019research,
  title={Research on K-value selection method of K-means clustering algorithm},
  author={Yuan, Chunhui and Yang, Haitao},
  journal={J},
  volume={2},
  number={2},
  pages={226--235},
  year={2019},
  publisher={MDPI}
}

@inproceedings{bao2024efficient,
  title={Efficient Target Propagation by Deriving Analytical Solution},
  author={Bao, Yanhao and Shibuya, Tatsukichi and Sato, Ikuro and Kawakami, Rei and Inoue, Nakamasa},
  booktitle={AAAI Conference on Artificial Intelligence (AAAI)},
  year={2024}
}

@inproceedings{jin2023poster,
  title={POSTER: ML-Compass: A Comprehensive Assessment Framework for Machine Learning Models},
  author={Jin, Zhibo and Zhu, Zhiyu and Hu, Hongsheng and Xue, Minhui and Chen, Huaming},
  booktitle={ACM Asia Conference on Computer and Communications Security (AsiaCCS)},
  year={2023}
}

@article{nguyen2021dataset,
  title={Dataset distillation with infinitely wide convolutional networks},
  author={Nguyen, Timothy and Novak, Roman and Xiao, Lechao and Lee, Jaehoon},
  journal={Advances in Neural Information Processing Systems (NeurIPS)},
  volume={34},
  pages={5186--5198},
  year={2021}
}

@article{gowal2021improving,
  title={Improving robustness using generated data},
  author={Gowal, Sven and Rebuffi, Sylvestre-Alvise and Wiles, Olivia and Stimberg, Florian and Calian, Dan Andrei and Mann, Timothy A},
  journal={Advances in Neural Information Processing Systems (NeurIPS)},
  volume={34},
  pages={4218--4233},
  year={2021}
}

@inproceedings{song2019privacy,
  title={Privacy risks of securing machine learning models against adversarial examples},
  author={Song, Liwei and Shokri, Reza and Mittal, Prateek},
  booktitle={ACM SIGSAC Conference on Computer and Communications Security (AsiaCCS)},
  year={2019}
}

@article{GUPTA20181,
title = {Distributed learning of deep neural network over multiple agents},
journal = {Journal of Network and Computer Applications},
volume = {116},
pages = {1-8},
year = {2018},
author = {Otkrist, Gupta and Ramesh, Raskar},
}

@inproceedings{gao2023pcat,
  title={$\{$PCAT$\}$: Functionality and data stealing from split learning by $\{$Pseudo-Client$\}$ attack},
  author={Gao, Xinben and Zhang, Lan},
  booktitle={32nd USENIX Security Symposium (USENIX Sec)},
  pages={5271--5288},
  year={2023}
}

@article{poirot2019split,
  title={Split learning for collaborative deep learning in healthcare},
  author={Poirot, Maarten G and Vepakomma, Praneeth and Chang, Ken and Kalpathy-Cramer, Jayashree and Gupta, Rajiv and Raskar, Ramesh},
  journal={arXiv preprint arXiv:1912.12115},
  year={2019}
}

@article{vepakomma2018split,
  title={Split learning for health: Distributed deep learning without sharing raw patient data},
  author={Vepakomma, Praneeth and Gupta, Otkrist and Swedish, Tristan and Raskar, Ramesh},
  journal={arXiv preprint arXiv:1812.00564},
  year={2018}
}

@article{zhang2023privacy,
  title={Privacy and efficiency of communications in federated split learning},
  author={Zhang, Zongshun and Pinto, Andrea and Turina, Valeria and Esposito, Flavio and Matta, Ibrahim},
  journal={IEEE Transactions on Big Data (TBD)},
  volume={9},
  number={5},
  pages={1380--1391},
  year={2023},
  publisher={IEEE}
}

@article{singh2019detailed,
  title={Detailed comparison of communication efficiency of split learning and federated learning},
  author={Singh, Abhishek and Vepakomma, Praneeth and Gupta, Otkrist and Raskar, Ramesh},
  journal={arXiv preprint arXiv:1909.09145},
  year={2019}
}

@article{wu2023split,
  title={Split learning over wireless networks: Parallel design and resource management},
  author={Wu, Wen and Li, Mushu and Qu, Kaige and Zhou, Conghao and Shen, Xuemin and Zhuang, Weihua and Li, Xu and Shi, Weisen},
  journal={IEEE Journal on Selected Areas in Communications},
  volume={41},
  number={4},
  pages={1051--1066},
  year={2023},
  publisher={IEEE}
}

@ARTICLE{9964015,
  author={Wu, Leijie and Guo, Song and Wang, Junxiao and Hong, Zicong and Zhang, Jie and Ding, Yaohong},
  journal={IEEE Network}, 
  title={Federated Unlearning: Guarantee the Right of Clients to Forget}, 
  year={2022},
  volume={36},
  number={5},
  pages={129-135},
}

@article{distil1Unl,
  title={Distill to Delete: Unlearning in Graph Networks with Knowledge Distillation},
  author={Sinha, Yash and Mandal, Murari and Kankanhalli, Mohan},
  journal={arXiv preprint arXiv:2309.16173},
  year={2023}
}

@inproceedings{dong2023rai2,
  title={RAI2: Responsible Identity Audit Governing the Artificial Intelligence.},
  author={Dong, Tian and Li, Shaofeng and Chen, Guoxing and Xue, Minhui and Zhu, Haojin and Liu, Zhen},
  booktitle={NDSS},
  year={2023}
}

@article{zhang2015character,
  title={Character-level convolutional networks for text classification},
  author={Zhang, Xiang and Zhao, Junbo and LeCun, Yann},
  journal={Advances in neural information processing systems},
  volume={28},
  year={2015}
}

@article{zhang1996birch,
  title={BIRCH: an efficient data clustering method for very large databases},
  author={Zhang, Tian and Ramakrishnan, Raghu and Livny, Miron},
  journal={ACM sigmod record},
  volume={25},
  number={2},
  pages={103--114},
  year={1996},
  publisher={ACM New York, NY, USA}
}

\appendix
\section*{Appendix}


\section{Proof of Theorem~\ref{theorem_DP-based label anonymization}}\label{appendix_proof}

\begin{prf}[Theorem~\ref{theorem_DP-based label anonymization}]

$ \forall  {V_{a}^*},  {V_{b}^*} \in  \mathbb{V}_{exp} $ and $\forall {V}_{t} \in \mathbb{V} $,

\begin{equation}
    \begin{aligned}
        & e^{-\varepsilon} \leq \frac{\Pr[ {V}_{t} | {V_{a}^*} ]}{\Pr[  {V}_{t} | {V_{b}^*}]} \leq e^{\varepsilon}. 
    \end{aligned}
\label{equation_4_2_2}
\end{equation}

Thus, the probability that any two elements in $\mathbb{V}_{exp}$ correspond to the same element in $\mathbb{V}$ can be evaluated.
Use $ \boxtimes $ to indicate that two masked elements correspond to the exact origin one.
$ \forall  {V_{a}^*},  {V_{b}^*} ,  {V_{c}^*}\in  \mathbb{V}_{exp} $, according to~\eqref{equation_4_2_2}, we have

\begin{equation}
    \begin{aligned}
        \frac{\Pr [{V_{a}^*} \boxtimes {V_{b}^*}]}{\Pr [{V_{a}^*} \boxtimes {V_{c}^*}]} & = \frac{\sum_{q=1}^{Q}{\Pr[ {V}_{t} | {V_{a}^*}] \Pr[ {V}_{t} | {V_{b}^*}] }}{\sum_{q=1}^{Q}{\Pr[ {V}_{t} | {V_{a}^*}] \Pr[ {V}_{t} | {V_{c}^*}] }} \\
        & \leq \frac{\sum_{q=1}^{Q}{\Pr[ {V}_{t} | {V_{a}^*}] ( (e^{\varepsilon} \Pr[ {V}_{t} | {V_{a}^*}] )}}{\sum_{q=1}^{Q}{\Pr[ {V}_{t} | {V_{a}^*}] ((e^{-\varepsilon} \Pr[ {V}_{t} | {V_{a}^*}] )}} \\ 
        & = e^{2\varepsilon}.
    \end{aligned}
\label{equation_4_2_4}
\end{equation}

Thus, the probability that any two elements in $\mathbb{V}_{exp}$ correspond to the same element in $\mathbb{V}$ satisfies $2 \varepsilon$-DP.

\end{prf}

As demonstrated in Theorem~\ref{theorem_DP-based label anonymization}, the server cannot infer the mapping $\mathcal{G}_V$ from the differences between any two DP-protected intermediate values. This inability also prevents the server from deducing the mapping $\mathcal{G}_Y$ from the intermediate values, thereby protecting the original label information from being derived from the masked labels. Within the \textsc{SplitWiper} framework, carefully selecting the parameter $\varepsilon$ is essential to balance privacy with maintaining data utility and model effectiveness.


\section{Complexity Analysis}
\label{appendix_complexity}

To further demonstrate the advantages of the \textsc{SplitWiper} and \textsc{SplitWiper+ }frameworks in reducing computational and communication overhead, we analyze its complexity in both learning and unlearning scenarios. 
We compare these metrics with traditional Vanilla and U-shaped SL, which adhere to the standard procedures of the usual round-robin SL processes detailed in~\cite{GUPTA20181, gao2023pcat}, as shown in Table~\ref{tab_overhead} and Table~\ref{tab_overhead_server}.

\subsection{Learning Scenarios}

\subsubsection{Computational Complexity}

\noindent\textbf{Client-side computational complexity.}
In \textsc{SplitWiper} framework, the primary computational cost is derived from the pre-training of local models on each client $k$, which is $ O(N \cdot \mathcal{H}_K) $, where $ N $ is the epochs for client model pre-training and $ \mathcal{H}_K $ is the complexity of a single training iteration.
While in \textsc{SplitWiper+}, beyond the pre-training computational cost, there is an additional component that arises from the computations required by the label expansion strategy implemented after the pre-training phase on the client side. 
This additional cost is $O(\gamma \;\mathcal{H}_{exp})$, where $\mathcal{H}_{exp}$ represents the complexity of the label expansion-and-masking scheme, and $\gamma$ is the average of the label expansion factors.

However, since the pre-training of client models is conducted independently of the server, it is excluded from our computational complexity analysis for the SL process. 
Instead, we focus on the final epoch of the pre-training process, disregarding the impact of $N$.
Due to the one-way-one-off propagation design, clients freeze their local models after pre-training and do not participate in the subsequent $M$ epochs of server training. 
Therefore, the computational complexity of the proposed frameworks is also unaffected by $M$.
For a total of $K$ clients, the computational complexity across all clients in \textsc{SplitWiper} is quantified as $O(K \cdot \mathcal{H}_K)$.
Correspondingly, the computational complexity of \textsc{SplitWiper+} is $O(K \cdot ( \mathcal{H}_K + \gamma \; \mathcal{H}_{exp}))$.

In contrast, traditional SL, such as Vanilla and U-shaped SL, requires all $ K $ clients to be involved across the entire learning process for $ M $ epochs, leading to higher computational complexity.
Specifically, the computational complexity for the Vanilla SL framework is $ O(M \cdot K \cdot \mathcal{H}_{K}) $.
As for the U-shaped framework, which allocates the output layer to the client side, an additional computation cost is required. Consequently, the total complexity for the U-shaped framework is $O(M \cdot K \cdot (\mathcal{H}_{K} + \mathcal{H}_{out}))$, with $\mathcal{H}_{out}$ representing the complexity associated with the output layer operations.

\begin{table}[t]
\centering
\caption{Client-side \textbf{Privacy} and \textbf{Overhead} in SL}
\label{tab_overhead}
\renewcommand{\arraystretch}{1.3} 

\resizebox{\linewidth}{!}{
\begin{tabular}{|c|cccc|} 
\toprule

& \textbf{Framework} & \textbf{\makecell{Label Privacy}} & \textbf{Computational} & \textbf{Communication}  \\
\midrule

 \multirow{4}{*}{\rotatebox{90}{\textbf{Learning}}} &  Vanilla SL  &  \textit{Low}  &    $O(M \cdot K  \cdot \mathcal{H}_{K} )$ &    $ O( M \cdot K \cdot (\mathcal{I}_{K} + \mathcal{I}_{Y})) $\\

&  U-shaped SL &  \textit{High}  &    $O(M \cdot K \cdot (\mathcal{H}_{K}+ \mathcal{H}_{out}))$ &    $ O( M \cdot K  \cdot \mathcal{I}_{K}) $ \\

\cmidrule{2-4}

&  \textbf{\textsc{SplitWiper}}  & \cellcolor{blue!10} \textit{Low} &  \cellcolor{blue!10}  $   O( K \cdot \mathcal{H}_{K} ) $ &\cellcolor{blue!10}   $  O(K \cdot (\mathcal{I}_{K} + \mathcal{I}_{Y})) $ \\

&  \textbf{\textsc{SplitWiper+}}  &     
\cellcolor{blue!10}    \textit{Medium} &     
\cellcolor{blue!10}     $  O( K \cdot (\mathcal{H}_{K} + \gamma \; \mathcal{H}_{exp})) $ &  \cellcolor{blue!10}     $ O(K \cdot  \gamma \; (\mathcal{I}_{K} + \mathcal{I}_{Y})) $ \\

\midrule

 \multirow{4}{*}{\rotatebox{90}{\textbf{Unlearning}}} &  Vanilla SL &  \textit{Low} &     $O(M \cdot K \cdot  \mathcal{H}_{K} )$ &    $ O( M \cdot K \cdot  (\mathcal{I}_{K} + \mathcal{I}_{Y})) $ \\

&  U-shaped SL & \textit{High} &    $O(M \cdot K \cdot  (\mathcal{H}_{K} + \mathcal{H}_{out}))$ &    $ O( M \cdot K \cdot  \mathcal{I}_{K}) $  \\

\cmidrule{4-5}

& \multirow{1}*{  \textbf{\textsc{SplitWiper}} } &   \cellcolor{yellow!25}  \multirow{1}*{ \textit{Low} }  &    \cellcolor{yellow!25}    $O(\mathcal{H}_{K}) $ &  \cellcolor{yellow!25}  $ O( \mathcal{I}_{K} + \mathcal{I}_{Y}) $\\

& \multirow{1}*{  \textbf{\textsc{SplitWiper+}} } &   \cellcolor{yellow!25}  \multirow{1}*{ \textit{Medium} }  &    \cellcolor{yellow!25}   $O(\mathcal{H}_{K} + \gamma \; \mathcal{H}_{exp}) $ &     \cellcolor{yellow!25}   $ O( \gamma \; (\mathcal{I}_{K} + \mathcal{I}_{Y})) $\\

\cmidrule{2-4}

\end{tabular}
}
\begin{tablenotes}

\item {\footnotesize \textit{High}: no label sharing; 
\textit{Low}: real label sharing; \textit{Medium}: masked label sharing; $ \mathcal{H}_K $: client-side training complexity; $\mathcal{H}_{exp}$: label expansion-and-masking scheme complexity; $\mathcal{H}_{out}$: output layer complexity; $ \mathcal{I}_{K}$: size of intermediate outputs; $ \mathcal{I}_{Y} $: size of labels ; $\mathcal{H}_{out}$: output layer complexity.}

\end{tablenotes}
\end{table}

\begin{table}[t]
\centering
\caption{Server-side \textbf{Overhead} in SL}
\label{tab_overhead_server}
\renewcommand{\arraystretch}{1.3} 

\resizebox{\linewidth}{!}{
\begin{tabular}{|c|cccc|} 
\toprule

& \textbf{Framework}  & \textbf{Computational} & \textbf{Communication} & \textbf{\makecell{Storage}}  \\
\midrule

 \multirow{4}{*}{\rotatebox{90}{\textbf{Learning}}} &  Vanilla SL   &    $ O(M \cdot \mathcal{H}_{S}) $ &    $O(M \cdot \mathcal{I}_{loss})$ &  $O( \mathcal{W}_s)$ \\

&  U-shaped SL  &    $ O(M \cdot (\mathcal{H}_{S} - \mathcal{H}_{out})) $ &    $O(M \cdot \mathcal{I}_{S})$ &  $O( \mathcal{W}_s)$ \\

\cmidrule{2-4}

&  \textbf{\textsc{SplitWiper}}  &  \cellcolor{blue!10}  $ O(M \cdot \mathcal{H}_{S}) $ &\cellcolor{blue!10}  \textit{NA} & \cellcolor{blue!10}  $O(K \cdot \mathcal{D}_{inter} + \mathcal{W}_s )$\\

&  \textbf{\textsc{SplitWiper+}}  &     
\cellcolor{blue!10}$ O(M \cdot \mathcal{H}_{S}) $ &  \cellcolor{blue!10}  \textit{NA}  & \cellcolor{blue!10}  $O(K \cdot \gamma \cdot \mathcal{D}_{inter} + \mathcal{W}_s)$\\

\midrule

 \multirow{4}{*}{\rotatebox{90}{\textbf{Unlearning}}} &  Vanilla SL &   $O(M \cdot K \cdot  \mathcal{H}_{K} )$ &    $ O( M \cdot K \cdot  (\mathcal{I}_{K} + \mathcal{I}_{Y})) $ &  $O( \mathcal{W}_s)$ \\

&  U-shaped SL &   $O(M \cdot K \cdot  (\mathcal{H}_{K} + \mathcal{H}_{out}))$ &    $ O( M \cdot K \cdot  \mathcal{I}_{K}) $ &  $O( \mathcal{W}_s)$ \\

\cmidrule{4-5}

& \multirow{1}*{  \textbf{\textsc{SplitWiper}} }  &    \cellcolor{yellow!25}    $O(\mathcal{H}_{K}) $ &  \cellcolor{yellow!25}  $ O( \mathcal{I}_{K} + \mathcal{I}_{Y}) $ &  \cellcolor{yellow!25} $O(K \cdot \mathcal{D}_{inter} + \mathcal{W}_s )$\\

& \multirow{1}*{  \textbf{\textsc{SplitWiper+}} } &     \cellcolor{yellow!25}   $O(\mathcal{H}_{K} + \gamma \; \mathcal{H}_{exp}) $ &     \cellcolor{yellow!25}   $ O( \gamma \; (\mathcal{I}_{K} + \mathcal{I}_{Y})) $ &  \cellcolor{yellow!25} $O(K \cdot \gamma \cdot \mathcal{D}_{inter} + \mathcal{W}_s)$\\

\cmidrule{2-4}

\end{tabular}
}
\begin{tablenotes}

\item {\footnotesize  $ \mathcal{H}_{S} $: server-side training complexity; $\mathcal{H}_{out}$: output layer complexity;  $\mathcal{I}_{loss}$: size of server-side loss; $\mathcal{I}_{S}$: size of server-side outputs; $ \mathcal{I}_{K}$: size of intermediate outputs; $ \mathcal{I}_{Y} $: size of labels; $\mathcal{W}_s$: size of server-side model weights; $\mathcal{D}_{inter}$: size of client-side intermediate values.}

\end{tablenotes}
\end{table}

In Algorithm~\ref{algo:label-expansion}, the complexity of expansion-and-masking $\mathcal{H}_{exp}$ scales linearly with the number of real labels and the size of intermediate values. This complexity remains within the same order of magnitude as $\mathcal{H}_{K}$, especially given the significant size of large models today. This computational load drives clients to use SL in collaboration with a server rather than independently handling the entire learning process. For security and efficiency, the expansion factor $\gamma$ is typically kept small, and training epochs $M$ are much larger, often ranging from several dozen to hundreds. As a result, the computational overhead of \textsc{SplitWiper+} is substantially lower than that of traditional Vanilla and U-shaped SL frameworks. Moreover, \textsc{SplitWiper} has only $1/M$ of the computational complexity of Vanilla SL, further highlighting its advantages.

\smallskip
\noindent\textbf{Server-side computational complexity.}
The server’s computational cost for updating its model in \textsc{SplitWiper} and \textsc{SplitWiper+} is $O(M \cdot \mathcal{H}{S})$, where $M$ denotes the number of training epochs, and $\mathcal{H}{S}$ represents the complexity of a single training iteration.
This complexity mirrors that in the Vanilla SL framework, where the server-side complexity is also $ O(M \cdot \mathcal{H}_{s}) $. 
Conversely, the U-shaped SL framework generally exhibits slightly lower server-side complexity, which is $ O(M \cdot (\mathcal{H}_{S} - \mathcal{H}_{out})) $, because clients undertake output layer operations, offloading computational demands from the server.

\subsubsection{Communication Overhead}

\noindent\textbf{Client-side communication complexity.}
In the \textsc{SplitWiper} and the \textsc{SplitWiper+} frameworks, the communication overhead is determined by transmitting the (masked) labels and corresponding intermediate outputs from each client to the server in a one-way propagation manner. 
This leads to a complexity of $ O(K \cdot (\mathcal{I}_{K}+\mathcal{I}_{Y})) $ for \textsc{SplitWiper} and $ O(K \cdot \gamma \;(\mathcal{I}_{K}+\mathcal{I}_{Y})) $ for \textsc{SplitWiper+}, where $ \mathcal{I}_{K}$ is the size of intermediate outputs and $ \mathcal{I}_{Y} $ is the size of labels on client-side. 

Compared to Vanilla SL, which incurs a per-epoch communication complexity of $ O(K \cdot (\mathcal{I}_{K}+\mathcal{I}_{Y}))$ and U-shaped SL that does not require label transmission, our label expansion strategy in \textsc{SplitWiper+} does introduce additional communication overhead. 
However, considering that traditional SL frameworks involve continuous communication between clients and the server across $ M $ epochs, the overall communication complexities are respectively $ O(M \cdot K \cdot (\mathcal{I}_{K} + \mathcal{I}_{Y})) $ and $ O(M \cdot K \cdot \mathcal{I}_{K}) $. 
Given that $M$ is significantly greater than $\gamma$, the \textsc{SplitWiper+} framework still maintains a substantial advantage regarding communication efficiency.

\smallskip
\noindent\textbf{Server-side communication complexity.}
In the proposed frameworks, the information transfer from clients to the server follows a one-way propagation manner, allowing us to consider that the server incurs no communication costs. 
In contrast, the Vanilla SL framework requires the server to send the loss back to each client to facilitate model training, leading to a communication complexity of $O(M \cdot \mathcal{I}_{loss})$ over $M$ epochs, where $\mathcal{I}_{loss}$ represents the size of the loss during each epoch.
Similarly, our U-shaped SL necessitates the server to return outputs back to clients for operations related to the final output layer, resulting in a communication complexity of $O(M \cdot \mathcal{I}_{S})$, 
where $\mathcal{I}_{S}$ is the size of the server-side outputs. 


\subsubsection{Server-side Storage Overhead} 

In addition to computational and communication overheads, our \textsc{SplitWiper} and \textsc{SplitWiper+} frameworks introduce additional storage overhead on the server to support the one-way-one-off propagation scheme. 
The server needs to store the intermediate values received from clients, ensuring that these values can be readily accessed during unlearning scenarios without disrupting other clients which do no request unlearning, whereas in Vanilla SL and U-shaped SL, storage overhead depends solely on $\mathcal{W}_s$, the size of the stored server-side model weights.

In \textsc{SplitWiper}, server's storage complexity is $O(K \cdot \mathcal{D}{inter}+\mathcal{W}_s)$, where $K$ is the number of clients and $\mathcal{D}{inter}$ is the average size of the intermediate values transmitted by each client. 
In \textsc{SplitWiper+}, the intermediate values are expanded, resulting in the storage complexity of $O(K \cdot \gamma \cdot \mathcal{D}_{inter}+\mathcal{W}_s)$ with the average of the label expansion factors $\gamma$.

\subsection{Unlearning Scenarios}

In the unlearning scenario, the retraining process in traditional Vanilla and U-shaped SL frameworks necessitates the participation of all clients. 
This implies that each client must re-initiate model updates and transmit intermediate values for an additional $M$ epochs, replicating the computational and communication costs observed in the learning scenario. 

Conversely, \textsc{SplitWiper} and \textsc{SplitWiper+} introduce an optimized approach, wherein unlearning requests are isolated to the specific client initiating the request, significantly reducing the overall burden. 
This isolation avoids the need for unrelated clients to engage in redundant computations and communications, thereby enhancing efficiency and reducing client-side total resource consumption. 
In terms of client-side computational complexity in \textsc{SplitWiper} framework, the primary costs come from the retraining of the local model on the initiating client $k$, which is $O(\mathcal{H}_{K})$.
In contrast, in \textsc{SplitWiper+}, client $k$ also incurs the additional cost of performing a new round of the expansion-and-masking scheme, resulting in a computational complexity of $O(\mathcal{H}_{K}+ \gamma \; \mathcal{H}_{exp})$.

\begin{algorithm}[t]
\linespread{1.1}
\footnotesize
\caption{Learning in \textsc{SplitWiper}}
\begin{algorithmic}[1]
\Require $K$  (\textcolor{teal}{Number of clients}),
 $N$ (\textcolor{teal}{number of epochs for client model training}),
 $M$  (\textcolor{teal}{number of epochs for server model training});
\Ensure Predicted label

\State \textcolor{violet!90}{\textbf{Clients:}} (Conducted in parallel for each client-$k$) 
\For{$k = 1$ to $K$} 
    \State Initialize client model $\mathcal{F}_o^k$   
    \For{$n = 1$ to $N$} \label{alg_1:pre-training_start}
        \State Train $\mathcal{F}_o^k$ on $D_o^k$, excluding server 
    \EndFor \label{alg_1:pre-training_end}
    \State Freeze model weights of $\mathcal{F}_o^k$, disabling gradient updating functions   \label{alg_1:freezing} 
    \State Send labels and intermediate outputs to the server
\EndFor


\State \textcolor{violet!90}{\textbf{Server:}}
\State Cache received labels and intermediate outputs
\State Initialize server model $\mathcal{F}_o^s$
\For{$m = 1$ to $M$}  \label{alg_1:training_start}
    \State Train $\mathcal{F}_o^s$ on cached labels and intermediate outputs   
\EndFor    \label{alg_1:training_end}
\State \textbf{RETURN} predicted label    

\end{algorithmic}
\label{algo:sisa-sl}
\end{algorithm}

\begin{algorithm}[t]
\linespread{1.1}
\footnotesize
\caption{Unlearning in \textsc{SplitWiper}}
\begin{algorithmic}[1]
\Require $N$ (\textcolor{teal}{number of epochs for client model retraining}),  $M$ (\textcolor{teal}{number of epochs for server model updating}), initiating client-$k$

\Ensure Predicted label

\State \textcolor{violet!90}{\textbf{Clients:}} (Only client-$k$ involved, others remain silent)
\State Unfreeze gradient updating functions  \label{alg_2:unfreezing_start} 
\State Initialize local model $ \mathcal{F}_u^k $
\State Remove intended unlearned samples from $D_o^k$, obtaining $D_u^k$ \label{alg_2:unfreezing_end}
\For{$n = 1$ to $N$} \label{alg_2:unlearning_start} 
    \State Train $\mathcal{F}_u^k$ on $D_u^k$, excluding server 
\EndFor
\State Freeze updated weights of $\mathcal{F}_u^k$ \label{alg_2:unlearning_end} 
\State Send intermediate outputs of $\mathcal{F}_u^k$ and corresponding labels to the server


\State \textcolor{violet!90}{\textbf{Server:}}
\State Update cache with intermediate outputs of $\mathcal{F}_u^k$ \label{alg_2:updating_start} 
\State  Initialize server model $ \mathcal{F}_o^s $ based on cache, $ \mathcal{F}_u^s \leftarrow \mathcal{F}_o^s$
\For{$m = 1$ to $M$}
    \State Train $\mathcal{F}_u^s$ on cached records from client-$k$ and other clients
\EndFor \label{alg_2:updating_end} 
\State \textbf{RETURN} predicted label 
\end{algorithmic}
\label{algo:sul-labels}
\end{algorithm}

\begin{algorithm}[t]
\linespread{1}
\footnotesize
\caption{Label Expansion Strategy}
\begin{algorithmic}[1]
\Require $K$  (\textcolor{teal}{number of clients}), $\{\mathbb{Y}_1, \cdots, \mathbb{Y}_K\}$ (\textcolor{teal}{ clients' local label sets}), $\mu$ (\textcolor{teal}{ maximum number of shared labels in each round }), $\lambda_1/\lambda_2$ (\textcolor{teal}{ threshold of client participation }), $\Gamma$ (\textcolor{teal}{selected range for label expansion factor}).

\Ensure Real label set $\mathbb{Y}$, Expanded label set $\mathbb{Y}_{exp}$, and corresponding map between these two set $\mathcal{G}_Y$

\State \textcolor{violet!90}{\textbf{Clients:}} 
\State Initialize two empty sets $\mathbb{Y}$ and $\mathbb{Y}_{exp}$, an empty mapping $\mathcal{G}_Y$, Initialize array $\mathbb{J}$ with $K$ zeros
\Repeat   \label{alg_3:label_sharing_start}
    \For{$k = 1$ to $K$} 
        \State \textbf{if} $\mathbb{J}[k] \neq 0$ \textbf{then} 
 \textbf{continue}
        \State  \textbf{if} generate a random number $r_1 < \lambda_1$ \textbf{then} 
            \State \hspace{4mm} Select $r_{\mu} \leftarrow [1,\min(\max(|\mathbb{Y}_k \setminus \mathbb{Y}|, 1), \mu)]$ elements from $\mathbb{Y}_k \setminus \mathbb{Y}$ and update $\mathbb{Y} \leftarrow \mathbb{Y} \bigcup $ the selected elements
        \State \textbf{if} generate a random number $r_2 < \lambda_2$ and $\mathbb{Y}_j \subseteq \mathbb{Y}$ \textbf{then} $\mathbb{J}[k] \leftarrow 1$

    \EndFor
\Until {$\sum_{k=1}^{K} \mathbb{J}[k] = K$}  \label{alg_3:label_sharing_end}

\State \textbf{for} $q = 1$ to $\left| \mathbb{Y} \right|$: Generate $\gamma_q$ from $\Gamma$ and add $\mathbb{Y}[q]$ for $\gamma_q$ times to $\mathbb{Y}_{exp}$ \label{alg_3:label_expansion_start} 
\State $\mathbb{Y}_{exp} \leftarrow$ Shuffle the elements in $\mathbb{Y}_{exp}$
\State \textbf{for} $q = 1$ to $\left| \mathbb{Y} \right|$: $\mathcal{G}_Y[\mathbb{Y}[q]] \leftarrow$ List($\mathbb{Y}_{exp}.\text{index}(\mathbb{Y}[q])$)

\State $\mathbb{Y}_{exp} \leftarrow$ Re-number the elements in $\mathbb{Y}_{exp}$ according to the index \label{alg_3:label_expansion_end} 
\State \textbf{RETURN} $\mathbb{Y}$, $\mathbb{Y}_{exp}$, $\mathcal{G}_Y$
\end{algorithmic}
\label{algo:label-sharing}
\end{algorithm}

\begin{algorithm}[!]
\linespread{1.1}
\footnotesize
\caption{DP-based Expansion-and-Masking Scheme}
\begin{algorithmic}[1]
\Require $\mathcal{G}_Y$ (\textcolor{teal}{map between $\mathbb{Y}$ and $\mathbb{Y}_{exp}$}), $\varepsilon$ (\textcolor{teal}{privacy budget of DP}), $\{Y_q, V_q^k \}$ (\textcolor{teal}{pairs of labels and intermediate values from client-$k$}).
\Ensure Output set of the expansion-and-masking scheme $\mathbb{U}_{exp} = \{\mathbb{Y}_{exp}, \mathbb{V}_{exp}\} = \bigcup \{Y_e^*, {V_e^*}^k \}$

\State \textcolor{violet!90}{\textbf{Clients:}} 
\State Initialize an empty set $\mathbb{U}_{exp}$
\State $ X \leftarrow  \mathcal{G}_Y[Y_s]$
\For{$x \in  X$ } \label{alg_4:expansion_start}
    \State $Y_e^* \leftarrow x $
    \State ${V_e^*}^k \leftarrow V_q^k + DP(\varepsilon)$
    \State $\mathbb{U}_{exp} \leftarrow \mathbb{U}_{exp} \bigcup \{Y_e^*, {V_e^*}^k \} $
\EndFor  \label{alg_4:expansion_end}
\State \textbf{RETURN} $\mathbb{U}_{exp}$
\end{algorithmic}
\label{algo:label-expansion}
\end{algorithm}

Communication overhead is determined by transmitting the intermediate outputs from the initiating client $k$ to the server, leading to a complexity of $ O(\gamma \;(\mathcal{I}_{K}+\mathcal{I}_{Y})) $ in the proposed frameworks, where $\gamma = 1$ represents \textsc{SplitWiper}, and $\gamma > 1$ represents \textsc{SplitWiper+}.
Additionally, in unlearning scenarios, the server only updates the stored intermediate values related to client $k$ without requiring additional storage costs, allowing us to consider server-side storage overhead during unlearning scenarios consistent with that of learning.
It satisfies \textbf{G1} (independent unlearning), improving the robustness and reducing the overhead of unlearning processes.


\section{Implementation}\label{appendix_algo}

The procedure steps in \textsc{SplitWiper} (cf. Sec.\ref{sec:sisa_design}) can be found from Algorithms~\ref{algo:sisa-sl}\&\ref{algo:sul-labels}, while that of its enhanced privacy-preserving option \textsc{SplitWiper+} (cf. Sec.\ref{sec:label-protection}) can be found from Algorithms~\ref{algo:label-sharing}\&\ref{algo:label-expansion}.

\begin{figure*}[htbp]
    \centering
    \subfigure[CIFAR-10 over VGG\label{fig_convergence_cifar10}]{
    \includegraphics[width=0.3\linewidth]{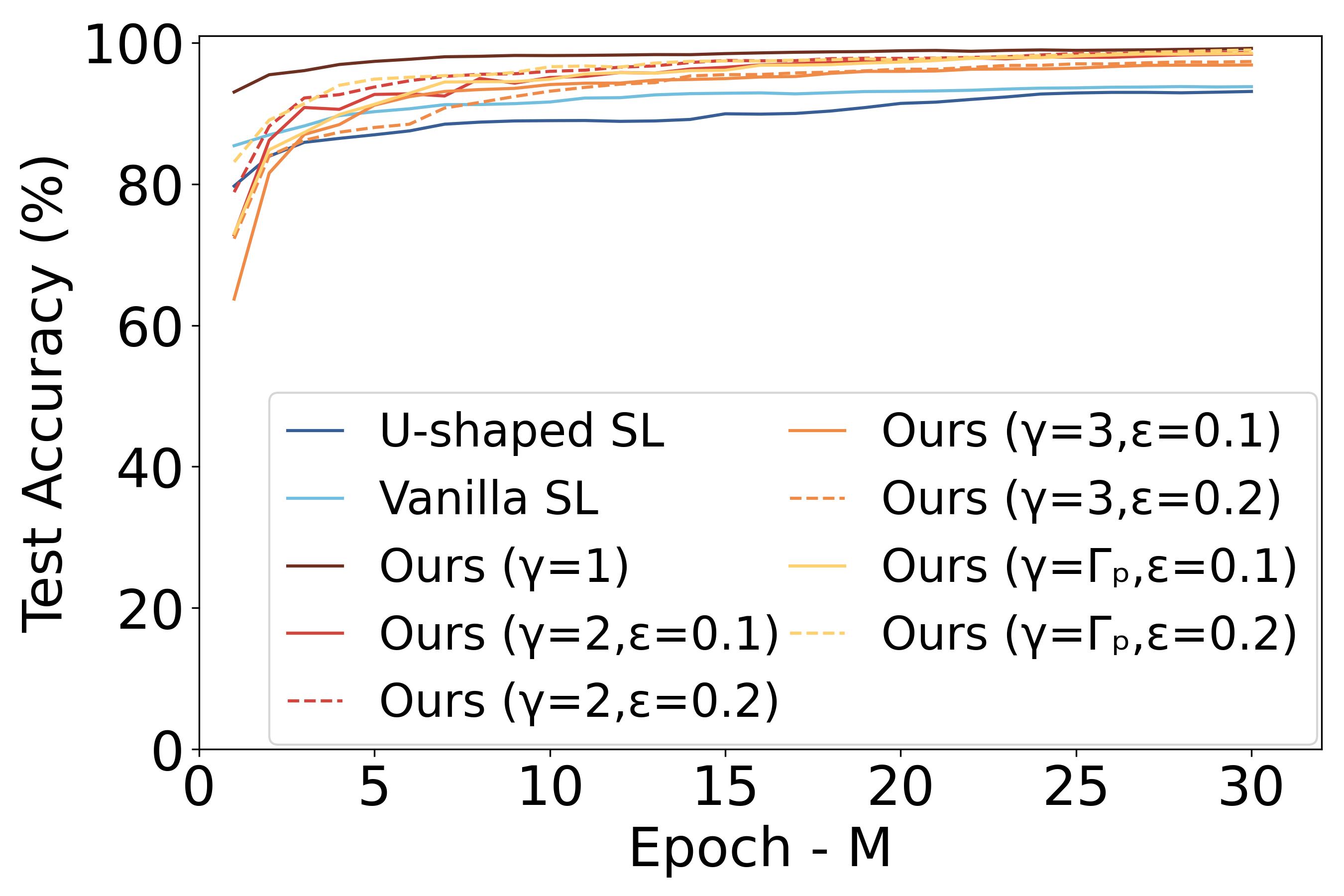}
    }
    \subfigure[CIFAR-100 over VGG\label{fig_convergence_cifar100}]{
    \includegraphics[width=0.3\linewidth]{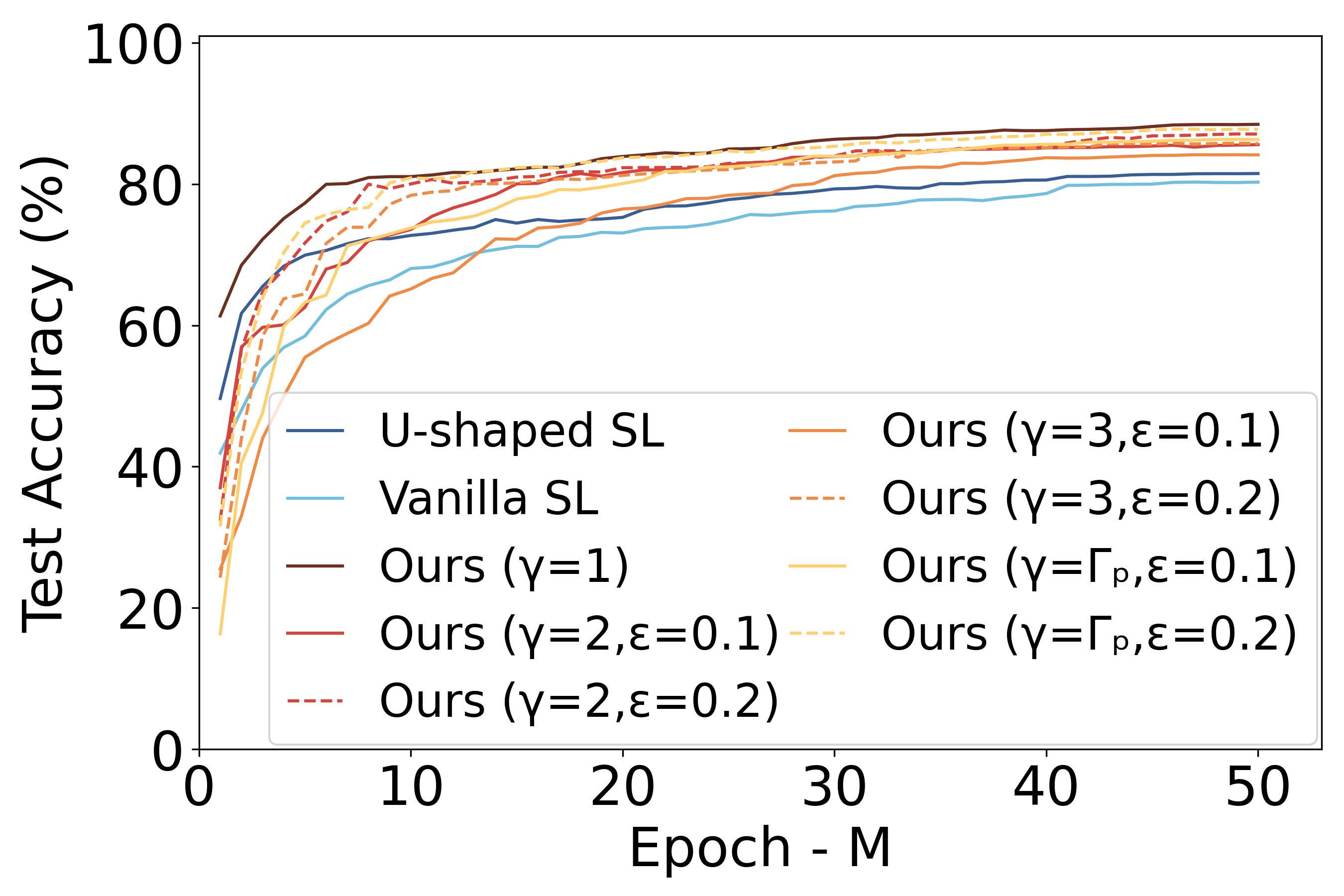}
    }
    \subfigure[CIFAR-10 + MNIST  over VGG\label{fig_convergence_mnistandcifar10}]{
    \includegraphics[width=0.3\linewidth]{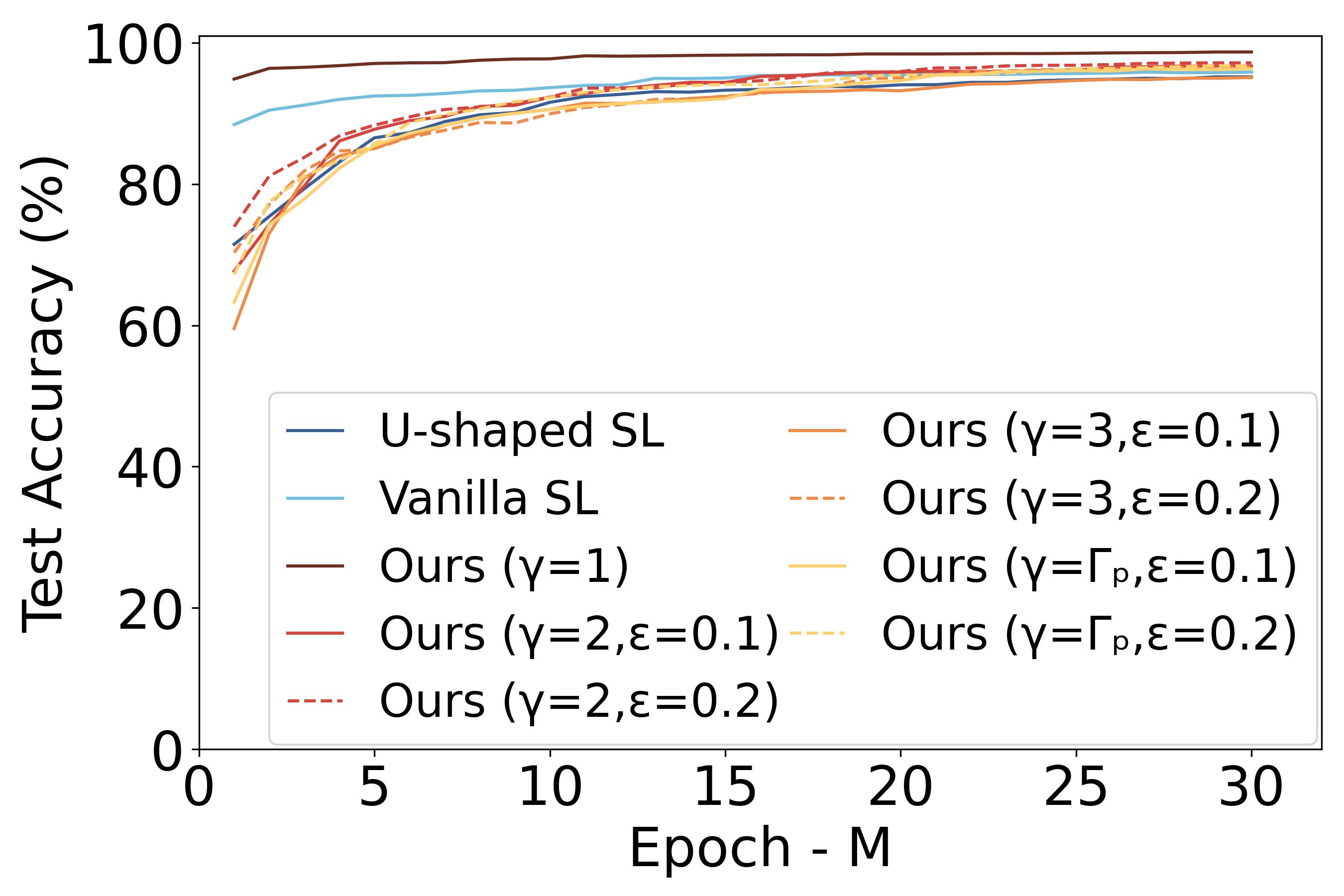}
    }
    \subfigure[CIFAR-10 over ResNet18\label{fig_convergence_cifar10}]{
    \includegraphics[width=0.3\linewidth]{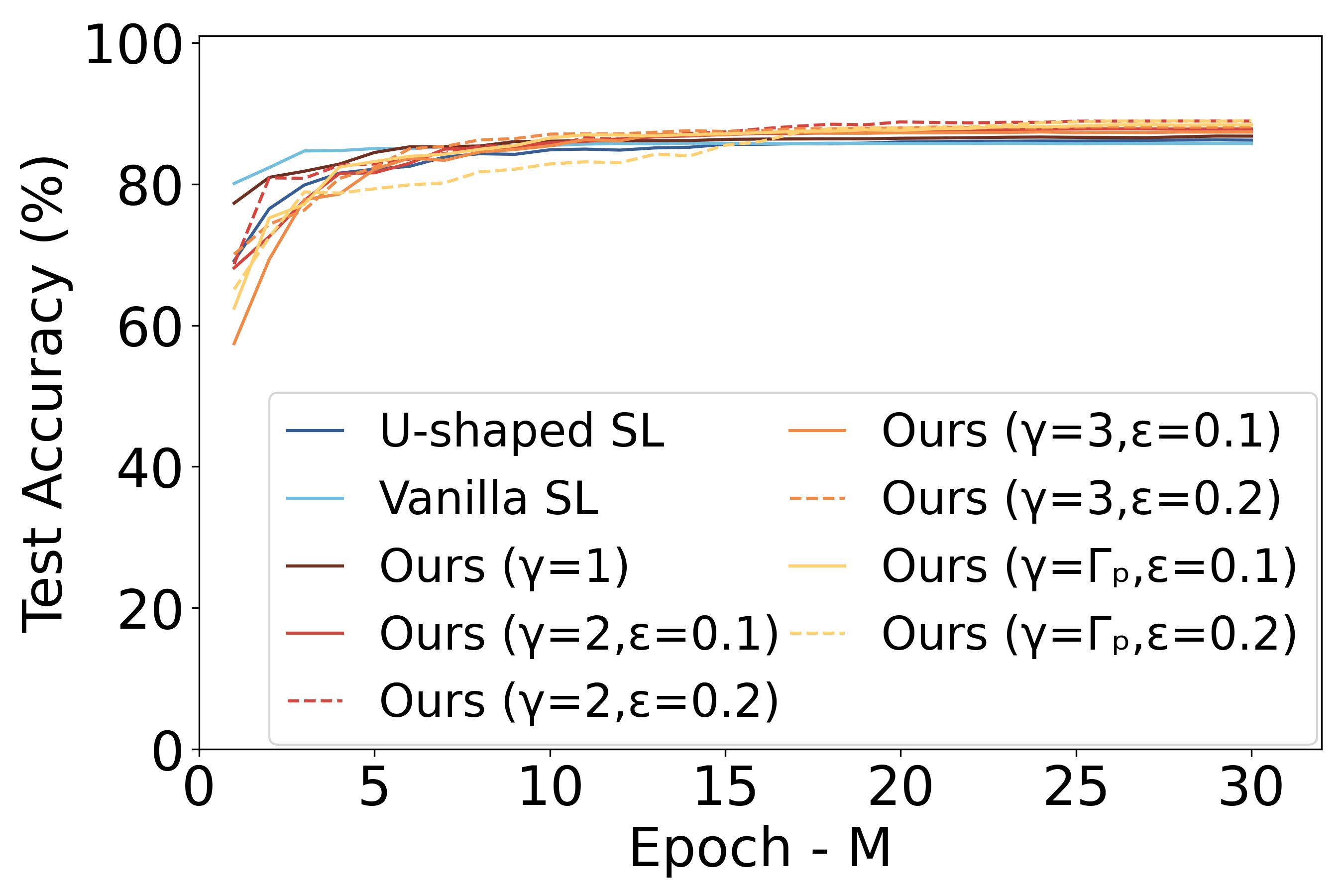}
    }
    \subfigure[CIFAR-100 over ResNet18\label{fig_convergence_cifar100}]{
    \includegraphics[width=0.3\linewidth]{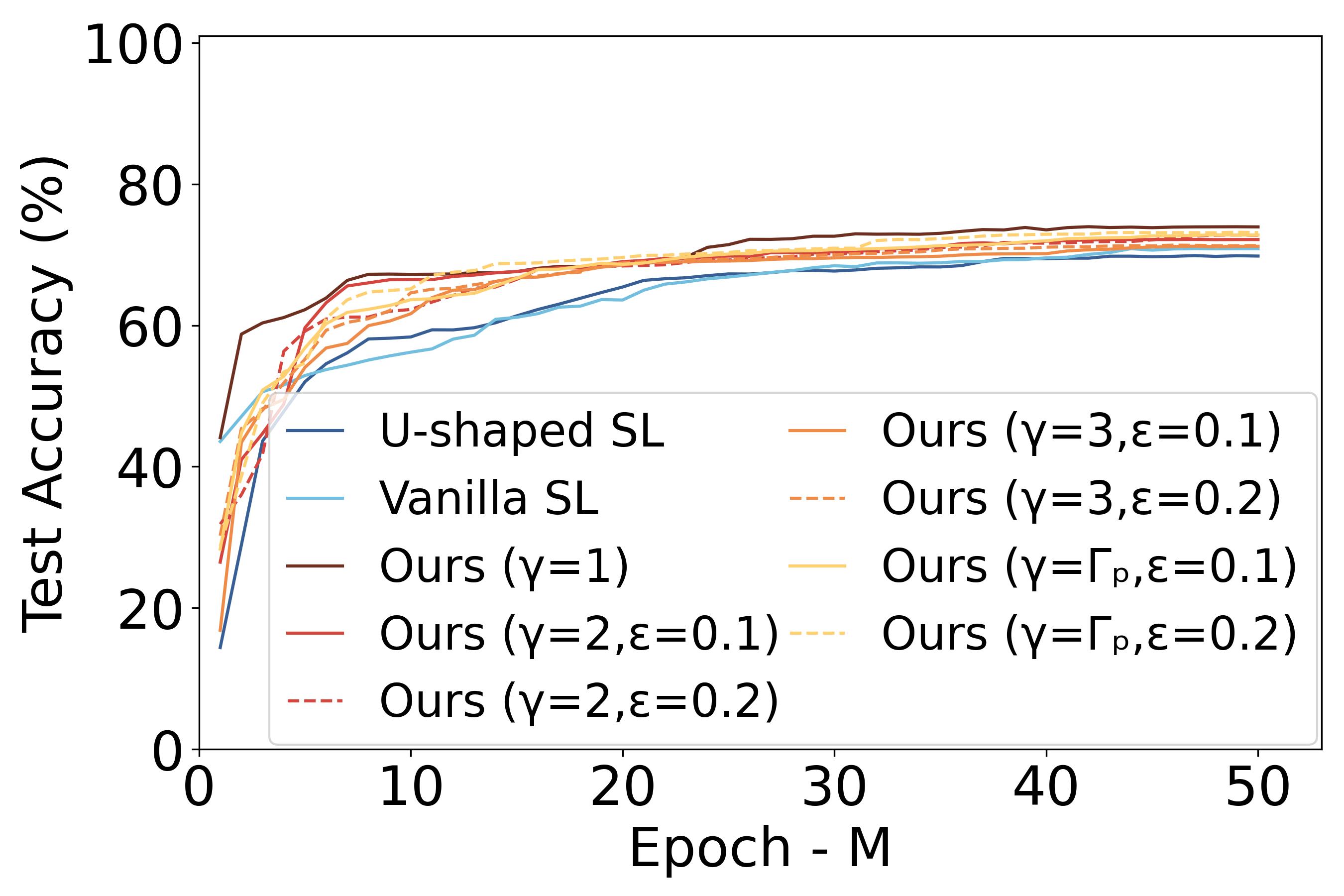}
    }
    \subfigure[CIFAR-10 + MNIST over ResNet18\label{fig_convergence_mnistandcifar10}]{
    \includegraphics[width=0.3\linewidth]{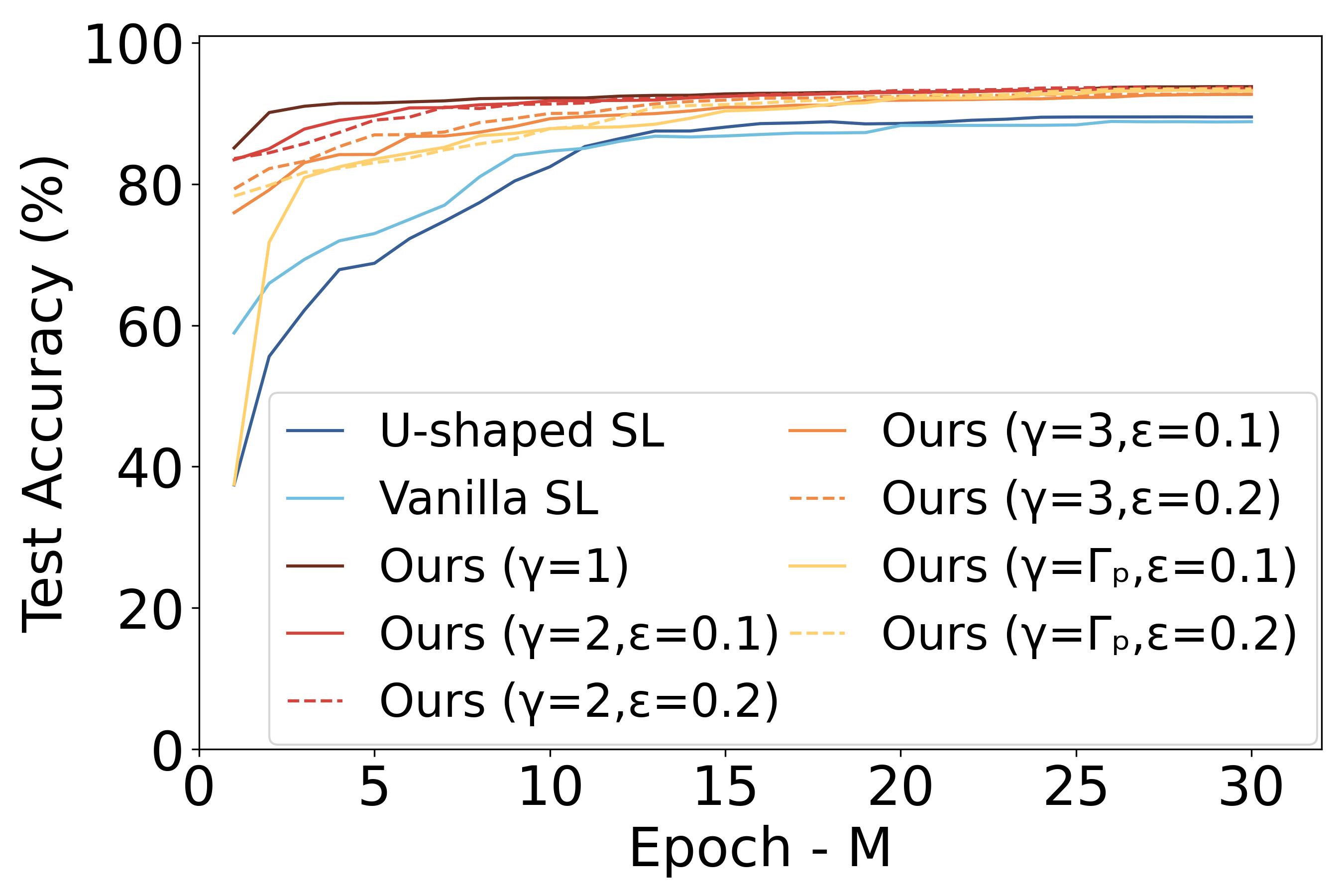}
    }
    \DeclareGraphicsExtensions.
    \caption{Test accuracy vs epochs for different SL frameworks over the VGG and ResNet18 architectures across various datasets with heterogeneous label distributions. As the number of training epochs ($M$) increases, the accuracy of \textsc{SplitWiper} and \textsc{SplitWiper+} converges.}
    \label{fig_convergence}

    \vspace{-0.5em}
\end{figure*}

\section{Evaluating Sample-Level Unlearning Tasks}
\label{appendix_exp_sample unlearning}

Considering sample-level unlearning tasks, the client with unlearning requests, denoted as $\mathbb{C}_{u}$, selects a subset of data samples corresponding to the given label $\mathbb{Y}_{u}$ to perform the unlearning task. The subset of samples designated for unlearning is represented as $\mathbb{X}_{u}$, which constitutes a proportion $\delta$ of the total data associated with $\mathbb{Y}_{u}$. 
Other clients unaffected by unlearning requests are denoted as $\mathbb{C}_{o}$, while labels not requiring unlearning are as $\mathbb{Y}_{o}$.

Consistent with the setup of the previous experiments in Section~\ref{subsection:exp_acc}, we utilize the simplified-VGG~\cite{bao2024efficient} and ResNet18~\cite{he2016deep} as the model architectures for the split (un)learning experiments, considering four different data distribution scenarios across three datasets.

To demonstrate the effectiveness of sample-level unlearning, we focus on the test accuracy of the unlearned label $\mathbb{Y}_{u}$. 
With increasing $\delta$ (Table~\ref{tab_sample_unlearning_acc}), the proportion of unlearned samples grows, leading to an accelerated decline in test accuracy for label $\mathbb{Y}_{u}$, while the performance on other labels $\mathbb{Y}_{o}$ remains largely unaffected. These results highlight the ability to effectively meet sample-level unlearning requirements without compromising the utility of other labels.

Additionally, we examine the impact on membership inference attacks targeting the unlearned samples $\mathbb{X}_{u}$. The success probability of performing membership inference against the requesting client $\mathbb{C}_{u}$ is measured using the \textit{hit rate} of $\mathbb{X}_{u}$.
As shown in Table~\ref{tab_sample_member_inference}, our frameworks significantly reduce the hit rate of $\mathbb{X}_{u}$, thereby exhibiting stronger resistance to membership inference attacks~\cite{10.1145/3460120.3484756,lu2022label} for $\mathbb{C}_{u}$. Compared to Vanilla SL, our frameworks provide enhanced protection against such attacks, particularly in \textsc{SplitWiper+}, where the hit rate of $\mathbb{X}_{u}$ is further minimized, demonstrating superior resilience to member inference vulnerabilities.

\begin{table*}[t]
\centering
\setlength\tabcolsep{2pt}
\renewcommand\arraystretch{1.15}
\caption{Evaluation on \textbf{Efficiency} and \textbf{Effectiveness} over \textcolor{violet}{BERT} across Different SL Frameworks}
\label{tab_time_BERT}

\resizebox{\linewidth}{!}{
\begin{tabular}{|cc|cc|  ccccccc|  cccccccccc|}
\toprule

\multicolumn{4}{c}{} &   \multicolumn{7}{c}{\textbf{(Client-side) Training / \textcolor{violet}{Transmission} Time (s)}} &  \multicolumn{10}{c}{\textbf{Accuracy (\%)}} \\

\cmidrule{6-11}\cmidrule{13-21}

&&&   &   & \multicolumn{2}{c}{\textbf{Existing SL}} & \textbf{\textsc{SplitWiper}} & \multicolumn{3}{c}{\textbf{\textsc{SplitWiper+}}} & &   \multicolumn{2}{c}{\textbf{Existing SL}} &  \textbf{\textsc{SplitWiper}} & \multicolumn{6}{c|}{\textbf{\textsc{SplitWiper+}}} \\

\cmidrule{6-7} \cmidrule{9-11} \cmidrule{13-14}  \cmidrule{16-21} 

&&&  &  &  U-shaped & Vanilla & $\gamma=1$ & \textbf{$\gamma=2$} & \textbf{$\gamma=3$} & \textbf{$\gamma=\Gamma_p$} & &  U-shaped & Vanilla   & $\gamma=1$ & \multicolumn{2}{c}{\textbf{$\gamma=2$}} & \multicolumn{2}{c}{\textbf{$\gamma=3$}}  & \multicolumn{2}{c|}{\textbf{$\gamma=\Gamma_p$}} \\

\midrule
 
 \multirow{4}*{\rotatebox{90}{\textbf{\makecell{Clients with \\ same task}}}} &  \multirow{4}*{\textbf{\rotatebox{90}{\makecell{Partially\\overlapping\\labels}}}} & \multirow{4}*{\rotatebox{90}{\makecell{AG-News}}} & \multirow{2}*{L.} & \makecell{\cellcolor{blue!9}$\mathbb{C}_{u}$   } & 34.77/\textcolor{violet}{51.92} & 33.98/\textcolor{violet}{52.08} & \cellcolor{teal!11} 1.34/\textcolor{violet}{1.74} & \cellcolor{teal!11} 3.71/\textcolor{violet}{3.44} & \cellcolor{teal!11} 5.80/\textcolor{violet}{5.52} & \cellcolor{teal!11} 2.95/\textcolor{violet}{3.02} & 
 \cellcolor{blue!9}$\mathbb{Y}_{u}$  & 81.94 & 88.47 & \cellcolor{yellow!14} 89.21 & \cellcolor{yellow!14} 90.31 & \cellcolor{yellow!14} 90.89 & \cellcolor{yellow!14} 90.37 & \cellcolor{yellow!14} 90.68 & \cellcolor{yellow!14} 91.16 & \cellcolor{yellow!14} 92.32 \\

 & &  &  & \makecell{\cellcolor{blue!9}$\mathbb{C}_{o}$   } & 30.55/\textcolor{violet}{49.26} & 30.30/\textcolor{violet}{49.80} & \cellcolor{teal!11} 1.03/\textcolor{violet}{1.65} & \cellcolor{teal!11} 3.25/\textcolor{violet}{3.26} & \cellcolor{teal!11} 5.12/\textcolor{violet}{4.86} & \cellcolor{teal!11} 2.81/\textcolor{violet}{3.05} & 
 \cellcolor{blue!9}$\mathbb{Y}_{o}$ & 83.14 & 85.68 & \cellcolor{yellow!14} 85.16 & \cellcolor{yellow!14} 86.24 & \cellcolor{yellow!14} 86.67 & \cellcolor{yellow!14} 85.28 & \cellcolor{yellow!14} 85.99 & \cellcolor{yellow!14} 85.08 & \cellcolor{yellow!14} 86.41 \\

 & & & \multirow{2}*{U.} & \makecell{\cellcolor{blue!9}$\mathbb{C}_{u}$   } & 16.45/\textcolor{violet}{40.47} & 16.16/\textcolor{violet}{40.56} & \cellcolor{teal!11} 0.53/\textcolor{violet}{1.35} & \cellcolor{teal!11} 1.78/\textcolor{violet}{2.57} & \cellcolor{teal!11} 3.01/\textcolor{violet}{4.39} & \cellcolor{teal!11} 1.76/\textcolor{violet}{2.26} & 
 \cellcolor{blue!9}$\mathbb{Y}_{u}$ & \textbf{0.00} & \textbf{0.00} & \cellcolor{yellow!14} \textbf{0.00} & \cellcolor{yellow!14} \textbf{0.00} & \cellcolor{yellow!14} \textbf{0.00} & \cellcolor{yellow!14} \textbf{0.00} & \cellcolor{yellow!14} \textbf{0.00} & \cellcolor{yellow!14} \textbf{0.00} & \cellcolor{yellow!14} \textbf{0.00}  \\

 & & &  & \makecell{\cellcolor{blue!9}$\mathbb{C}_{o}$   } & 30.25/\textcolor{violet}{49.03} & 32.57/\textcolor{violet}{49.44} & \cellcolor{teal!11} \textbf{0.00}/\textcolor{violet}{\textbf{0.00}} & \cellcolor{teal!11} \textbf{0.00}/\textcolor{violet}{\textbf{0.00}} & \cellcolor{teal!11} \textbf{0.00}/\textcolor{violet}{\textbf{0.00}} & \cellcolor{teal!11} \textbf{0.00}/\textcolor{violet}{\textbf{0.00}} &  
 \cellcolor{blue!9}$\mathbb{Y}_{o}$ & 82.21 & 85.19 & \cellcolor{yellow!14} 85.82 & \cellcolor{yellow!14} 86.46 & \cellcolor{yellow!14} 86.71 & \cellcolor{yellow!14} 86.07 & \cellcolor{yellow!14} 86.47 & \cellcolor{yellow!14} 87.19 & \cellcolor{yellow!14} 87.32 \\

\midrule

\multicolumn{2}{|c}{\ding{172}}  & \ding{173} &  \multicolumn{1}{c}{\ding{174}} &  \multicolumn{1}{c|}{\ding{175}} &  \multicolumn{6}{c}{}  & \multicolumn{1}{c|}{\ding{176}}  &   \multicolumn{3}{c|}{} & $\varepsilon=0.1$ & $\varepsilon=0.2$  & $\varepsilon=0.1$ & $\varepsilon=0.2$ & $\varepsilon=0.1$ & $\varepsilon=0.2$ \\


\end{tabular}
}

\begin{tablenotes}
      \scriptsize
      \item[] \textit{Notation:}  \ding{172} \textbf{Data distribution}; \ding{173} \textbf{Dataset}; \ding{174} \textbf{Scenario} (L. for learning, U. for unlearning); \ding{175} \textbf{Client type} ($\mathbb{C}_{u}$ for clients involved in unlearning, $\mathbb{C}_{o}$ for clients excluded from unlearning); \ding{176} \textbf{Label type} ($\mathbb{Y}_{u}$ for to-be-unlearned labels, $\mathbb{Y}_{o}$ for retained labels).
     \end{tablenotes}

\end{table*}

\begin{table*}[!]
\centering
\setlength\tabcolsep{2pt}
\renewcommand\arraystretch{1.15}
\caption{Accuracy of the Sample-level Unlearning (measured in other/\textcolor{violet}{unlearn} labels)}
\label{tab_sample_unlearning_acc_BERT}

\resizebox{\linewidth}{!}{
\begin{tabular}{ |c | c|cc|ccc| c cc cc cc| c |}
\toprule

& &&   &   & \multicolumn{2}{c}{\textbf{Existing SL }}  & \multicolumn{1}{c}{\textbf{\textsc{SplitWiper}}}& \multicolumn{6}{c}{\textbf{\textsc{SplitWiper+}}}  &  \\

\cmidrule{8-14}  

& & &  & \multicolumn{1}{c}{\quad$\delta$\quad}  & U-shaped   & Vanilla  & $\gamma=1$ & \textbf{$\gamma=2$} & \textbf{$\gamma=2$} & \textbf{$\gamma=3$}  &  \textbf{$\gamma=3$}  &  \textbf{$\gamma=\gamma_s$}    & \textbf{$\gamma=\gamma_s$} &  \\ 

\midrule



 \multirow{4}*{ \rotatebox{0}{\textbf{\makecell{Clients\\ with\\ same task}}} } &  \multirow{4}*{\rotatebox{90}{\textbf{\makecell{Partially
\\overlapping \\ labels}}}} & \multirow{4}*{\rotatebox{0}{AG-News}} &  L. \quad  &  \cellcolor{teal!9} & \, 
 82.95/\textcolor{violet}{82.19} & 
 85.68/\textcolor{violet}{89.08} & \cellcolor{yellow!15}85.54/\textcolor{violet}{89.39} &
 \cellcolor{yellow!15} 85.77/\textcolor{violet}{90.12} & \cellcolor{yellow!15}86.89/\textcolor{violet}{91.10}& \cellcolor{yellow!15}85.04/\textcolor{violet}{90.89} & 
 \cellcolor{yellow!15}85.69/\textcolor{violet}{90.50} & \cellcolor{yellow!15}85.47/\textcolor{violet}{91.11} & \cellcolor{yellow!15}86.39/\textcolor{violet}{91.74}   
 & \multirow{4}*{\textbf{ BERT }} 
 \\

 \cline{5-5}
  
 & &  & \multirow{3}*{U.} & \cellcolor{teal!9} 0.1 & \, 
 82.68/\textcolor{violet}{78.21} & 
 85.14/\textcolor{violet}{82.05} & \cellcolor{yellow!15}85.35/\textcolor{violet}{82.16} &\cellcolor{yellow!15} 84.96/\textcolor{violet}{82.46} & \cellcolor{yellow!15}86.04/\textcolor{violet}{82.73}& \cellcolor{yellow!15}86.15/\textcolor{violet}{81.67} & 
 \cellcolor{yellow!15}86.06/\textcolor{violet}{81.44} & \cellcolor{yellow!15}86.00/\textcolor{violet}{83.26}  & \cellcolor{yellow!15}86.52/\textcolor{violet}{83.20}  & \\

 & & & & \cellcolor{teal!9} 0.5 & \, 
 82.33/\textcolor{violet}{72.68} & 
 85.62/\textcolor{violet}{78.20} & \cellcolor{yellow!15}85.95/\textcolor{violet}{78.17} &\cellcolor{yellow!15} 85.69/\textcolor{violet}{77.95} & \cellcolor{yellow!15} 86.84/\textcolor{violet}{77.93}& 
 \cellcolor{yellow!15}85.90/\textcolor{violet}{77.68} & 
 \cellcolor{yellow!15}86.32/\textcolor{violet}{77.71} & \cellcolor{yellow!15}86.32/\textcolor{violet}{78.05}  & \cellcolor{yellow!15}87.68/\textcolor{violet}{78.26}  & \\
  
 & & &  & \cellcolor{teal!9} 0.9 & \,
 82.21/\textcolor{violet}{62.84} & 
 85.26/\textcolor{violet}{76.32} & \cellcolor{yellow!15}86.10/\textcolor{violet}{75.84} &\cellcolor{yellow!15} 86.07/\textcolor{violet}{69.21} & \cellcolor{yellow!15} 86.71/\textcolor{violet}{70.11}& 
 \cellcolor{yellow!15}86.21/\textcolor{violet}{67.16} & 
 \cellcolor{yellow!15}86.28/\textcolor{violet}{68.47} & \cellcolor{yellow!15}86.94/\textcolor{violet}{71.74}  & \cellcolor{yellow!15}87.33/\textcolor{violet}{71.97}  & \\

\midrule

\multicolumn{2}{|c}{\ding{172}}  & \ding{173} &  \multicolumn{1}{c}{\ding{174}} &  \multicolumn{1}{c|}{\ding{175}} &  \multicolumn{4}{c}{}  $\varepsilon=0.1$ & $\varepsilon=0.2$  & $\varepsilon=0.1$ & $\varepsilon=0.2$ & $\varepsilon=0.1$ & $\varepsilon=0.2$  & \ding{176} \\


\end{tabular}
}

\begin{tablenotes}
      \scriptsize
      \item[] \textit{Notation:}  \ding{172} \textbf{Data distribution}; \ding{173} \textbf{Dataset}; \ding{174} \textbf{Scenario} (L. for learning, U. for unlearning);  \ding{175} 
      \textbf{Unlearned sample proportion} (i.e., $\delta$); \ding{176} \textbf{Model}.
\end{tablenotes}

\end{table*}

\begin{table*}[!]
\centering
\setlength\tabcolsep{2pt}
\renewcommand\arraystretch{1.15}
\caption{Hit Rate (\%) of Member Inference in Sample-level Unlearning Tasks}
\label{tab_sample_member_inference_BERT}

\resizebox{\linewidth}{!}{
\begin{tabular}{ | c | c|cc |c cc | c cc cc cc| c| }
\toprule

\multicolumn{2}{|c}{\multirow{2}{*}{\textbf{Data distribution}}} & \multicolumn{1}{c}{\multirow{2}{*}{\textbf{Dataset}}}  & 
\multicolumn{1}{c}{\multirow{2}{*}{\textbf{Scenario}}}  &   
& \multicolumn{2}{c}{\,\textbf{Existing SL}} & \multicolumn{1}{c}{\textbf{\textsc{SplitWiper}}}  &
\multicolumn{6}{c}{\textbf{\textsc{SplitWiper+}}}  &  
\multicolumn{1}{c|}{\multirow{2}{*}{\,\textbf{Model}}} 
\\

\cmidrule{8-14}  

& & &  & \multicolumn{1}{c}{\,\quad$\delta$\quad\,}  &  U-shaped & Vanilla & $\gamma=1$ & \textbf{$\gamma=2$} & \textbf{$\gamma=2$} & \textbf{$\gamma=3$}  &  \textbf{$\gamma=3$}  &  \textbf{$\gamma=\gamma_s$}    & \textbf{$\gamma=\gamma_s$} &  \\ 

\midrule



 \multirow{4}*{ \rotatebox{0}{\textbf{\makecell{Clients  with\\ same task}}} } &  \multirow{4}*{\rotatebox{0}{\textbf{\makecell{Partially overlapping labels}}}} & \multirow{4}*{\rotatebox{0}{AG-News}} &  \quad Learning \quad  &  \cellcolor{teal!9} & \,
 97.32 & 
 99.97 & 
 \cellcolor{yellow!15}  99.96&
 \cellcolor{yellow!15}  99.79& 
 \cellcolor{yellow!15}  99.86& 
 \cellcolor{yellow!15}  99.90& 
 \cellcolor{yellow!15}  99.99& 
 \cellcolor{yellow!15}  99.97& 
 \cellcolor{yellow!15}  99.97& 
 \multirow{4}*{ \textbf{ BERT }} 
 \\

 \cline{5-5}
  
 & &  & \multirow{3}*{Unlearning} & \cellcolor{teal!9} 0.1 & \,
 90.58 & 
 99.07 & 
 \cellcolor{yellow!15}  98.40&
 \cellcolor{yellow!15}  97.15& 
 \cellcolor{yellow!15}  97.77& 
 \cellcolor{yellow!15}  96.18& 
 \cellcolor{yellow!15}  96.22& 
 \cellcolor{yellow!15}  97.63& 
 \cellcolor{yellow!15}  97.83& 
 \\

 & & & & \cellcolor{teal!9} 0.5 & \,
 79.95 & 
 90.83 & 
 \cellcolor{yellow!15}  90.37&
 \cellcolor{yellow!15}  89.26& 
 \cellcolor{yellow!15}  89.10& 
 \cellcolor{yellow!15}  88.63& 
 \cellcolor{yellow!15}  89.00& 
 \cellcolor{yellow!15}  89.13& 
 \cellcolor{yellow!15}  90.08& 
 \\
  
 & & &  & \cellcolor{teal!9} 0.9  & \,
 63.26 & 
 76.40 & 
 \cellcolor{yellow!15}  78.03&
 \cellcolor{yellow!15}  71.36& 
 \cellcolor{yellow!15}  72.37& 
 \cellcolor{yellow!15}  69.53& 
 \cellcolor{yellow!15}  70.64& 
 \cellcolor{yellow!15}  72.53& 
 \cellcolor{yellow!15}  73.45& 
 \\

\midrule

\multicolumn{2}{|c}{ }  &  &  \multicolumn{1}{c}{} &  \multicolumn{1}{|c|}{\ding{172}} &  \multicolumn{4}{c}{}  $\varepsilon=0.1$ & $\varepsilon=0.2$  & $\varepsilon=0.1$ & $\varepsilon=0.2$ & $\varepsilon=0.1$ & $\varepsilon=0.2$  &  \\


\end{tabular}
}
\begin{tablenotes}
      \scriptsize
      \item[] \textit{Notation:}  
      \ding{172} 
     \textbf{Unlearned sample proportion} (i.e., $\delta$). 
     \end{tablenotes}

\end{table*}

\begin{table*}[t]
\centering
\setlength\tabcolsep{2pt}
\renewcommand\arraystretch{1.15}
\caption{Accuracy of the Sample-level Unlearning (measured in other/\textcolor{violet}{unlearn} labels)}
\label{tab_sample_unlearning_acc}

\resizebox{\linewidth}{!}{
\begin{tabular}{ |c | c|cc|ccc| c cc cc cc| c |}
\toprule

& &&   &   & \multicolumn{2}{c}{\textbf{Existing SL }}  & \multicolumn{1}{c}{\textbf{\textsc{SplitWiper}}}& \multicolumn{6}{c}{\textbf{\textsc{SplitWiper+}}}  &  \\

\cmidrule{8-14}  

& & &  & \multicolumn{1}{c}{\quad$\delta$\quad}  & U-shaped   & Vanilla  & $\gamma=1$ & \textbf{$\gamma=2$} & \textbf{$\gamma=2$} & \textbf{$\gamma=3$}  &  \textbf{$\gamma=3$}  &  \textbf{$\gamma=\gamma_s$}    & \textbf{$\gamma=\gamma_s$} &  \\ 

\midrule



 \multirow{24}*{ \rotatebox{0}{\textbf{\makecell{Clients\\ with\\ same task}}} } &  \multirow{8}*{\rotatebox{90}{\textbf{\makecell{Homo labels\\Hetero volumes}}}} & \multirow{4}*{\rotatebox{0}{CIFAR-10}} &  L. \quad  &  \cellcolor{teal!9} & \, 
 77.04/\textcolor{violet}{90.30} & 
 78.08/\textcolor{violet}{88.60} & \cellcolor{yellow!15}79.86/\textcolor{violet}{89.60} &
 \cellcolor{yellow!15} 80.65/\textcolor{violet}{89.10} & \cellcolor{yellow!15}81.08/\textcolor{violet}{90.05}& \cellcolor{yellow!15}80.4/\textcolor{violet}{88.20} & 
 \cellcolor{yellow!15}80.82/\textcolor{violet}{90.10} & \cellcolor{yellow!15}80.73/\textcolor{violet}{89.30} & \cellcolor{yellow!15}81.06/\textcolor{violet}{90.80}   
 & \multirow{28}*{\textbf{ Resnet }} 
 \\
\end{tabular}
}

\begin{tablenotes}
      \scriptsize
      \item[] \textit{Notation:}  \ding{172} \textbf{Data distribution}; \ding{173} \textbf{Dataset}; \ding{174} \textbf{Scenario} (L. for learning, U. for unlearning);  \ding{175} 
      \textbf{Unlearned sample proportion} (i.e., $\delta$); \ding{176} \textbf{Model}.
     \end{tablenotes}

\end{table*}

\section{Experiments on Text Modality: AG-News with BERT}
\label{appendix_exp_BERT}

To further demonstrate the generality of our framework, \textsc{SplitWiper} and \textsc{SplitWiper+}, we extend our evaluation beyond vision tasks. 
Unlike methods tailored to specific datasets or architectures, our frameworks are designed to support a wide range of data modalities and can be seamlessly integrated with state-of-the-art models. 
While Section~\ref{sec:evaluation} presents strong performance on standard image benchmarks such as CIFAR-10, CIFAR-100, and MNIST, we supplement our analysis with experiments on a non-vision dataset: AG-News~\cite{zhang2015character}, using the mini-BERT~\cite{dong2023rai2} architecture.

Following Section~\ref{subsec: exp_setting}, we conduct class-level unlearning experiments to evaluate the efficiency and effectiveness of our methods. 
Both \textsc{SplitWiper} and \textsc{SplitWiper+} (Table~\ref{tab_time_BERT}), achieve 0\% accuracy on the unlearned label, indicating complete forgetting. 
Importantly, during the unlearning phase, only the client requesting unlearning is involved, while all other clients remain unaffected. 

\begin{figure}[b]
\setlength\abovecaptionskip{3pt} 
\setlength\belowcaptionskip{-0.3cm}
    \centering
    \subfigure[Perfect clustering accuracy.\label{fig_birch_perfect}]{
    \includegraphics[width=0.47\linewidth]{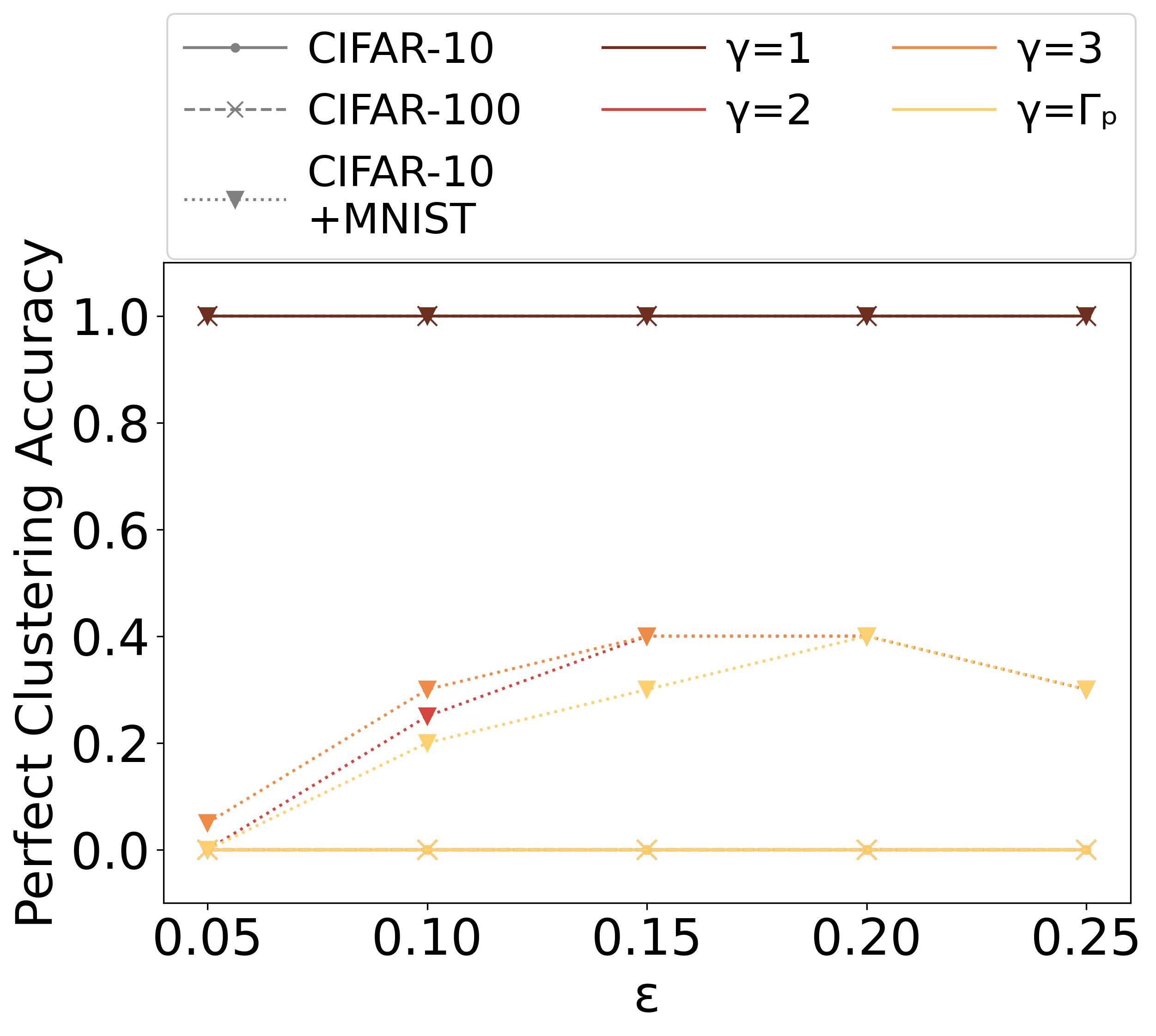}
    }
    \subfigure[Inclusive clustering accuracy.\label{fig_birch_inclusive}]{
    \includegraphics[width=0.47\linewidth]{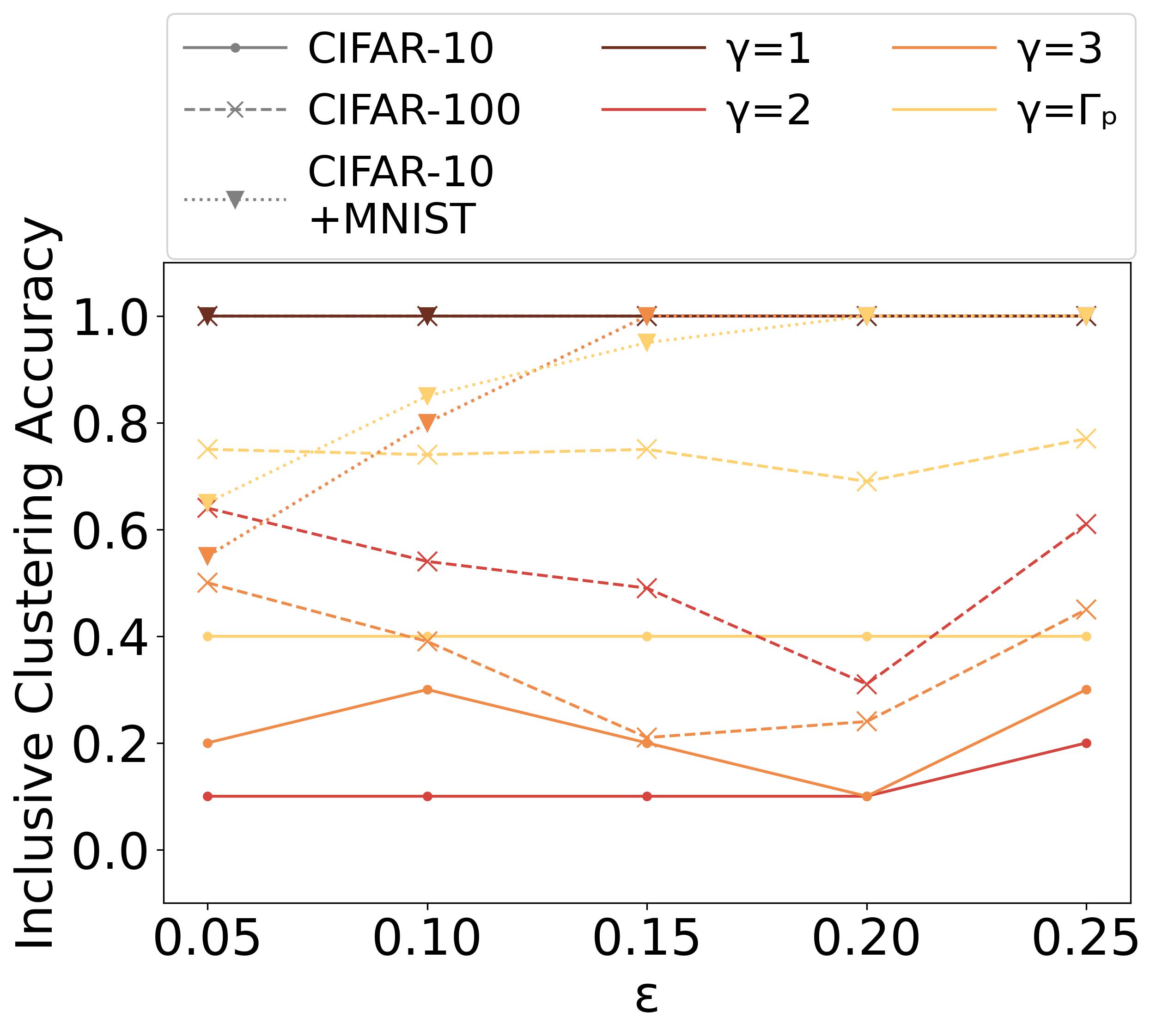}
    }
\DeclareGraphicsExtensions.
    \caption{BIRCH perfect clustering accuracy and inclusive clustering accuracy with Variable $\gamma$ and $\varepsilon$.}
    \label{fig_birch}
\end{figure}

We further evaluate sample-level unlearning through two lenses: unlearning accuracy and member inference hit rate.
As shown in Table~\ref{tab_sample_unlearning_acc_BERT}, both \textsc{SplitWiper} and \textsc{SplitWiper+} remain effective even when unlearning individual data samples. Compared to traditional SL retraining-based methods, our frameworks achieve competitive unlearning accuracy with reduced cost. 
Moreover, Table~\ref{tab_sample_member_inference_BERT} shows that our \textsc{SplitWiper+} substantially lowers the hit rate of membership inference attacks, approaching the level of U-shaped SL. This highlights the effectiveness of our label protection mechanism in enhancing privacy against inference attacks.

These supplementary experiments validate the generalizability of our frameworks across different data modalities and model architectures. By extending from image to text domains and integrating with different model types like BERT, we demonstrate that \textsc{SplitWiper} and \textsc{SplitWiper+} are not confined to specific tasks or datasets. 
The results reinforce practical advantages and flexibility of our design, which makes our frameworks well-suited for real-world applications where model diversity and data heterogeneity are common.

\section{Experiments on Label Privacy under BIRCH-based Inference Attacks}
\label{appendix_exp_BIRCH}

To further validate the robustness of \textsc{SplitWiper+} in protecting label privacy and the privacy of intermediate representations, we extend our evaluation beyond K-Means by introducing BIRCH~\cite{zhang1996birch}.
BIRCH is a hierarchical clustering method and provides a complementary perspective with different clustering assumptions. This allows us to better assess whether \textsc{SplitWiper+} can defend against a broader class of unsupervised inference attacks.

To maintain consistency with the clustering-based analysis in Section~\ref{subsection:exp_privacy}, we continue to evaluate label privacy under the BIRCH clustering algorithm using the same two metrics: perfect clustering accuracy and inclusive clustering accuracy. 
These metrics capture both strict and relaxed privacy leakage scenarios, where perfect clustering requires exact group alignment, and inclusive clustering tolerates partial overlap.

As shown in Figure~\ref{fig_birch}, the results obtained with BIRCH clustering are consistent with those reported using KMeans. Specifically, when no label expansion or differential privacy is applied, such as in \textsc{SplitWiper}, vanilla SL, and U-shaped SL, the attacker can cluster intermediate outputs with near-perfect accuracy. This indicates significant vulnerability in privacy of those frameworks.
By contrast, \textsc{SplitWiper+}, which employs label expansion strategies and injects DP noise into intermediate values, substantially reduces both perfect and inclusive clustering accuracy.
Notably, perfect  clustering accuracy drops to 0\% on CIFAR-10 and CIFAR-100, indicating that no original label group could be correctly reconstructed. 
This result confirms that effectiveness of \textsc{SplitWiper+} in mitigating clustering-based inference attacks, rendering label quantities and identities unrecognizable.

\begin{table*}[t]
\centering
\setlength\tabcolsep{2pt}
\renewcommand\arraystretch{1.15}
\caption{Hit Rate (\%) of Member Inference in Sample-level Unlearning Tasks}
\label{tab_sample_member_inference}

\resizebox{\linewidth}{!}{
\begin{tabular}{ | c | c|cc |c cc | c cc cc cc| c| }
\toprule

\multicolumn{2}{|c}{\multirow{2}{*}{\textbf{Data distribution}}} & \multicolumn{1}{c}{\multirow{2}{*}{\textbf{Dataset}}}  & 
\multicolumn{1}{c}{\multirow{2}{*}{\textbf{Scenario}}}  &   
& \multicolumn{2}{c}{\,\textbf{Existing SL}} & \multicolumn{1}{c}{\textbf{\textsc{SplitWiper}}}  &
\multicolumn{6}{c}{\textbf{\textsc{SplitWiper+}}}  &  
\multicolumn{1}{c|}{\multirow{2}{*}{\,\textbf{Model}}} 
\\

\cmidrule{8-14}  

& & &  & \multicolumn{1}{c}{\,\quad$\delta$\quad\,}  &  U-shaped & Vanilla & $\gamma=1$ & \textbf{$\gamma=2$} & \textbf{$\gamma=2$} & \textbf{$\gamma=3$}  &  \textbf{$\gamma=3$}  &  \textbf{$\gamma=\gamma_s$}    & \textbf{$\gamma=\gamma_s$} &  \\ 

\midrule




 \multirow{24}*{ \rotatebox{0}{\textbf{\makecell{Clients  with\\ same task}}} } &  \multirow{8}*{\rotatebox{0}{\textbf{\makecell{Homogeneous labels, \\ heterogeneous volumes}}}} & \multirow{4}*{\rotatebox{0}{CIFAR-10}} &  \quad Learning \quad  &  \cellcolor{teal!9} & \,
 86.94 & 
 97.48 & 
 \cellcolor{yellow!15} 90.58 &
 \cellcolor{yellow!15} 89.42 & 
 \cellcolor{yellow!15} 90.62 & 
 \cellcolor{yellow!15} 88.68 & 
 \cellcolor{yellow!15} 89.86 & 
 \cellcolor{yellow!15} 90.54 & 
 \cellcolor{yellow!15} 90.88 & 
 \multirow{28}*{ \textbf{ Resnet }} 
 \\

 \cline{5-5}
  
 & &  & \multirow{3}*{Unlearning} & \cellcolor{teal!9} 0.1 & \,
 68.60 & 
 78.40 & 
 \cellcolor{yellow!15} 77.00 &
 \cellcolor{yellow!15} 73.60 & 
 \cellcolor{yellow!15} 74.40 & 
 \cellcolor{yellow!15} 70.80 & 
 \cellcolor{yellow!15} 72.00 & 
 \cellcolor{yellow!15} 71.80 & 
 \cellcolor{yellow!15} 76.40 & 
 \\
\end{tabular}
}
\begin{tablenotes}
      \scriptsize
      \item[] \textit{Notation:}  
      \ding{172} 
     \textbf{Unlearned sample proportion} (i.e., $\delta$). 
     \end{tablenotes}

\end{table*}

\textsc{SplitWiper+} is a dedicated framework for protecting the true number of labels, which is an often overlooked yet critical privacy issue in SL. 
It introduces a novel label expansion strategy that obscures the original label structure, making it significantly harder for the server to infer sensitive class-level information. 
In combination with DP applied to intermediate values, \textsc{SplitWiper+} delivers comprehensive privacy protection by simultaneously shielding both semantic label information and representational features. This dual-defense approach positions \textsc{SplitWiper+} as a robust and generalizable solution for privacy-preserving SL.

\section{Evaluating Convergence}
\label{appendix_exp_convergence}

Establishing theoretical proof of convergence is challenging~\cite{chen2022graph} for \textsc{SplitWiper} and \textsc{SplitWiper+}. Therefore, we rely on empirical experiments to assess convergence by tracking the number of nodes that change in each iteration.
The experimental data indicates that the proportion of moved nodes steadily approaches zero within 30 epochs when using the CIFAR-10 dataset or a combination of CIFAR-10 and MNIST; for CIFAR-100, this occurs within 50 epochs.

Fig.~\ref{fig_convergence} illustrates how test accuracy varies with increasing epochs for our frameworks compared to traditional SL frameworks, using the VGG and ResNet18 architectures across different datasets with heterogeneous label distributions among clients. The proposed \textsc{SplitWiper} and \textsc{SplitWiper+} with various $\gamma$ and $\varepsilon$ consistently achieve higher accuracy than Vanilla SL and U-shaped SL across all datasets. It is also evident that test accuracy for all SL frameworks converges after a certain number of epochs, with CIFAR-100 requiring more epochs to reach convergence due to its larger label set. 
Since this trend is consistent across all scenarios and model architectures, we present one representative case with the highest degree of non-IID data distribution in Fig.~\ref{fig_convergence} for analysis.

\section*{Ethics Considerations}

Our experiments are conducted on a local testing platform using open-source libraries and public datasets, without connecting to any external or live systems. These experiments do not involve any issues related to animals, human beings, the environment, healthcare, or military factors. As of this writing, these studies have not had real-world impact as they have only been utilized in our experiments. Consequently, we have addressed numerous ethical considerations in our experimental design, strictly adhering to the ethical principles outlined in the Menlo Report.


\end{document}